\documentstyle[prl,twocolumn,epsf,floats,aps]{revtex}
\renewcommand{\narrowtext}{\begin{multicols}{2} \global\columnwidth20.5pc}
\renewcommand{\widetext}{\end{multicols} \global\columnwidth42.5pc}

\newcommand{\be}{\begin{equation}}
\newcommand{\ee}{\end{equation}}
\newcommand{\bea}{\begin{eqnarray}}
\newcommand{\eea}{\end{eqnarray}}
\newcommand{\non}{\nonumber}
\newcommand{\bk}{{\bf k}}
\newcommand{\br}{{\bf r}}
\newcommand{\up}{\uparrow}
\newcommand{\down}{\downarrow}
\newcommand{\ktxx}{\tilde{\kappa}_{xx}}
\newcommand{\ktxy}{\tilde{\kappa}_{xy}}
\newcommand{\bE}{{\bf E}}

\begin{document}

\twocolumn[\hsize\textwidth\columnwidth\hsize\csname
@twocolumnfalse\endcsname

\title{Paired states of fermions in two dimensions with breaking of parity
and time-reversal symmetries, and the fractional quantum Hall effect}
\author{N. Read and Dmitry Green}
\address{Departments of Physics and Applied Physics, Yale
University,\\
P.O. Box 208120, New Haven, Connecticut 06520-8120}
\date{\today}
\maketitle
\begin{abstract}
We analyze pairing of fermions in two dimensions for fully-gapped cases with
broken parity (P) and time-reversal (T), especially cases in which the gap
function is an orbital angular momentum ($l$) eigenstate, in particular $l=-1$
(p-wave, spinless or spin-triplet) and $l=-2$ (d-wave, spin-singlet). For
$l\neq0$, these fall into two phases, weak and strong pairing, which may be
distinguished topologically. In the cases with conserved spin, we derive
explicitly the Hall conductivity for spin as the corresponding topological
invariant. For the spinless p-wave case, the weak-pairing phase has a pair
wavefunction that is asympototically the same as that in the Moore-Read
(Pfaffian) quantum Hall state, and we argue that its other properties (edge
states, quasihole and toroidal ground states) are also the same, indicating
that nonabelian statistics is a {\em generic} property of such a paired phase.
The strong-pairing phase is an abelian state, and the transition between the
two phases involves a bulk Majorana fermion, the mass of which changes sign at
the transition. For the d-wave case, we argue that the Haldane-Rezayi state is
not the generic behavior of a phase but describes the asymptotics at the
critical point between weak and strong pairing, and has gapless fermion
excitations in the bulk. In this case the weak-pairing phase is an abelian
phase which has been considered previously. In the p-wave case with an unbroken
U(1) symmetry, which can be applied to the double layer quantum Hall problem,
the weak-pairing phase has the properties of the 331 state, and with nonzero
tunneling there is a transition to the Moore-Read phase. The effects of
disorder on noninteracting quasiparticles are considered. The gapped phases
survive, but there is an intermediate thermally-conducting phase in the
spinless p-wave case, in which the quasiparticles are extended.

\end{abstract}
]

%%%%%%%%%%%%%%%%%%%%%%%%%%%%%%%%%%%%%%%%%%%%%%%%%%%%%
\section{Introduction}
\label{introduction}

Most theories of superconductivity, or more generally of superfluidity in
fermion systems, depend on the concept of a paired ground state introduced
by Bardeen, Cooper and Schrieffer (BCS) in 1957 \cite{bcs,schrieffer}. The
ground state may be thought of loosely as a Bose condensate of pairs of
particles, since such a pair can be viewed as a boson. Within BCS mean field
theory, such a state forms whenever the interaction between the particles is
attractive. For weak attractive interaction the elementary excitations are
fermions (BCS quasiparticles), which can be created by adding or removing
particles {}from the system, or in even numbers by breaking the pairs in the
ground state, and the minimum excitation energy occurs at fermion wavevector
near $k_F$, the Fermi surface that would exist in the normal Fermi liquid
state at the same density of particles. There is also a collective mode, which
is a gapless phonon-like mode in the absence of long-range interactions
between the particles. This mode would also be present if one considered the
pairs as elementary bosons, and would be the only elementary low-energy
excitation in that case. If the attractive interaction becomes strong, the
energy to break a pair becomes large, and at all lower energies the system
behaves like a Bose fluid of pairs. In the original BCS treatment, each pair of
particles was in a relative s-wave ($l=0$) state, and the minimum energy for
a fermion excitation is then always nonzero. No phase transition occurs as the
coupling strength is increased to reach the Bose fluid regime.

Not long after BCS, the theory was generalized to nonzero relative angular
momentum ($l$) pairing, and after a long search, p-wave pairing was observed
in He$^3$ \cite{vollwolf}. It is believed that d-wave pairing occurs in heavy
fermion and high $T_c$ superconductors. Some nonzero $l$ paired states
generally have vanishing energy gap at some points on the Fermi surface
(for weak coupling), while others do not. While the absence of a transition is
well-known in the s-wave case, it seems to be less well known that in some
of these other cases, there is a phase transition as the coupling becomes
more strongly attractive. One reason for this is that the strong-coupling
regime must have a gap for all BCS quasiparticle excitations. But even when
the weak coupling regime is fully gapped, there may be a transition, and
these will be considered in this paper, in two dimensions.

In the paired states with non-zero $l$ there are many exotic
pheneomena, especially in the p-wave case, due to the breaking of
spin-rotation and spatial-rotation symmetries. These include
textures in the order parameters for the pairing, such as domain
walls, and quasiparticle excitations of vanishing excitation
energy on these textures (zero modes) (these are reviewed in Ref.\
\cite{vollwolf}). In transport, there may be Hall-type
conductivities for conserved quantities, such as spin and energy,
which are possible because of the breaking of both parity (P) and
time reversal (T) symmetries. The breaking of these symmetries,
and topological aspects of the paired state, are more crucial for
the ocurrence of these effects than is the angular momentum of the
pairing; the pairing need not be in a definite angular momentum
state. Many of these effects have been discussed in remarkable
papers by Volovik, of which a few are Refs.\
\cite{volovik88,volyak,volovik90,vol90lett,volovik92}. These are
related to effects we will explore in this paper. We will make an
effort to separate the effects related to breaking continuous
symmetries spontaneously, which leads to familiar Goldstone mode
physics, {}from those connected with topological effects,
quasiparticle zero modes and Hall-type responses for unbroken
symmetries.

In this paper we will make extensive use of the methods for BCS paired
states, and consider the transitions between the weak and strong coupling
regimes in two dimensions. In the weak-coupling regime, exotic phenomena
are possible when parity and time reversal are broken. The results are applied
to the fractional quantum Hall effect (FQHE) by using the composite fermion
approach, to be reviewed below. We also consider effects of disorder on
the phases and transitions, also within BCS mean field theory. In each
Section, we try to make the initial discussion general and accessible for
workers in many-body theory and superconductivity, before specializing to
applications to the FQHE. In the remainder of this Introduction we will
give an overview of the background and of the results of this paper.

We now review some background in the FQHE \cite{book}. The original Laughlin
states \cite{laugh} occur at filling factors $\nu=1/q$, with $q$ odd
(the filling factor is defined as $\nu=n \Phi_0/B$, where $n$ is the density
in two dimensions, $B$ is the magnitude of the magnetic field, and
$\Phi_0=hc/e$ is the flux quantum). An early idea of Halperin
\cite{halp83} was to generalize the Laughlin states by assuming
that under some conditions, the electrons could form pairs, which as charge-2
bosons could form a Laughlin state. A variety of such states were proposed
\cite{halp83,dmh,hr,mr}. Since the Laughlin states for bosons occur at filling
factors for the lowest Landau level (LLL) $\nu_b=1/m$, with $m$ even, and the
filling factor for the electrons is related to that for the bosons by
$\nu=4\nu_b$, the electron filling factor is either of the form $\nu=1/q$ ($q$
an integer), or $\nu=2/q$, $q$ odd. In particular, these values include
$\nu=1/2$, $1/4$, \ldots, which do not correspond to incompressible states in
the usual hierarchy scheme \cite{hald83}, because the filling factors in the
latter always have odd denominators (when common factors have been removed).

The relation of the paired states in the FQHE to those in superfluidity theory
becomes much closer once one introduces the notion of composite particles
\cite{g,gm,read87,laughan,zhk,read89,jain,fishlee,lopfrad,hlr,read94,sm,dhlee,ph,read98}.
A simple, direct formal approach is to use a flux attachment or Chern-Simons
transformation (see e.g.\ Ref.\ \cite{hlr} in particular), which represents
each particle as (in the case of most interest) a fermion plus an integer
number $\tilde{\phi}$ of $\delta$-function flux tubes. After the
transformation, the system can be represented by an action that includes a
Chern-Simons term for a U(1) gauge field, that couples to the Fermi field.
We refer to this as the CS fermion approach. The net magnetic field seen by
the fermions is the sum of the external field and the $\delta$-function fluxes
on the other particles. In a mean field treatment, given a uniform density of
particles, this produces a net average field that vanishes when the filling
factor is $\nu=1/\tilde{\phi}$. In this case the fermions can form a Fermi sea
\cite{hlr}, or they could form a BCS paired state. Some of the existing paired
FQHE trial wavefunctions can be interpreted this way, as pointed out in
Ref.\ \cite{mr}, and others can be constructed by analogy \cite{mr}.

A more physical way of looking at the formation of the composite particles,
particularly when they are considered as the elementary excitations of the
system, is as bound states of one of the underlying particles (or particles
for short), and $\tilde{\phi}$ vortices in the particle wavefunction
\cite{read89,read94}. The bound states, which correspond to the CS fermions,
again behave as fermions in zero net field if the particles obey Fermi
statistics and $\tilde{\phi}$ is even, (or if the particles obey Bose
statistics and $\tilde{\phi}$ is odd) and $\nu=1/\tilde{\phi}$. Note that we
will consistently use the term ``particle'' for the underlying particles, and
``fermion'' for the CS or composite fermions (bound states). Some
statements apply also when the transformed particles are bosons (obtained by
interchanging the words ``even'' and ``odd'' in the preceding definitions), in
which case we refer to composite particles. It is generally more important to
keep track of the statistics and net magnetic field seen by the composite
particles than those of the underlying particles. Recent work has formalized
the bound state picture, and improved our understanding
\cite{sm,dhlee,ph,read98}. However, the results of the CS approach remain
valid, and because that approach is simple to use, and we will mainly require
only a mean field picture here, we assume that that is the approach we are
implicitly using.

The Laughlin states can be viewed as Bose condensates of composite bosons in
zero net magnetic field \cite{read87,zhk,read89}. Because the bosons are
coupled to a gauge field (in the CS approach, the CS gauge field), vortex
excitations cost only a finite energy, but there is still an effective
Meissner effect for the CS gauge field. Because the flux of the CS gauge
field is related to the particle density, a vortex carries a fractional charge
and corresponds to Laughlin's fractionally-charged quasiparticles \cite{laugh}
(we refer to such excitations as FQHE quasiparticles). Hence the Meissner
effect in the superfluid becomes the incompressiblity of the FQHE state (there
is of course no Meissner effect or superfluidity in the response to the
electromagnetic field). Similarly, when pairing of composite fermions occurs
in zero net magnetic field, the state becomes incompressible \cite{mr}. In
contrast, the Fermi liquid state of Halperin, Lee and one of the authors
\cite{hlr} has no Meissner effect for the CS field and is compressible
\cite{hlr}.

The wavefunction proposed by Haldane and Rezayi (HR) in Ref.\ \cite{hr} is
a spin-singlet paired state, which can be interpreted as a d-wave paired
state of composite fermions \cite{mr}. Moore and Read (MR) \cite{mr}
proposed a p-wave paired state (the ``Pfaffian state'') of spinless
electrons with a structure analogous to the HR state. Both states can
occur for filling factors $\nu=1/2$, $1/4$, \ldots. The HR state was
proposed as an explanation for the observed $\nu=5/2$ QH state
\cite{willett}, which collapses when a parallel component of the magnetic
field is applied, suggesting that it is a spin singlet. However, it was
also proposed later that the $5/2$ state is the MR state \cite{gww}. In
both proposals, the LLL is filled with electrons of both spins, and the
paired FQHE state describes only the electrons in the first excited LL.
The latter proposal is supported by recent numerical work \cite{morf,rhnew}.
The collapse of the state under a parallel magnetic field must then be due to
another mechanism, involving the effects of the finite thickness of the
single-particle wavefunctions in the direction perpendicular to the
two-dimensional layer, which is poorly understood at present. Another
paired state with a similar interpretation is the 331 state \cite{halp83},
which can be viewed as a p-wave paired state of two-component composite
fermions \cite{hr,gww2,halpnewport}. It is likely that this is closely
related to a FQHE state observed in double-layer and single-thick-layer
systems at $\nu=1/2$ \cite{expt,he}.

Moore and Read \cite{mr} suggested that nonabelian statistics
might occur in QH states, and the Pfaffian state was proposed as an
example. Nonabelian statistics means that the space of states for a
collection of quasiparticles at fixed positions and quantum numbers is
degenerate, and when quasiparticles are exchanged adiabatically (for which
we need the system to have an energy gap for all excitations), the effect
is a matrix operation on this space of degenerate states. This generalizes
the idea of fractional statistics, in which the effect of an exchange is a
phase factor and the states as specified are nondegenerate; when the
phase for an elementary exchange is $\pm 1$, one has bosons or fermions.
The arguments in MR that this would occur were based heavily on
the identification of the many-particle wavefunctions in the FQHE
as chiral correlators (conformal blocks) in two-dimensional conformal
field theory, which possess similar properties under monodromy (analytic
continuation in their arguments). It was expected that an effective field
theory description of these effects would be based on nonabelian
Chern-Simons theories, which are known to be connected with conformal
field theory, and to lead to nonabelian statistics \cite{wit89}.
In the MR state, and other paired states, there are, apart {}from the
usual Laughlin quasiparticles which contain a single flux quantum, also
finite energy vortex excitations containing a half flux quantum \cite{mr},
and it is these which, in the MR state, are supposed to possess nonabelian
statistics properties.

Evidence for nonabelian statistics in the MR (Pfaffian) state accumulated
in later work \cite{wen3,wwh,milr,nayak,rr}, which investigated the
spectrum of edge states, quasihole states, and ground states on the torus
(periodic boundary conditions), all of which were obtained as the
zero-energy states for the three-body Hamiltonian of Greiter {\it et al.},
for which the MR state is the exact unique ground state \cite{gww}.
The states found agreed precisely with the expectations based on conformal
field theory. There was also evidence for similar effects in the HR state
\cite{wwh,milr,rr}, however the interpretation was problematic because the
natural conformal field theory for the bulk wavefunctions is nonunitary
and therefore cannot directly describe the edge excitations, as it does in
other cases such as the MR state. Some solutions to this were proposed
\cite{milr,gfn}. Explicit derivations of nonabelian statistics and of
effective theories have been obtained later for the MR state
\cite{gn,frad}. The 331 state \cite{halp83} is an abelian state, which
can be viewed as a generalized hierarchy state \cite{read90,blokwen}, as
is evident {}from the plasma form of the state, and these two descriptions
are related by a bosonization mapping \cite{milr,rr}.
The hierarchy states and their generalizations possess abelian statistics
properties, which can be characterized by a (Bravais) lattice
\cite{read90,blokwen}. Thus incompressible FQHE states in general can be
divided into two classes, termed abelian and nonabelian. It is clear that
in Halperin's picture \cite{halp83} of bound {\em electron} pairs which
form a Laughlin state of charge 2 bosons, the properties will be abelian
and are simply described by a one-dimensional lattice, in the language of
Ref.\ \cite{read90} (they are the simplest examples of a class of abelian
states in which the objects that Bose condense contain more than one
electron plus some vortices, while the hierarchy states, and all
generalizations considered in Ref.\ \cite{read90}, have condensates
involving single electrons).

In spite of the work that has been done, one may still ask questions such
as what is the microscopic mechanism, in terms of composite fermions, for
the degeneracies of FQHE quasiparticle states that is the basis for
nonabelian statistics, and whether it is robust against changes in the
Hamiltonian. A similar question is about the effects of disorder. These
will be addressed in this paper, by a direct and simple analysis of the
paired states using BCS mean field theory, and developments such as the
Bogoliubov-de Gennes equations \cite{degennes}. We find that the
nontrivial paired FQHE states are related to the weak-coupling regime (or
more accurately, the weak-pairing phase); in particular the MR and 331 states
have wavefunctions that contain the generic long-distance behavior in
spinless and spin-triplet p-wave weak-pairing phases, respectively. In
contrast, the strong-coupling regimes, or strong-pairing phases, lead in
the FQHE to the Halperin type behavior. There is a similar picture in the
spin-singlet d-wave case, except that the HR state, which might have been
expected to represent the weak-pairing phase, is in fact at the phase
transition, and therefore is gapless in the bulk. The weak-pairing phases
are topologically nontrivial, and possess edge states and nontrivial
quantum numbers on the vortices (FQHE quasiparticles); the spinless p-wave
case (MR phase) is nonabelian, while the spinful p-wave case (with
unbroken U(1) symmetry), and the spin-singlet d-wave case are abelian
states, which we characterise. We also consider the effect of tunneling on
the double layer system which is the best candidate for realising the 331
state, and show that the phase diagram includes a MR phase.
The theory also leads to a description of the critical theories at the
transitions, at least within a mean field picture. The role of
fluctuations, and the full effective field theories at these transitions,
remain to be understood.

We also consider disorder within the same approximation, making
use of recent results on noninteracting BCS quasiparticles with
disorder
\cite{oppermann,az,sfbn,bundschuh,senthil1,chalker,senthil2,glr},
and in particular we find that in the spinless p-wave case, there
can be an intermediate phase with the thermal properties of a
metal, between the two localized phases which correspond to the
weak (MR) and strong-pairing phases of the pure case. A disordered
version of the MR phase still exists in the presence of disorder,
though its properties, including nonabelian statistics, may become
more subtle.

One further issue which we discuss is the transport coefficients
of Hall type for various conserved quantities, especially spin and
energy. Concentrating on the quantities that are related to
unbroken symmetries, we derive explicitly the values of these
conductivities in the spin case, for spin-singlet and triplet
states, and show that they are quantized in the sense of being
given by topological invariants (in the disordered cases, we do
not prove this directly). There have been claims that, in some
sense or other, the ordinary Hall conductivity for charge
(particle number) transport takes a nonzero quantized value in a
He$^3$ film in the A phase \cite{volovik88,volovik92} and in a
d$_{x^2-y^2}+i$d$_{xy}$ (i.e. P and T violating d-wave)
superconductor \cite{laughlin98,ish}. It seems unlikely to us that
these claims are correct, if the Hall conductivity is defined in
the usual way, as the current response to an external (or better,
to the total) electric field, taking the wavevector to zero before
the frequency. While one can set up a detailed calculation, using
a conserving approximation as in the Appendix, which duly includes
the collective mode effect in this case where the symmetry
corresponding to the transported quantity is broken, we prefer to
give here a more direct and appealing argument. This works in the
case where pairing is assumed to occur in a system of interacting
fermions of mass $m$ with Galilean invariance in zero magnetic
field, as in most models of pairing. If one considers the linear
response to an imposed uniform but finite frequency ($\omega$)
electric field, then there are well-known arguments that the
conductivity is simply
$\sigma_{\mu\nu}(\omega)=ne^2\delta_{\mu\nu}/(m\omega+i\eta)$
($\mu$, $\nu=x$, $y$, and $\eta$ is a positive infinitesimal).
This arises {}from the so-called diamagnetic term, and the
contribution of the two-point current-current function vanishes in
this limit. This is independent of interactions, and hence also of
whether the interactions produce pairing or not. The result can be
understood as the contribution of the center of mass, which is
accelerated by the applied uniform electric field, while the
relative motion of the particles is unaffected, as a consequence
of Galilean invariance. There is clearly no Hall conductivity.
(However, the similar calculation in a magnetic field produces the
standard Hall conductivity, and is an aspect of Kohn's theorem.)
We may be curious about non-Galilean invariant models, and whether
a paired state could have a quantized Hall conductivity as
claimed. But if it were quantized, it would be invariant under any
continuous change in the Hamiltonian that preserves the gap.
Hence, in the ground state of any model that can be continuously
connected to the Galilean-invariant models, the Hall conductivity
must either, if quantized, vanish, or else vary continuously and
not be quantized.

The plan of the remainder of this paper is as follows. In Section
\ref{spinless}, we first consider the ground state in a system of spinless
fermions with p-wave pairing, for the infinite plane and for periodic
boundary conditions (a torus). We show that a transition occurs between
weak and strong-pairing phases, which can be distinguished topologically
in momentum space, or by the number of ground states on the torus for even and
odd particle number. In Subsec.\ \ref{majorana}, we consider the system in
the presence of edges and vortices. We argue that there are chiral fermions
on an edge, and degeneracies due to zero modes on vortices, when these
occur in the weak-pairing phase. In Subsec.\ \ref{geom},
we show how the results for ground states can be extended to other
geometries, such as the sphere. Section \ref{spinless} as a whole shows
that the properties of the weak-pairing phase, the ground state
degeneracies, chiral edge states and degeneracies of vortices agree with
those expected in the MR phase in the FQHE. The strong-pairing phase has
the properties expected in the Halperin paired states. In Sec. \ref{331},
we consider the case of spin-triplet pairing, with applications to the
double-layer FQHE system. There is a weak-pairing phase with the
properties of the 331 state, and also a distinct phase with the properties
of the MR state. In Sec.\ \ref{dwave}, we consider spin-singlet d-wave
pairing. In Subsec.\ \ref{hr}, we argue that the HR state corresponds to
the transition point between weak- and strong-pairing, and so has gapless
fermions in the bulk. Then we analyse the generic weak-pairing d-wave
phase, and argue that it corresponds to an abelian FQHE state, with a
spin-1/2 doublet of chiral Dirac fermions on the edge, which has also been
constructed previously. We also discuss here (in Subsec.\ \ref{spinhall})
general arguments for the existence of the edge states and other effects,
based on Hall-type conductivities and induced CS actions in the bulk, for
all the paired states. In Sec.\ \ref{disorder}, we discuss the effects of
disorder on all the transitions and phases. The explicit calculations of
the Hall spin conductivity, in the pure systems, are given in the
Appendix.

A brief announcement of our results for pure systems was made in Ref.\
\cite{readgreen}. Some of the results have also been found independently
by others \cite{sfw}.

%%%%%%%%%%%%%%%%%%%%%%%%%%%%%%%%%%%%%%%%%%%%%%%%%%%%%
\section{Complex p-wave pairing of spinless or spin-polarized fermions}
\label{spinless}

In this Section, we first set up the BCS effective quasiparticle
Hamiltonian, and review its solution by Bogoliubov transformation. We show
that this leads to the existence of a transition between distinct phases,
which we label weak- and strong-pairing. They are distinguished
topologically. The weak-pairing phase is tentatively identified with the
MR phase because of its behavior in position space. This is extended to
the torus (periodic boundary conditions), where we find three ground
states for $N$ even, one for $N$ odd, in the weak-pairing phase, in
agreement with the MR state. In Subsec.\ \ref{majorana}, we show that the
BCS quasiparticles at long wavelengths near the transition are
relativistic Majorana fermions, and use this to analyze the Bogoliubov-de
Gennes equations for domain walls (edges) and vortices, again arguing
that the results agree with those obtained for the MR state. In Subsec.\
\ref{geom}, which may be omitted on a first reading, we show how p-wave
pairing on a general curved surface can be handled mathematically, and
that the ground states agree with conformal blocks, as expected {}from MR.

%%%%%%%%%%%%%%%%%%%%%%%%%%%%%%%%%%%%%%%%%%
\subsection{Weak and strong pairing phases}
\label{weakstrong}

First we recall the relevant parts of BCS mean field theory \cite{schrieffer}.
The effective Hamiltonian for the quasiparticles is
\be
K_{\rm eff}=\sum_{\bk}\left[\xi_\bk c_\bk^\dagger c_\bk +\frac{1}{2}\left(
\Delta_\bk^\ast c_{-\bk}c_\bk+\Delta_\bk c_\bk^\dagger
c_{-\bk}^\dagger\right)\right],
\ee
where $\xi_\bk=\varepsilon_\bk-\mu$ and $\varepsilon_\bk$ is the
single-particle kinetic energy and $\Delta_\bk$ is the gap function. For the
usual fermion problems, $\mu$ is the chemical potential, but may not have this
meaning in the FQHE applications. At small $\bk$, we assume
$\varepsilon_\bk\simeq k^2/2m^\ast$ where $m^\ast$ is an effective mass, and so
$-\mu$ is simply the small $\bk$ limit of $\xi_\bk$. For complex p-wave
pairing, we take $\Delta_\bk$ to be an eigenfunction of rotations in $\bk$ of
eigenvalue (two-dimensional angular momentum) $l=-1$, and thus at small $\bk$
it generically takes the form
\be
\Delta_\bk\simeq\hat{\Delta}(k_x-ik_y),
\ee
where $\hat{\Delta}$ is a constant. For large $\bk$, $\Delta_\bk$ will go to
zero. The $c_\bk$ obey $\{c_\bk,c_{\bk'}^\dagger\}=\delta_{\bk\bk'}$; we work
in a square box of side $L$, and consider the role of the boundary conditions
and more general geometries later.

The normalized ground state of $K_{\rm eff}$ has the form
\be
|\Omega\rangle={\prod_\bk}'(u_\bk+v_\bk c_\bk^\dagger
c_{-\bk}^\dagger)|0\rangle,
\label{OmegaNeven}
\ee
where $|0\rangle$ is the vacuum containing no fermions. The prime on the
product indicates that each distinct pair $\bk$, $-\bk$ is to be taken once.
(We will later consider the precise behavior at $\bk={\bf 0}$.) The functions
$u_\bk$ and $v_\bk$ are complex and obey $|u_\bk|^2+|v_\bk|^2=1$ to ensure
$\langle\Omega|\Omega\rangle=1$. They are determined by considering the
Bogoliubov transformation
\bea
\alpha_\bk&=&u_\bk c_\bk -v_\bk c_{-\bk}^\dagger,\non\\
\alpha_\bk^\dagger&=&u_\bk^\ast c_\bk^\dagger - v_\bk^\ast c_{-\bk},
\eea
so that $\{\alpha_\bk,\alpha_{\bk'}^\dagger\}=\delta_{\bk\bk'}$ and
$\alpha_\bk|\Omega\rangle=0$ for all $\bk$. By insisting that
$[\alpha_\bk,K_{\rm eff}]=E_\bk \alpha_\bk$ for all $\bk$, which implies that
\be
K_{\rm eff}=\sum_\bk E_\bk \alpha_\bk^\dagger \alpha_\bk + \hbox{\rm const},
\ee
with $E_\bk\geq0$, one obtains (the simplest form of) the Bogoliubov-de Gennes
(BdG) equations,
\bea
E_\bk u_\bk &=& \xi_\bk u_\bk - \Delta_\bk^\ast v_\bk\non\\
E_\bk v_\bk &=& -\xi_\bk v_\bk - \Delta _\bk u_\bk.
\eea
These imply that
\bea
E_\bk &=& \sqrt{\xi_\bk^2+|\Delta_\bk|^2},\label{Ek}\\
v_\bk/u_\bk &=& -(E_\bk-\xi_\bk)/\Delta_\bk^\ast,\label{voveru}\\
|u_\bk|^2 &=& \frac{1}{2}\left(1+\frac{\xi_\bk}{E_\bk}\right),\label{moduksq}\\
|v_\bk|^2 &=& \frac{1}{2}\left(1-\frac{\xi_\bk}{E_\bk}\right).\label{modvksq}
\eea
The functions $u_\bk$ and $v_\bk$ are determined only up to an overall phase
for each $\bk$, so they can be multiplied by a $\bk$-dependent phase,
$u_\bk\rightarrow e^{i\phi_\bk}u_\bk$, $v_\bk\rightarrow e^{i\phi_\bk}v_\bk$
without changing any physics. One may adopt a convention that one of $u_\bk$
and $v_\bk$ is real and positive; in either case the other must be odd and of
p-wave symmetry under rotations. We do not use such a convention explicitly
because there is no single choice that is convenient for all that follows.

Because of Fermi statistics, which imply $(c_\bk^\dagger)^2=0$, we
can rewrite the ground state (up to a phase factor) as
\be
|\Omega\rangle=\prod_\bk|u_\bk|^{1/2}\exp(\frac{1}{2}\sum_\bk
g_\bk c_\bk^\dagger c_{-\bk}^\dagger)|0\rangle,
\ee
where $g_\bk=v_\bk/u_\bk$. Then the wavefunction for the component of the state
with $N$ fermions ($N$ even) is, up to an $N$-independent
factor,
\bea
\Psi(\br_1,\ldots,\br_N)&=&\frac{1}{2^{N/2}(N/2)!}\sum_P {\rm
sgn}\, P\non\\
&&\mbox{}\times \prod_{i=1}^{N/2}g(\br_{P(2i-1)}-\br_{P(2i)}),
\label{genpfwfn}
\eea
where $g(\br)$ is the inverse Fourier transform of $g_\bk$,
\be
g(\br)=L^{-2}\sum_\bk e^{i\bk.\br}g_\bk,
\ee
and $P$ runs over all permutations of $N$ objects. (For
fermions with spin, this appears on p.\ 48 in Ref.\ \cite{schrieffer}.) The
right-hand side of Eq.\ (\ref{genpfwfn}) is a Pfaffian, which for a general
$N\times N$ matrix with elements $M_{ij}$ ($N$ even) is defined by
\be
{\rm Pf}\, M=\frac{1}{2^{N/2}(N/2)!}\sum_P {\rm
sgn}\, P \prod_{i=1}^{N/2}M_{P(2i-1)P(2i)},
\ee
or as the square root of the determinant, ${\rm Pf}\, M=\sqrt{\det M}$, for $M$
antisymmetric.

We now consider the form of the solutions to the above equations. In the usual
BCS problem, the functions $\Delta_\bk$ and $\varepsilon_\bk$ are found
self-consistently {}from the gap equation (including Hartree-Fock effects), and
$\mu$ would be determined by specifying the fermion density. However, we are
not interested in all these details, but in the nature of the possible phases
and in the transitions between them. We expect that the phases can be accessed
by changing the interactions and other parameters of the problem, but we will
not address this in detail. In particular, some phases may require that the
interactions be strong, while BCS theory is usually thought of as weak
coupling. We will nonetheless continue to use the BCS mean field equations
presented above, as these give the simplest possible view of the nature of the
phases.

{}From Eqs.\ (\ref{moduksq},\ref{modvksq}), we see that since
$E_\bk-|\xi_\bk|\rightarrow 0$ as $\bk\rightarrow{\bf 0}$, we will have one of
three possibilities for the behavior at small $\bk$, which will turn out to
govern the phases. As $\bk\rightarrow 0$, either (i) $\xi_\bk >0$, in which
case $|u_\bk|\rightarrow 1$, $|v_\bk|\rightarrow 0$, or (ii) $\xi_\bk<0$,
in which case $|u_\bk|\rightarrow 0$, $|v_\bk|\rightarrow 1$, or (iii)
$\xi_\bk\rightarrow0$, in which case $|u_\bk|$ and $|v_\bk|$ are both nonzero.
We will term the first case the strong-pairing phase, the second case
weak-pairing, while the third case, in which the dispersion relation of the
quasiparticles is gapless, $E_\bk\rightarrow 0$ as $\bk\rightarrow0$, is the
phase transition between the weak and strong pairing phases. Thus {\em for
$\mu$ positive, the system is in the weak pairing phase, for $\mu$ negative,
the strong-pairing phase}, and the transition is at $\mu=0$ within our
parametrization. The reason for these names will be discussed below.

We now discuss the two phases and the transition in more detail. We expect that
the large $\bk$ behavior of $\xi_\bk$ and $\Delta_\bk$ that would be produced
by solving the full system of equations will not be affected by the occurrence
of the transition which involves the small $\bk$'s only. Note that, at large
$\bk$, $v_\bk\rightarrow0$ and $|u_\bk|\rightarrow 1$, which ensures in
particular that the fermion number, which is governed by
$\overline{N}=\sum_\bk\overline{n}_\bk$, with
\be
\langle c_\bk^\dagger c_\bk\rangle =\overline{n}_\bk=|v_\bk|^2,
\ee
will converge. Also we assume that $E_\bk$ does not vanish at any other $\bk$,
which is generically the case in the $l=-1$ states. Thus within our mean field
theory we can ignore the dependence of the functions $\xi_\bk$ and
$\Delta_\bk$ on the distance {}from the transition, which we can represent by
$\mu$.

In the strong-pairing phase, $\mu<0$, we have $v_\bk\propto k_x-ik_y$, as
$\bk\rightarrow0$. Then the leading behavior in $g_\bk=v_\bk/u_\bk$ is
$\propto k_x-ik_y$, which is real-analytic in $k_x$ and $k_y$. If $g_\bk$
is real-analytic in a neighborhood of $\bk=0$, then $g(\br)$ will fall
exponentially for large $\br$, $g(\br)\sim e^{-r/r_0}$, but even if not it
will fall rapidly compared with the other cases below. Thus we term this phase
the strong-pairing phase because the pairs in position space are tightly bound
in this sense. Note that this region is $\mu<0$, which would only be reached
for strongly attractive coupling of the fermions (we disregard the possibility
of other non-pairing phases for such couplings). Because
$\overline{n}_\bk=|v_\bk|^2$, there is little occupation of the small $\bk$
values in this phase.

In the weak-pairing phase, $u_\bk\propto k_x+ik_y$ for $\bk\rightarrow0$, and
so $g_\bk\propto 1/(k_x+ik_y)$, which gives
\be
g(\br)\propto 1/z
\ee
for large $\br$, where $z=x+iy$. This long tail in $g(\br)$ is the reason for
the term ``weak pairing''. {\em This is exactly the behavior of $g(\br)$ in the
Moore-Read (MR) Pfaffian state in the FQHE.} In the latter this form, $1/z$, is
valid for all distances. We will therefore try to argue that all the universal
behavior associated with the MR state is generic in the weak-pairing phase,
when the theory is applied to the paired FQHE states. Notice that in the
weak-pairing phase, the occupation numbers of the small $\bk$ states approaches
1. Of course, this is also the behavior of the Fermi sea. When the attractive
coupling is weak, one would expect the weak pairing phase. If we imagine that
only the magnitude of the coupling is varied, then when it is small and
negative, the BCS weak-coupling description is valid, and $E_\bk$ has a
minimum at $k_F$. This is {\em not} the weak-strong transition. Close to the
latter transition, $E_\bk$ has a minimum at $\bk={\bf 0}$. As the coupling
weakens, a point is reached at which the minimum moves away {}from $\bk={\bf
0}$, and eventually reaches $|\bk|=k_F$ when the coupling strength is zero and
the transition to the Fermi sea (or Fermi liquid phase) takes place. Thus the
weak-{\em pairing} phase does not require that the coupling be weak, but is
continuously connected to the weak-{\em coupling} BCS region.

At the weak-strong transition, $\mu=0$, we find at small $\bk$ that
$E_\bk=|\Delta_\bk|$, $|u_\bk|^2$, $|v_\bk|^2\rightarrow 1/2$, and
$g_\bk=(k_x-ik_y)/|\bk|$. The singular behavior of $g_\bk$ leads to
\be
g(\br)\propto \frac{1}{z|z|}
\ee
for large $\br$. This behavior is intermediate between those of the two phases.
Because, like the strong-pairing behavior, it is not a rational function of
$z$, it does not correspond to a ``nice'' LLL FQHE wavefunction.
Also as $\mu\rightarrow0$, the length scale on which the asympototics of the
two phases is valid ($r_0$ on the strong pairing side) diverges. This length
should not necessarily be identified with the coherence length $\xi_0$ of the
paired state. If the latter is viewed as ``the size of a pair'', it may be
better associated with the decay of the inverse Fourier transform of
$\langle c_\bk c_{-\bk}\rangle=u_\bk^\ast v_\bk=-\Delta_\bk/E_\bk$, which
develops long-range behavior only at the transition. Also, for some purposes
the relevant function may be $\Delta_\bk$, which is always nonsingular at
$\bk=0$, and does not vanish at the transition. For the purpose of making
contact with the existing FQHE wavefunctions, $g(\br)$ is the appropriate
quantity.

The distinction between the two phases, which we can view as requiring the
existence of the phase transition, does not lie in symmetries, unlike many
phase transitions; instead, it is topological in nature \cite{volovik88}.
Within the mean field treatment, this can be seen in terms of the topology
of the functions $u_\bk$, $v_\bk$, or of $E_\bk$, $\Delta_\bk$. We recall
that $u_\bk$, $v_\bk$ obey $|u_\bk|^2+|v_\bk|^2=1$, and that multiplying
them both by the same phase is irrelevant. These conditions imply that
they can be viewed as spinor (or ``homogeneous'') coordinates for a
2-sphere $S^2$, for which two real coordinates are sufficient for a
non-redundant parametrization of any neighborhood. As usual, we view
$u=1$, $v=0$ as the north pole $\cal N$, $u=0$, $v=1$ as the south pole
$\cal S$. We can alternatively parametrize the sphere using the unit
vector ${\bf n}_\bk=({\rm Re}\,\Delta_\bk,-{\rm Im}\,
\Delta_\bk,\xi_\bk)/E_\bk$, and this agrees with the parametrization by
$u_\bk$, $v_\bk$. This is essentially the ``pseudospin'' point of view of
Anderson \cite{anderson58}, who described the s-wave BCS state as a spherical
domain wall in the pseudospin in $\bk$-space, of radius $k_F$.
Because $v_\bk\rightarrow0$ as $\bk\rightarrow\infty$ in any
direction, we can add a point at infinity in $\bk$ space and view it also as
topologically a 2-sphere, with the point at infinity at $\cal N$. The functions
$u_\bk$, $v_\bk$ thus describe a mapping {}from $S^2$ ($\bk$-space) to $S^2$
(spinor space), with $\cal N$ always mapping to $\cal N$. Such maps are
classified topologically into equivalence classes, called homotopy classes,
such that maps in the same class can be continuously deformed into each other.
The special cases of maps {}from $S^n$, $n=1$, $2$, \ldots, to any space $X$
define the homotopy groups $\pi_n(X)$ (using a general method of producing a
group structure on the equivalence classes with base points, which here are
$\cal N$). In our case, $\pi_2(S^2)={\bf Z}$, the group of integers. By
inspection, we find that in the strong-pairing phase, the map is topologically
trivial. In this phase the map can be deformed to the topologically-trivial
map that takes all $\bk$ to $\cal N$. In any topologically nontrivial map,
as $\bk$ varies over the plane plus point at infinity, $u$, $v$ (or $\bf n$)
range over the whole sphere. Indeed, the number of times that a given point
$\neq {\cal N}$, such as $\cal S$, is taken by $u$, $v$ must be at
least $|m|$, for a map in the class labelled by the integer $m$. In our case,
in the weak-pairing phase the map passes through $\cal S$ at least once, namely
when $\bk=0$, and possibly (most likely) only once, and we can choose to
identify the class with $m=1$. This nontrivial topology in $\bk$-space
associated with non-s-wave weak-pairing ground states in two dimensions was
pointed out by Volovik \cite{volovik88}. We also note that the p-wave
weak-pairing phase map is a topologically non-trivial texture of the
pseudospin that is also familiar in physics as a two-dimensional instanton,
or a skyrmion in a $2+1$ dimensional system, in {\em position} space,
in the O(3) nonlinear sigma model. Since it is impossible to pass
smoothly between $m=0$ and $m=1$, the map $u_\bk$, $v_\bk$ must be
discontinuous at the transition, which is what we found. In fact, $|u|^2$ and
$|v|^2$ tend to $1/2$ as $\bk\rightarrow0$ at the transition, which
corresponds to points on the equator. The simplest form of such a map is one
that covers just the northern hemisphere, centered on $\cal N$.

The topological distinction between weak- and strong-pairing phases is typical
for the fully-gapped complex non-s-wave paired states, and not only when
the gap function is an angular momentum eigenstate. For contrast, note
that any s-wave state yields a topologically-trivial ($m=0$) map. For the
s-wave case, it is well known that $u_\bk$ and $v_\bk$ can be chosen real,
and continuously interpolate between the weak and strong coupling limits
without a phase transition. Also, in three dimensions, the relevant
homotopy group for the present spinless case is $\pi_3(S^2)={\bf Z}$, and
so nontrivial states do exist in principle, the simplest of which is the
Hopf texture in $\bk$-space. The usual s-wave state is again
topologically trivial. Note that in the usual BCS model
with a separable interaction, the gap function vanishes outside a thin shell
around the Fermi surface, which makes the maps slightly discontinuous. We
consider only interactions that are continuous in $\bk$-space, for which the
gap function and the map are continuous except at the transition.

To close this subsection, we consider (as promised earlier) the effect on the
ground states of the boundary conditions and the total fermion number. That is,
we use a two-dimensional (Bravais) lattice, and consider a system on a plane
with points differing by a lattice vector identified, generally described as
periodic or generalized periodic boundary conditions, or by saying that the
system is a torus.

To obtain a low-energy state in a translationally-invariant system when $\Delta$
itself is viewed as a dynamical parameter, we will assume that $\Delta$ is
position-independent, and thus we can still use $\Delta_\bk$ in the
quasiparticle Hamiltonian in $\bk$-space. To be consistent with this, the
fermions must obey either periodic or antiperidic boundary conditions for each
of the two primitive directions of the lattice (or fundamental cycles on the
torus). For a rectangle with $x$ and $x+L_x$, $y$ and $y+L_y$ identified, this
means the boundary conditions for the $x$ and $y$ directions. These choices of
boundary conditions are well-known in the description of flux quantization in
superconductors (see e.g.\ Schrieffer \cite{schrieffer}). We may imagine that
either zero or one-half of the flux quantum $hc/e$ threads either of the
``holes'' (fundamental cycles) in the torus. The half-flux quanta could be
represented either by a vector potential, with peridic boundary conditions
on the fermions in both directions, or by no vector potential and an
antiperiodic boundary condition for each direction that wraps around a flux, or
by a combination of these. The different choices are related by gauge
transformations. We choose to use boundary conditions and no vector potentials,
so that $\Delta$ is always position-independent. We should be aware that if the
gauge field (fluxes) are viewed as external, they are fixed as part of the
definition of the problem, and there will be a single ground state for each of
the four possible choices of boundary conditions, $++$, $+-$, $-+$, and $--$, in
a notation that should be obvious. However, if the gauge field is viewed as
part of the internal dynamics of the system and can fluctuate quantum
mechanically (as in highly correlated systems, including the FQHE, where it is
not interpreted as the ordinary electromagnetic field, and also in the usual
superconductors where it is) then the four sectors we consider correspond to
four ground states of a single physical system, in a single Hilbert space,
albeit treated within a mean field approximation. The latter is the view we
will take.

For each of the four boundary conditions for the fermions, the allowed $\bk$
values run over the usual sets, $k_x=2\pi\nu_x/L_x$ for $+$,
$2\pi(\nu_x+1/2)/L_x$ for $-$, where $\nu_x$ is an integer, and similarly
for $k_y$. In particular, $\bk=(0,0)$ is a member of the set of allowed $\bk$
only in the case $++$. For a large system, $\xi_\bk$ and $\Delta_\bk$ will be
essentially the same functions of $\bk$ for all four boundary conditions, but
evaluated only at the allowed values. In the paired ground states, $\bk$ and
$-\bk$ will be either both occupied or both unoccupied, to take advantage of
the pairing ($\Delta_\bk$) term in $K_{\rm eff}$. When $\bk={\bf 0}$
is in the set of allowed $\bk$, $\bk={\bf 0}$ and $-\bk={\bf 0}$ cannot both
be occupied, because of Fermi statistics. However, $\Delta_\bk$ vanishes at
$\bk={\bf 0}$, so $\bk={\bf 0}$ will be occupied or not depending solely on the
sign of $\xi_{\bk={\bf 0}}$. That is, it will be occupied for $\mu>0$ (in the
weak-pairing phase), and unoccupied for $\mu<0$ (in the strong pairing phase),
and this is entirely consistent with the limiting behavior of
$\overline{n}_\bk=|v_\bk|^2$ as $\bk\rightarrow0$ in the two phases. At the
transition, $\mu=0$, the occupied and unoccupied states are degenerate.

We conclude that in either the weak- or strong-pairing phases, there is a
total of four ground states, three for boundary conditions $+-$, $-+$, $--$
which are linear combinations of states with even values of the fermion number
in both phases, but for $++$ boundary conditions the ground state has odd
fermion number in the weak-pairing phase, even fermion number in the
strong-pairing phase, because of the occupation of the $\bk={\bf 0}$ state.
In most cases, the ground state is as given in Eq.\ (\ref{OmegaNeven}),
but in the weak pairing phase for $++$ boundary conditions, it is
\be
|\Omega\rangle={\prod_{\bk\neq{\bf 0}}}'(u_\bk+v_\bk c_\bk^\dagger
c_{-\bk}^\dagger)c_{\bf 0}^\dagger|0\rangle.
\ee
The ground states specified, whether for $N$ even or odd, will have the same
energy in the thermodynamic limit (not just the same energy density).
Note that if the $\bk={\bf 0}$ state is occupied in the strong-pairing phase,
or unoccupied in the weak-pairing phase, this costs an energy $E_{\bk={\bf 0}}$
which we are assuming is nonzero, and all states where quasiparticles are
created on top of our ground states cost a nonzero energy, since we assume that
$E_\bk$ is fully gapped in both phases. However, at the transition $\mu=0$, the
ground states for $++$ with odd and even particle number are degenerate, and
there is a total of five ground states.

If we now compare with results for the MR state on the torus
\cite{gww,rr}, which were derived as exact zero-energy ground states of a
certain Hamiltonian, then we see that the weak-pairing phase for {\em even}
fermion number agrees with the exact result that there are three ground states.
On the other hand, it was stated in Ref.\ \cite{rr} that there are no
zero-energy ground states for $N$ odd. Unfortunately, that result was in
error; there is just one such state for $++$ boundary conditions, which can be
constructed by analogy with that for the 331 state in Ref.\ \cite{rr}. Before
turning to the wavefunctions of these states, we also mention that the
behavior found in the present approach in the strong-pairing phase agrees
with that expected in the Halperin point of view \cite{halp83} on the paired
states, as Laughlin states of charge 2 bosons. That point of view predicts
four ground states for $N$ even, none for $N$ odd. Note that in comparing
with FQHE states, we factor out the center of mass degeneracy which is always
the denominator $q$ of the filling factor $\nu=p/q$ (where $p$ and $q$ have no
common factors) \cite{hald85}. The remaining degeneracy in
a given phase is independent of $\nu$ in the sense that it does not change
under the operation of vortex attachment, which maps a state to another in which
$1/\nu$ is increased by 1, and for generic Hamiltonians this degeneracy is
exact only in the thermodynamic limit. We note that Greiter {\it et al.}
\cite{gww} claimed that the special Hamiltonian for which the MR state is
exact should have four ground states on the torus for $N$ even, though they
found only three. They did not distinguish the weak- and strong-pairing phases,
and by assuming that the Halperin point of view is always valid, they in effect
ascribed the properties of the strong-pairing phase to the MR state. In fact,
there is a total of four ground states in the weak-pairing phase, but the
fourth is at odd fermion number! They also claimed that the statistics would
be abelian, even though they considered the MR state, and suggested that the
prediction of nonabelian statistics by Moore and Read \cite{mr} would hold
only at some special point.

If we consider the position-space wavefunctions of the paired states on the
torus, then for even $N$ in the weak-pairing phase we once again find exact
agreement of the long-distance behavior of $g(\br_i-\br_j)$ with that which
holds for all distances in the eigenstates of the special Hamiltonian
\cite{gww,rr}. Here long-distance means $g(\br)$ where $\br$ is not close to a
lattice point $(mL_x,nL_y)$, $m$, $n$ integers. The long-distance form $g\sim
1/z$ in the plane is replaced by an elliptic function (or ratio of Jacobi theta
functions) \cite{gww,rr} for $+-$, $-+$, $--$. The ground state for $N$ odd
with $++$ boundary conditions can be written, similarly to one for $N$ even in
the 331 case \cite{rr}, as
\bea
\Psi_{++}(\br_1,\ldots,\br_N)&=&\frac{1}{2^{(N-1)/2}((N-1)/2)!}\sum_P
{\rm sgn}\, P\non\\
&&\mbox{}\times \prod_{i=1}^{(N-1)/2}g_{++}(\br_{P(2i)}-\br_{P(2i+1)})
\eea
Here we can take the torus to have sides $L_x$, $L_x\tau$ in the complex plane
(${\rm Im}\, \tau>0$, and $\tau=iL_y/L_x$ for the rectangle), and
\be
g_{++}(\br)=\frac{\theta_1'(z/L_x|\tau)}{\theta_1(z/L_x|\tau)}+\frac{2\pi
iy}{L_x{\rm Im}\,\tau}
\ee
where $\theta_1$ is a Jacobi theta function, and
$\theta_1'(z|\tau)=d\theta_1(z|\tau)/dz$. $g_{++}(\br)$ is periodic
because of properties of the function $\theta_1$ mentioned for
example in Ref.\ \cite{rr}, and has a simple pole $g_{++}\propto 1/z$ as
$z\rightarrow0$. For the ground state with $N$ odd, the
non-(complex)-analytic dependence on $y$ in $g_{++}$ cancels. Notice that
the unpaired fermion with $i=P(1)$ in the terms in $\Psi_{++}$ occupies the
constant, $\bk={\bf 0}$ single-particle state. When used as part of a
wavefunction in the LLL, the present function is the zero energy state for $++$
boundary conditions on the torus for $N$ odd, which was omitted in Ref.\
\cite{rr}.

%%%%%%%%%%%%%%%%%%%%%%%%%%%%%%%%%%%%%%%%%%%%
\subsection{Majorana fermions, edges and vortices}
\label{majorana}

In this Subsection, we consider the problems of edges and of vortices
(which correspond to FQHE quasiparticles), on which we argue there are
chiral fermions and zero modes, respectively, in the weak-pairing phase.
Again, this supports the identification with the MR state.

We begin by considering in more detail the low-energy spectrum near the
transition at $\mu=0$. When $\mu$ and $\bk$ are small, we can use
\bea
\xi_\bk&\simeq&-\mu, \non\\
\Delta_\bk&\simeq&\hat{\Delta}(k_x-ik_y),
\eea
where $\hat{\Delta}$ can be complex, and find
\be
E_\bk\simeq\sqrt{|\hat{\Delta}|^2|\bk|^2+\mu^2},
\ee
which is a relativistic dispersion with $|\hat{\Delta}|$ playing the role of
the speed of light. Further, using the same approximation, the BdG equations
become in position space
\bea
i\frac{\partial u}{\partial t}&=&\mbox{}-\mu u +\hat{\Delta}^\ast i
\left(\frac{\partial}{\partial x}+i\frac{\partial}{\partial y}\right)v,\non\\
i\frac{\partial v}{\partial t}&=&\mu v +\hat{\Delta} i
\left(\frac{\partial}{\partial x}-i\frac{\partial}{\partial y}\right)u,
\label{dirac}
\eea
which is a form of the Dirac equation for a spinor $(u,v)$. The BdG equations
are compatible with $u(\br,t)=v(\br,t)^\ast$, and this is related to the
existence of only a single fermion excitation mode for each $\bk$. Thus the
quasiparticles are their own antiparticles; Dirac fermions with this property
are known as Majorana fermions. Near the transition, the BCS
quasiparticles make up a single Majorana fermion quantum field, and at the mean
field level the critical theory is a single massless Majorana fermion. There is
a diverging length scale at the transition, $\sim \hat{\Delta}/|\mu|$, and the
diverging lengths mentioned in the last subsection should all diverge
proportionately to this, at least within mean field theory.

Next we wish to consider the behavior near an edge; outside the edge the
particle number density should go to zero. In the Hamiltonian, this can be
arranged by having a potential $V(\br)$ that is large and positive outside the
edge. In the quasiparticle effective Hamiltonian, this can be viewed as making
$\mu$ large and negative outside the edge, and we will use this notation.

In general, the problem with the edge should be solved self-consistently, which
involves solving the gap equation for $\Delta$ in the presence of the edge. We
are interested in the generic properties of the solution, and wish to avoid the
complexities. Accordingly we will consider only a simplified problem, which is
the effective Hamiltonian with a given gap function. Since $\mu$ becomes
negative outside the edge, it must change sign near the edge if it is positive
inside the bulk of the system. But in the bulk at least, a change in sign
across a line represents a domain wall between the weak- and strong-pairing
phases, since $\mu=0$ is the point at which the transition occurs in our
treatment. Thus we are arguing that the weak-pairing phase (where $\mu$ is
positive) must have a domain wall at an edge, while the strong-pairing phase
need not. We will consider a domain wall in the bulk as a model for the edge of
the weak-pairing phase. In the latter the region of strong pairing between the
domain wall and the actual edge may be extremely narrow and we might say there
is no well-defined strong-pairing region. But the point is the topological
distinction between the phases. The strong-pairing phase has the same topology
as the vacuum, and can be continuously connected to it. The weak-pairing phase
is nontrivial and the generic properties of an edge should be captured by a
domain wall \cite{hopriv}.

We consider a straight domain wall parallel to the $y$-axis, with
$\mu(\br)=\mu(x)$ small and positive for $x>0$ and small and negative for
$x<0$, and also varying slowly such that the above long-wavelength
approximation can be used, with $\mu$ now $x$-dependent.

We can consider solutions with definite $k_y$, which at first we set to zero.
Then we have a $1+1$-dimensional Dirac equation. We assume that
$|\mu|\rightarrow\mu_0$, a constant, as $|x|\rightarrow\infty$. At $E=0$ there
is a normalizable bound state solution first obtained by Jackiw and Rebbi
\cite{jackreb}. The equations are (with $\hat{\Delta}$ real and $>0$)
\bea
\hat{\Delta}i\partial v/\partial x&=&\mu(x)u,\non\\
\hat{\Delta}i\partial u/\partial x&=&-\mu(x) v.
\eea
By putting $v=iu$, we find the unique normalizable solution
\be
u(x)\propto e^{-i\pi/4}\exp -\frac{1}{\hat{\Delta}}\int^x\mu(x)dx,
\ee
where the phase was inserted to retain $v=u^\ast$. Solutions at finite $E$
should exhibit a gap, as in the bulk in either phase.

At finite $k_y$, the equations become
\bea
Eu&=&-\mu u +\hat{\Delta}i\left(\frac{\partial v}{\partial x}-k_y v\right),
\non\\
Ev&=&\mu u +\hat{\Delta}i\left(\frac{\partial u}{\partial x}+k_y u\right).
\eea
For $E=-\hat{\Delta}k_y$, these have solutions that are bound to the domain
wall, and clearly they propagate in one direction along the wall. There is only
one fermion mode for each $k_y$, and so we have a chiral Majorana (or
Majorana-Weyl) fermion field on the domain wall.

It will be important to consider also a pair of domain walls. We consider two
walls, lying at $x=0$ and $x=W$, with $\mu>0$ in $0<x<W$ and $\mu<0$ outside.
Again we assume $k_y=0$ initially. This time \cite{jackreb} there are no
$E=0$ modes for finite $W$. Clearly as $W\rightarrow\infty$ we expect to find
an $E=0$ mode on either wall, so we expect bound solutions for small $E$ when
$W$ is large but finite. For non-zero $E$ we can replace the pair of
first-order equations with a single second-order equation for either of $u\pm
iv$, (with $k_y\neq0$ for generality),
\be
(E^2-\hat{\Delta}^2k_y^2)(u\pm iv)=\left(-\hat{\Delta}^2
\frac{\partial^2}{\partial x^2}+\mu^2\pm \hat{\Delta}\frac{\partial \mu}
{\partial x}\right)(u\pm iv).
\ee
When $\mu$ varies slowly compared with its magnitude $\mu_0$ far {}from the
walls, we may study the equations by the WKB method. We may view the equation
for $u\pm iv$ as a Schrodinger equation with potential
\be
V_{\pm}(x)=\mu^2\pm \hat{\Delta}\frac{\partial \mu}{\partial x}.
\ee
If $\partial \mu/\partial x$ has extrema at $x=0$, $W$, as is reasonable, then
$V_-(x)$ has minima at $x=0$, $W$, but that at $x=0$ is deeper than that at
$x=W$. The reverse is true for $V_+(x)$. As $W\rightarrow\infty$, there will be
a $k_y=0$ solution for $E^2$ which $\rightarrow0$ exponentially and
which corresponds to a normalizable eigenfunction for $u-iv$ that is
concentrated at $x=0$ with negligible weight at $x=W$, and similarly an
eigenfunction for $u+iv$ concentrated at $x=W$. The subtle but important point
is that these solutions are not independent, because they are related by the
original first-order system, for any non-zero $E$. There is only a single
normalizable solution for the pair $u$, $v$ for $E$ small positive, and
another for $E<0$. Consequently, there is only a single fermion mode,
shared between the two domain walls, not one on each. For non-zero $k_y$, the
eigenfunctions for $E>0$ become concentrated on one wall or the other,
depending on the sign of $k_y$. Thus the set of low-energy states can be
viewed as a single non-chiral Majorana fermion theory, with the left-moving
modes on one wall, the right-moving modes on the other, and the $k_y=0$ mode
shared between the two. {\em This agrees precisely with the results of Ref.\
\cite{milr} on the edge states of the MR state on the cylinder}.

We next consider the quasiparticle spectrum in the presence of vortices of the
order parameter; in two dimensions vortices are point objects. These
necessarily contain an integer number of half flux quanta in the gauge
field; without the gauge field the vorticity is quantized but the total energy
of an isolated vortex diverges logarithmically. We will not see these effects
in the energy here because we only consider the quasiparticle excitation
spectrum in the presence of a given gap function and gauge field
configuration. Up to now we have ignored the U(1) gauge field except in
discussing the boundary conditions on the torus and cylinder. It could be the
standard electromagnetic field in a superconductor, or the CS field in the
FQHE. We will only consider vortices of the minimal flux, namely a half flux
quantum, because addition of any integer number of flux quanta can be viewed,
on scales larger than the penetration depth, as a gauge transformation, which
does not affect the spectrum. Outside the vortex core, which we assume is
small, the covariant derivative of the gap function must vanish. As in the
case of the torus, we will choose a gauge in which the gap function is
single-valued and independent of the angle relative to the position of the
nearby vortex, but the Fermi fields are double-valued on going around the
vortex.

The basic idea is to consider a vortex as a small circular edge, with vacuum
(vanishing density) at the center. Accordingly, we expect that nothing
interesting happens at sufficiently low energies for vortices in the
strong-pairing phase. But a vortex in the weak-pairing phase must include a
concentric circular domain wall to separate the vacuum at the center {}from the
weak-pairing phase outside. We now study this using the Majorana fermion
equations near the transition, assuming that the wall has large enough radius;
the vortex core (where $\Delta$ vanishes) can be taken to have negligible size
and the boundary condition at $r\rightarrow 0$ is unimportant. With our
choice, the BdG equations for a single vortex and for $E=0$ becomes in polar
coordinates $r$, $\theta$
\bea
\hat{\Delta}ie^{i\theta}\left(\frac{\partial}{\partial r}
+\frac{i}{r}\frac{\partial}{\partial \theta}\right)v&=&\mu u,\non\\
\hat{\Delta}ie^{-i\theta}\left(\frac{\partial}{\partial r}
-\frac{i}{r}\frac{\partial}{\partial \theta}\right)u&=&-\mu v;\non\\
\eea
$u$ obeys $u(r,\theta+2\pi)=-u(r,\theta)$, and similarly for $v$.
We can assume $\mu\rightarrow \mu_0>0$ as $r\rightarrow\infty$,
$\mu\rightarrow-\mu_0$ as $r\rightarrow0$. The normalizable solutions
have the form
\bea
u&=&(i\overline{z})^{-1/2}f(r),\non\\
v&=&(-iz)^{-1/2}f(r)=u^\ast,
\label{vortsol1}
\eea
where $f(r)$ is a real function. The equations reduce to
\be
df/dr=-\mu(r)f(r)/\hat{\Delta},
\ee
with solution
\be
f(r)\propto\exp(-\int^r\mu(r')dr'/\hat{\Delta}). \label{vortsol2}
\ee Thus we find just one normalizable bound state at zero energy.
Again we expect this to persist as we relax our assumptions, as
long as the bulk outside the vortex is in the weak-pairing phase.
We point out that our result should be contrasted with the known
result for a vortex in an s-wave superconductor, which has bound
quasiparticle modes at energies that are low in the weak coupling
limit, but not generally zero as ours are here \cite{boundstates}.
Since we mainly work at moderate or strong coupling, analogous
modes are not important for our purposes. We note that a zero mode
on a vortex in an A-phase p-wave paired state was first found in
Ref.\ \cite{kopnin}.

For the case of $2n$ well-separated vortices, we have not obtained
analytic solutions for the bound states. However, we can give a
simple argument. The problem is analogous to a double-well
potential. We take a set of $2n$ $E=0$ solutions like Eqs.\
(\ref{vortsol1}), (\ref{vortsol2}) centered at each vortex, and
use these as a basis set (we must introduce additional branch
points into each basis state to satisfy the boundary conditions at
all the other vortices); at finite separation there is mixing of
the states, and the energies split away from zero. Since the
solutions to the Dirac equation are either zero modes or $E$, $-E$
pairs, we expect to obtain $n$ $E>0$ solutions, $n$ $E<0$
solutions. In general, each $E>0$ solution of the Dirac or BdG
equation corresponds to a creation operator, and the related $E<0$
solution to the adjoint (destruction) operator, while an $E=0$
solution would correspond to a real (or Majorana) fermion
operator. In our case, this means that there are $n$ modes in
which we may create fermions, with energies $E$ tending to zero as
the separation diverges. (A similar picture applies for $2n$
domain walls.) {\em This is in agreement with the results for the
special Hamiltonian} \cite{rr}. This result is crucial for the
nonabelian statistics we expect in the FQHE case, since by
occupying the zero modes one obtains a total of $2^n$ degenerate
states, or $2^{n-1}$ for either even or odd fermion number $N$,
when there are $2n$ vortices ($n>0$); this was found for the
special Hamiltonian in Ref.\ \cite{nayak,rr}.

We may also consider here the edge states of a system in the form of a disk of
radius $R$, by studying a large circular domain wall enclosing the
weak-pairing phase, and strong-pairing phase or vacuum outside. In this case,
there is no flux enclosed by the wall, and $u$ and $v$ are single-valued.
One does not find $E=0$ states, but instead there is a set of chiral fermion
modes with angular momentum $m$ quantized to half-integral values,
$m\in{\bf Z}+1/2$, and $E\propto m/R$ (this fixes the definition of $m=0$).
These are just the modes expected for the
chiral Majorana fermion on such a domain wall with the ground state inside,
since an antiperiodic boundary condition is natural for the ground state
sector. If a half-flux quantum is added at the center of the disk,
the quantization is $m\in {\bf Z}$, and this extends the result for the zero
mode $m=0$ of a single vortex. {\em These results agree with the results
of Ref.\ \cite{wen3,milr} for a disk of the MR state.}

We also note that the form of the modes near a vortex, containing $z^{-1/2}$
or its conjugate, is similar to the form of the fermion zero mode functions
found in Ref.\ \cite{rr}, once the factors associated with the charge sector
are removed, though the factors $f(r)$ are not. However, the most appropriate
comparison to make is that between the {\em many-fermion} wavefunctions, as we
already made for the ground states on the plane and torus. We will not
consider this further here for the vortex or the edge (or domain wall) states,
though we expect that these should correspond at long distances to those found
in Ref.\ \cite{nayak,rr,milr}, as for the ground states. However, we are able
to find the ground states on other geometries, namely the sphere and Riemann
surfaces of genus (number of handles) greater than one (the sphere is genus
zero, the torus genus one). We consider this briefly in Subsection \ref{geom}
below.

%%%%%%%%%%%%%%%%%%%%%%%%%%%%%%%%%%%%%%%%%%%%
\subsection{Other geometries and conformal field theory}
\label{geom}

In this Subsection, we briefly introduce some general connections of pairing
theory for p-wave states to relativistic fermions, which enables us to discuss
geometries other than the plane and torus, such as the sphere, and to make
more explicit connections with conformal field theory ideas \cite{mr}. This
Subsection can be omitted on a first reading, but some of the formalism is
mentioned again later.

In the preceding Subsection, we used the fact that the BdG equations at long
wavelengths become the Dirac equation, with a reality condition so that the
Fermi field is a Majorana fermion. We also mentioned the coupling of the gap
function and Fermi fields to a U(1) vector potential (which in the FQHE context
would be the CS vector potential), which is of a standard form. But the
interpretation of the fermion as Majorana would seem to raise a problem,
because for a single Majorana there is no continuous symmetry of the Yang-Mills
type, and so apparently no way to minimally couple it to a vector potential.
We will see that there is nonetheless a natural way to incorporate the
vector potential and still give an interpretation in terms of the Dirac
equation, and this will also enable us to discuss the ground states on
curved surfaces.

The most general form for the p-wave gap function in Fourier space,
retaining once again only the long-wavelength part, can be written
\be
\Delta_\bk=\Delta_x k_x-i\Delta_y k_y.
\ee
Here $\Delta_{x,y}$ are two complex coefficients, or equivalently four
real numbers, which we will arrange into a $2\times2$ matrix $e$. In
position space, the $\bk$ can be replaced by $-i\nabla$. Then in a general
coordinate system $x^i$, with corresponding partial derivatives
$\partial_i$, ($i=x$, $y$), the BdG equations become
\be
(\partial_t +\frac{1}{2}i\omega_t^{bc}\Sigma_{bc})\psi +
e^{ia}\alpha_a(\partial_i
+\frac{1}{2}i\omega_i^{bc}\Sigma_{bc})\psi
+i\beta m\psi=0,
\ee
where we use a spinor $\psi=(v,u)$, and $m=\mu$ in previous notation.
The indices $a$, $b$, $c$ take the values $x$, $y$, and the matrices are
$\alpha_x=\sigma_x$, $\alpha_y=\sigma_y$, $\beta=\sigma_z$; we use the
summation convention. Here we have also reinstated the
vector potential $A_\mu=\omega_\mu^{xy}/2$ (where $\mu=t$, $x$, $y$),
using the matrix
\be
\Sigma_{xy}=\sigma_z.
\ee
The equation is therefore invariant under U(1) gauge transformations
$\psi\to e^{i\Lambda^{xy}\Sigma_{xy}} \psi$, and a corresponding
transformation of $\omega_\mu$, with a real scalar parameter
$\Lambda_{xy}(x^\mu)$.

If we multiply through by $\beta$, then we obtain the more
covariant form of the Dirac equation,
\be
e^{\mu
a}\gamma_a(\partial_\mu+\frac{1}{2}\omega_\mu^{bc}\Sigma_{bc})\psi+im\psi=0,
\ee with $e^{\mu t}=\delta_{\mu t}$ in our case. This has the form
of the general Dirac equation suitable for use in general
coordinate systems on general curved spaces or spacetimes
\cite{gsw}. A similar form was obtained in Ref.\ \cite{vol90lett}.
In general, $\psi$ is a Dirac spinor (with two components in the
$2$ and $2+1$ cases of interest here); $\mu=1$, $2$, \ldots $d$ is
a spacetime index, while $a$, $b$, $c=1$, $2$, \ldots, $d$ is an
internal ``local Lorentz'' index; the vielbein $e$ is a tensor
with indices as shown; $\gamma_a$ are a set of Dirac matrices
satisfying $\{\gamma_a,\gamma_b\}=2\eta_{ab}$, where $\eta$ is the
Minkowski or Euclidean metric, and $\Sigma_{ab}=\frac{1}{2}i
[\gamma_a, \gamma_b]$ are the generators of SO$(d-1,1)$ Lorentz
transformations (or simply SO$(d)$ rotations, in the Euclidean
case); the spin connection $\omega$ is a tensor  field with one
spacetime and two internal Lorentz indices, and is antisymmetric
in the latter. Spacetime indices are raised and lowered using
$g^{\mu\nu}$ and $g_{\mu\nu}$, while internal Lorentz indices are
raised and lowered using $\eta^{ab}$ and $\eta_{ab}$. This form of
the Dirac equation is covariant under coordinate transformations
(diffeomorphisms), under which the spinor is viewed as
transforming as a scalar function of position, and $e$ and
$\omega$ as tensors. It is also covariant under SO$(d-1,1)$ (or
SO$(d)$) local Lorentz transformations, which act like gauge
transformations, with $\psi$ transforming in the spinor
representation, the vielbein transforming as a vector in the $a$
index, and $\omega$ transforming inhomogeneously as a nonabelian
vector potential or connection for the gauge transformations. This
formalism (also known, in four dimensions, as the vierbein or
tetrad formalism) can be used to reformulate, for example, general
relativity in a form equivalent to the usual one involving
Christoffel symbols; this involves imposing relations \bea e_\mu^a
e_{\nu a}&=&g_{\mu\nu},\non\\ e^\mu_a e_{\mu b}&=&\eta_{ab}. \eea
On the other hand, it is the only known way to couple Dirac fields
to curved spacetime; more details can be found in Ref.\
\cite{gsw}. In our case, we have a distinguished choice of time
coordinate, we consider only the restricted form with $e^{\mu
t}=\delta_{\mu t}$, and require covariance under only the
SO(2)$\cong$U(1) subgroup that describes ``internal spatial
rotations''. Then the Dirac equation becomes precisely the BdG
equation. This relation with the vielbein formalism suggests that
the vector potential will play a natural role when we consider
pairing of nonrelativistic fermions on a curved surface.

The problem of spinless p-wave pairing of
nonrelativistic fermions moving on a general curved manifold should be
formulated as follows. The manifold has a metric and a corresponding Riemann
curvature tensor, which for two dimensions reduces to a curvature scalar.
We will consider only the case in which this curvature is constant on the
manifold, and we will also introduce a U(1) or SO(2) gauge potential.
For a large system, the radius of curvature of the manifold is large, and
locally the solution to the gap equation should resemble that in flat
space, which we assume is of the $l=-1$ form. In order to minimise the
energy density, we expect that the gap function should be as constant as
possible. In more invariant language, this means that the vielbein should
be covariantly constant (note that the covariant derivative of the
vielbein must be covariant under both coordinate and U(1)
gauge transformations), and there should be no vortices. Since the
Riemannian geometry (the metric and the
Levi-Civita or metric connection) of the manifold are assumed given, this
condition relates the spin connection $\omega$ to derivatives of the
vielbein, a result analogous to the usual requirement (for s-wave
pairing, in flat space) that the vector potential be the gradient of the
phase of the gap function. This covariant-constancy condition also appears
when formulating general relativity \cite{gsw}.

This condition can be satisfied globally (we assume the surface on which
we are working is compact) only if the field strength in the SO(2) vector
potential is related to the Riemann curvature of the manifold. The integral
of the latter over the surface, divided by $4\pi$, is a topological
invariant, the Euler invariant, equal to $2(1-g)$ for a Riemann surface of
genus $g$ (one with $g$ handles). In our usual units for flux quanta, the
number of flux quanta in the SO(2) or U(1) vector potential must be $g-1$.
Otherwise, we will have vortices somewhere on the surface, at which
$\hat{\Delta}$ goes to zero. In particular, for the sphere, this agrees with
the familiar fact for the MR state that the number of flux $N_\phi$ seen by
the underlying particles is one less than in the Laughlin state at the same
filling factor, so the composite fermions see a net flux of $-1$. Physically,
the nonzero angular momentum of the pairs causes them to see the curvature of
the manifold on which they move as a gauge field, the field strength of which
is cancelled (locally, not just globally) by the imposed U(1) (i.e. SO(2))
gauge field, so that a uniform condensate is possible, much like the
condition of vanishing field strength for uniform s-wave condensates in
ordinary superconductors.

It can be shown that the long-wavelength wavefunction involves the inverse of
(or Green's function for) part of the (covariant) Dirac operator we have
discussed, namely the part $\Delta^\dagger$, where $\Delta$ is the part of the
Dirac operator, including the vector potential, that acts on $c^\dagger$
and maps it to $c$ (like the earlier gap function) in the Dirac equation.
$\Delta^\dagger$ contains derivatives like $\partial/\partial
\overline{z}$ in local coordinates. On the sphere, in stereographic
coordinates the Green's function is known to be essentially
$1/(z_i-z_j)$ for particles $i$ and $j$. For any surface, this description
in terms of inverting the massless Dirac operator is identical to the problem
of finding the correlators of two-dimensional chiral Majorana fermions
(in Euclidean spacetime), and so it is not surprising that this agrees with
the conformal block for $N$ two-dimensional massless fermions on the sphere
in conformal field theory. We note that the paired ground state on the sphere
can be described in angular momentum space, in terms of single-particle
angular-momentum eigenstates with eigenvalues $j$, $m$ (and $j=1/2$, $3/2$,
\ldots, due to the single flux quantum), as BCS pairing of $j$, $m$ with
$j$, $-m$; antisymmetry and vanishing total angular momentum for each pair
require that the $j$'s be half-odd-integral, as they are for p-wave.

We may also consider the cases of Riemann surfaces of genus
greater than one. Here the explicit functions, which again are
built out of $1/\Delta^\dagger$, are more difficult to find, but
certainly exist and describe the MR state on these surfaces. The
required number of flux seen by the fermions is $g-1$. The only
aspect we wish to discuss further here is the number of distinct
ground states for $g>1$. When handles are present, the vector
potential is determined only up to addition of a pure gauge piece
describing holonomy around the $2g$ fundamental cycles of the
surface. The holonomy, or phase picked up when a fermion is
parallel-transported around a cycle, can only be $\pm 1$, since it
comes {}from the double cover of SO(2) by Spin(2), the group which
possesses the spinor representation. This effect, which is a
restatement of flux quantization, agrees with and generalizes the
discussion of boundary condition sectors for the torus. There is
thus a set of $2^{2g}$ possible boundary condition sectors, which
in the present differential geometry set-up are known as spin
structures. The spin structures on a genus $g$ surface can be
divided into two sets, known as the even and odd spin structures.
The difference between these, for our purposes, is that the odd
spin structures possess a single zero mode for the Dirac operator,
and the even spin structures possess none. Then the BCS ground
states in the weak-pairing phase will include one fermion
occupying the zero mode when one exists, and since the other
fermions are all paired, we conclude that the odd spin structures
give rise to ground states with $N$ odd, and the even spin
structures to $N$ even, and these ground states will be degenerate
in the thermodynamic limit. It is known that there are
$2^{g-1}(2^g+1)$ even spin structures, and $2^{g-1}(2^g-1)$ odd
spin structures, so these formulas give the number of ground
states for $N$ even and odd, for all $g \geq 0$. These numbers
(and the long-distance wavefunctions) agree with the conformal
blocks for a correlator on the genus $g$ surface with $N$ Majorana
fields inserted. All of this is in beautiful agreement with the
CFT picture of Ref.\ \cite{mr}. We note that the U(1) charge
sector which is present in the FQHE states gives another factor
$q^g$ in the degeneracy for filling factor $\nu=p/q$, in the
thermodynamic limit \cite{wn}, which generalizes the
center-of-mass degeneracy $q$  of the torus \cite{hald85}.

%%%%%%%%%%%%%%%%%%%%%%%%%%%%%%%%%%%%%%%%%%%%%%%%%%
\section{Spin-triplet complex p-wave pairing}
\label{331}

In this Section we consider spin-triplet p-wave pairing. Since the general
classification of such states is complicated (compare the three-dimensional
version in Ref.\ \cite{vollwolf}), we concentrate on a particular case directly
related to the FQHE. The FQHE system we have in mind is the double-layer system
at $\nu=1/2$ \cite{expt}. This is assumed to be spin-polarized, but the layer
index of the electrons plays the role of a spin, which we refer to as
isospin, to avoid confusion with the pseudospin discussed in the previous
Section. The $S_z$-values of the isospin will be denoted $\up$, $\down$
for the two layers.

In the double-layer FQHE system, as in the other systems we discuss, we go to
a CS fermion representation by using layer-independent fluxes attached to the
CS fermions. Because interactions between electrons in the same and in
different layers are different (though the two layers are on an equal footing),
the Hamiltonian will not have SU(2) symmetry. However, in the absence of a
tunneling term and of interactions that transfer electrons between layers,
the number $N_\up-N_\down$ is conserved, and since this quantity is twice the
total $S_z$ of the isospin, there is a U(1) $\cong$ SO(2) symmetry that
rotates the isospin about the $z$ axis. Also interchange of the two layers
(or reflection in the plane midway between the two layers) is a ${\bf Z}_2$
symmetry. Together these make up an O(2) symmetry. We also consider the effect
of a tunneling term $-t\sigma_x$ for each particle; $t$ is the tunneling
amplitude and $\sigma_x$, etc, denote the Pauli matrices. Nonzero $t$ breaks
the symmetry to the ${\bf Z}_2$ of layer exchange.

The FQHE system at $\nu=1/2$ has a possible ground state which is a 331 state,
which can be viewed as complex p-wave pairing of composite fermions \cite{rr}.
The pairing and the effective quasiparticle Hamiltonian are best considered
in terms of isospin states which are eigenstates of $\sigma_x$, namely
$e=(\up+\down)/\sqrt{2}$, $o=(\up-\down)/\sqrt{2}$, which are respectively even,
odd under the ${\bf Z}_2$. As in the earlier work on this problem, we assume
that the ${\bf Z}_2$ symmetry, and for $t=0$ the O(2) symmetry, are not broken
spontaneously. Then symmetry dictates that the effective quasiparticle
Hamiltonian has the form
\bea
K_{\rm eff}&=&\sum_{\bk}\left[(\xi_\bk-t) c_{\bk e}^\dagger c_{\bk e}+
\frac{1}{2}\left(\Delta_{\bk e}^\ast c_{-\bk e}^{\vphantom\dagger}c_{\bk
e}\right.\right.\non\\
&&\mbox{}\left.\left.+\Delta_{\bk e}c_{\bk e}^\dagger c_{-\bk e}^\dagger\right)
+(\xi_\bk+t) c_{\bk o}^\dagger c_{\bk o}\right.\non\\
&&\mbox{}\left.+\frac{1}{2}\left(\Delta_{\bk o}^\ast c_{-\bk o}c_{\bk o}
+\Delta_{\bk o}c_{\bk o}^\dagger c_{-\bk o}^\dagger\right)\right].
\label{331ham}
\eea
We have taken the same kinetic term $\xi_\bk\simeq k^2/2m^\ast-\mu$ for both $e$
and $o$ since a difference here is unimportant (and forbidden by symmetry when
$t=0$). For $t=0$, $\Delta_{\bk e}=\Delta_{\bk o}$, and in general we assume
both have p-wave symmetry, with
\be
\Delta_{\bk e}\simeq\hat{\Delta}_e(k_x-ik_y)
\ee
at small $\bk$, and similarly for $\Delta_{\bk o}$. We have also neglected the
possibility of many-body renormalization of the splitting $2t$ between $e$ and
$o$ (such as an exchange enhancement).

We see that the unbroken ${\bf Z}_2$ symmetry has led to decoupled $e$ and $o$
Hamiltonians. These are the same as for the spinless p-wave case. Consequently,
we see that separate transitions {}from weak to strong pairing are possible when
$t\neq0$. For $t=0$, these coincide. The pairing function $g(\br)$ in the
wavefunction is now a four-component object because of the isospin
variables. We write it as a vector in the tensor product space of the two
spinors. When at least one of the two components is in its weak-pairing phase,
the pairing function at long distances has the form
\be
\frac{\cos(\theta-\pi/4)e_i e_j +\sin(\theta-\pi/4)o_i o_j}{z_i-z_j}.
\label{hostate}
\ee
This is the form that was assumed for all distances in Refs.
\cite{halpnewport,ho,rr}. For $t=0$, we would put $\theta=0$, in which case it
reduces to
\be
\frac{\up_i\down_j+\down_i\down_j}{z_i-z_j},
\ee
which is the form in the 331 state. As $t$ increases, neglecting the likely
change in $\Delta_{\bk e}$, $\Delta_{\bk o}$ momentarily, a point is reached at
which the $o$ spins have an effective chemical potential $\mu-t=0$, and undergo
a transition to strong pairing. Then the long distance behavior is $e_i
e_j/(z_i-z_j)$, so $\theta=\pi/4$ at the transition and remains at that value
thereafter. The resulting phase is expected to have the statistics properties
of the MR state, unaffected by the strong-pairing $oo$ pairs present in the
ground state.

\begin{figure}
\epsfxsize=3.375in
\centerline{\epsffile{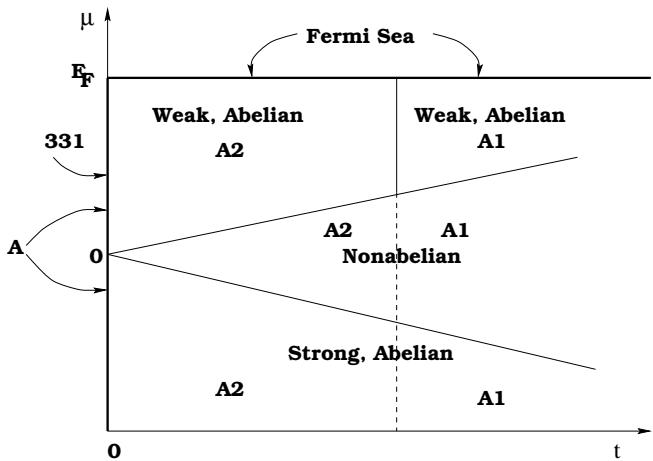}}
\vspace{0.1in}
\caption{Schematic phase diagram for the p-wave phases, as discussed in
the text. The A-phase with unbroken U(1) symmetry appears as the vertical
axis $t=0$, with the region $\mu>0$ being the 331 phase. Similarly, the
Fermi-liquid phase in which pairing disappears is identified with the line
$\mu=E_F$, since that is the value of $\mu$ there at fixed density, neglecting
Hartree-Fock corrections.} \label{fig331}
\end{figure}

When $\mu$ is decreased, there will be a transition {}from the MR phase to
strong pairing in both components when $\mu+t=0$. Thus we obtain the phase
diagram shown in Fig.\ \ref{fig331}. We have also included labels of the
analogous phases in He$^3$ \cite{vollwolf}. In He$^3$, the roles of $e$, $o$
are played by $\up$, $\down$, and that of $t$ is played by the Zeeman
splitting due to a field $h$ along the $z$ direction. In He$^3$, there is full
SU(2) symmetry of spin rotations when $h=0$, that is broken spontaneously in
any spin-triplet phase, but this distinction is unimportant here. The state
for $t=0$ has the structure of the ABM state or A phase, adapted to two
dimensions, while for $t\neq0$, we expect that self-consistent solution of the
gap equation would give $\hat{\Delta}_e > \hat{\Delta}_o$, and this state has
the structure of the A2 phase of He$^3$. As $t$ increases, a point may be
reached at which $\Delta_{\bk o}=0$ for all $\bk$, which gives the A1 phase
(we ignore intermediate possibilities in which $\Delta_{\bk o}$ vanishes only
in some region of $\bk$ space). For $\mu-t>0$, which would be the case in
He$^3$, the A1 phase is a distinct phase, which would have a Fermi surface for
$o$ spins. On the other hand, for $\mu-t<0$, no excitations become gapless at
the point where $\Delta_{\bk o}$ vanishes, and the change is merely the
disappearance of $oo$ pairs {}from the ground state, so this is not a true
phase transition; this is indicated by the dashed line in Fig.\ \ref{fig331}.
The position of this boundary, if it occurs at all, is very uncertain. We
simply wish to emphasize that the boundaries between A, A2 and A1 in general
do {\em not} coincide with the weak- to strong-pairing transitions. However,
for the wavefunctions that are of the form ${\rm Pf}\, g$ with $g$ given by
Eq.\ (\ref{hostate}) for {\em all} distances, the transition at $\theta=\pi/4$
to the MR state can also be considered as the A2-A1 transition. We discuss this
further below.

For $t=0$, the quasiparticle excitations near $\bk=0$ and $\mu\simeq0$ form a
Dirac relativistic fermion spectrum, which has two distinct but degenerate
particle and antiparticle excitations. These are eigenstates of the U(1)
symmetry of eigenvalues $+1$, $-1$, respectively, so the particle is associated
with one layer, the antiparticle with the other. A Dirac fermion field is
equivalent to two Majorana fields, which can be thought of roughly as its real
and imaginary components. For $t\neq0$, these Majorana fermions have different
masses (and different velocities if $\hat{\Delta}_e\neq \hat{\Delta}_o$), and
the transitions occur when one or other mass changes sign.

The ground-state degeneracies, edge state and vortex (quasiparticle) properties
in the different phases can now be read off {}from the above and the results for
the spinless p-wave case. For $\mu>0$ and $t=0$, we find the same results as
for the 331 state \cite{milr,rr}. The FQHE state is abelian and can be
described in Coulomb plasma language because of the Cauchy determinant
identity, as explained in Ref.\ \cite{milr,rr}. The equivalence of the results
for the neutral edge excitations, or for the quasihole states, in terms of
bosonic fields, or the two-component plasma mapping, and Dirac fermion fields,
corresponds to bosonization \cite{milr,rr} and will not be repeated here in
full. We will mention only an instructive example, consisting of a pair of
vortices in the weak pairing phase $\mu>0$. Given that there is a
single $E=0$ mode for the pair, which can here be occupied by either type of
fermion, then there are four states. Since the fermions in the $\up$-$\down$
basis have $S_z$ quantum numbers $\pm 1/2$, the four states have $S_z$ values
$1/2$, $0$, $0$, $-1/2$ (half of these states have odd total fermion number
$N$). These states must be interpreted as saying that each of these elementary
vortices carries $S_z$ of $\pm 1/4$, so there is a fractionalization of the
$S_z$ quantum number. For $N$ even we have only two states, both with
$S_z=0$, if there are no other excitations of the system, due to global
selection rules related to the total quantum numbers. This agrees with the
two-component plasma description for the 331 states, on including the charge
degree of freedom in the incompressible FHQE system \cite{milr,rr}. In the
latter formulation, the fractional $S_z$ is analogous to the fractional charge
of the Laughlin quasiparticles.

For $t\neq 0$, the U(1) symmetry is lost, and the quantum field theories,
whether the massive theory in the bulk or the chiral theory on the edge or
at the vortices, in the phase labelled ``weak-pairing abelian'' must be
described as two Majorana fermions. However, the counting of the edge
excitations, or of the vortex states just discussed, will be the same (though
unimportant degeneracies among edge excitations, that previously were due to
the U(1) symmetry, may be lost), and the universal statistics properties in this
phase, which are abelian, are the same as in the U(1) symmetric or Dirac
fermion case. As already mentioned, when the transition to strong-pairing occurs
for the $o$ spins, the system enters the MR nonabelian phase, and when the $e$
spins are also in the strong-pairing phase, the FQHE system has the rather
trivial abelian statistics of the Laughlin state of charge 2 bosons.

Ideas of Ho \cite{ho} involving a two-body pseudopotential Hamiltonian,
the ground state of which interpolates between the 331 state at $\theta=0$
and the MR state at $\theta=\pi/4$ were discussed critically in Ref.\
\cite{rr}. The model Hamiltonian contains a parameter corresponding to $\theta$
which directly determines the $\theta$ that describes the ground state
wavfunction. Except at $\theta=\pi/4$, the results of the model agree with the
above discussion of the abelian phase, though we do not now believe that the
U(1) symmetry is ``reappearing in the low-energy properties'' \cite{rr}.
At $\theta=\pi/4$, Ho's model is pathological, and the upshot of the
discussion was that if the three-body interaction of Greiter {\it et al.}
is also added as a isospin-independent interaction, then this pathology
of Ho's model is removed, and the ground state is nondegenerate
(on the sphere, without quasiholes) for all $\theta$ between $0$ and $\pi/4$.
It was noted that, even for this model, there are still peculiarities of the
$\theta=\pi/4$ point. In fact, there are unexpected degeneracies, larger than
those of the 331 or $0\leq \theta <\pi/4$ states, in the edge, quasihole and
toroidal ground states at $\theta=\pi/4$. These arise because the zero modes
can be occupied by either $e$ or $o$ fermions, even though there are no $o$s
in the ground state. In particular, on the torus in the $++$ sector, there are
degenerate states that differ only in the presence or absence of a $\bk=0$ $o$
fermion (these extra degeneracies in this model at this point were overlooked
in Ref.\ \cite{rr}, but can be obtained by the same methods used there).
This clearly suggests that the dispersion relation $E_{\bk o}$ for the $o$
Majorana fermions is gapless at this point, and seems to confirm that
{\em this model at $\theta=\pi/4$ is actually at the transition point between
the weak-pairing abelian and MR nonabelian phases}. This is consistent with
the result of the analysis here that when the leading long-distance part of
$g$ is described by $\theta=\pi/4$, the system is either at the transition or
in the MR phase. This is certainly not the case for $\theta<\pi/4$, as
indicated by the degeneracies found for the Ho Hamiltonian plus three-body
in Ref.\ \cite{rr}. However, the fact that the ground state of the model at
$\theta=\pi/4$ contains no $o$ fermion pairs at all, suggests that this point
is at the A2-A1 boundary as well as at the weak-MR transition. In the
quasiparticle effective Hamiltonian, this would correspond to vanishing
$\Delta_o$ as well as $\mu$, and therefore we would expect the dispersion
relation for the $o$ fermions to be $E_\bk \propto |\bk|^2$. Clearly this
is nongeneric behavior. When tunneling $t$ is also included, the value of
$\theta$ in the ground state cannot be read off the Hamiltonian in general,
but we expect that the weak to MR transition, which should now be generic,
is pushed to another value of the parameter in the Hamiltonian, so the region
in the MR phase with $\theta=\pi/4$ widens, while the A2-A1 boundary is still
at the value corresponding to $\theta=\pi/4$ originally, where the ground
state is known exactly \cite{rr}, and again contains no $o$ fermions.

%%%%%%%%%%%%%%%%%%%%%%%%%%%%%%%%%%%%%%%%%%%%%%%%%%%%
\section{Complex d-wave spin-singlet pairing}
\label{dwave}

In this Section we consider $l=-2$ complex d-wave pairing of
fermions, and more generally ${\rm d}_{x^2-y^2}+i{\rm d}_{xy}$,
which are necessarily spin-singlet. This has been considered
recently \cite{laughlin98,senthil1,chalker,senthil2,glr}. We argue
that the Haldane-Rezayi (HR) state \cite{hr}, which has this
symmetry, is {\em at} the transition {}from weak to strong
pairing. The weak pairing phase, like the strong, is abelian; we
work out its properties and its lattice description {}from the
pairing point of view, and identify the universality class as one
which has been obtained before by various methods, including a
trial wavefunction approach by Jain
\cite{jain,belkhir,milr2,leekane}. We relate these properties to
the Hall conductivity for spin, which we calculate explicitly in
an Appendix.

%%%%%%%%%%%%%%%%%%%%%%%%%%%%%%%%%%%%%%%%%%
\subsection{Weak- and strong-pairing phases, and the Haldane-Rezayi FQHE state}
\label{hr}

The basic structure of the problem is once again similar to the spinless p-wave
case, or (since it is spin-singlet) to the original BCS treatment, except for
being d-wave. The quasiparticle effective Hamiltonian is
\be
K_{\rm eff}=\sum_{\bk\sigma}\left(\xi_\bk c_{\bk\sigma}^\dagger c_{\bk\sigma}
+\Delta_\bk^\ast c_{-\bk\down}c_{\bk\up}+\Delta_\bk
c_{\bk\up}^\dagger
c_{-\bk\down}^\dagger\right),
\ee
where $\Delta_{\bk}\simeq\hat{\Delta}(k_x-ik_y)^2$ at small $\bk$.
The structure of the solution is similar to the spinless case.
However, the dispersion relation is now
\be
E_\bk=\sqrt{(k^2/2m^\ast-\mu)^2+|\hat{\Delta}_\bk|^2 k^4},
\ee
so at the transition point ($\mu=0$),
$E_\bk=k^2(|\hat{\Delta}|^2+(2m^\ast)^{-2})^{1/2}$. For $\mu$ nonzero,
\be
E_\bk\simeq |\mu|-\frac{k^2}{2m^\ast}{\rm sgn}\,\mu,
\ee
at small $\bk$, which implies there is a minimum at nonzero $\bk$ in the
weak-pairing phase, $\mu>0$.

The position-space wavefunction for $N$ particles, of which
$N/2$ have spin $\up$, $N/2$ have $\down$, has the form
\be
\Psi\propto\det g(\br_{i\up}-\br_{j\down}),
\ee
where $g$ is the inverse Fourier transform of $g_\bk=v_\bk/u_\bk$ (this is
equivalent to a result in Ref.\ \cite{schrieffer}, p.\ 48).
In the strong-pairing phase, $v_\bk/u_\bk\sim(k_x-ik_y)^2$, and $g(\br)$ falls
rapidly with $\br$. In the weak-pairing phase, $v_\bk/u_\bk \sim
(k_x+ik_y)^{-2}$, we find
\be
g(\br)\propto \overline{z}/z
\ee
for large $\br$. Thus $|g|\sim$ constant, and $g$ is very long-range. At the
critical point, $v_\bk/u_\bk\sim|\bk|^2/(k_x+ik_y)^2$ (with a coefficient that
depends on $\hat{\Delta}$ and $m^\ast$, unlike the p-wave case) and
\be
g(\br)\propto 1/z^2.
\ee
This is the same behavior as in the Haldane-Rezayi (HR) state \cite{hr},
when the latter is interpreted in terms of pairing of composite fermions
\cite{mr}. Therefore we suggest that {\em the HR state is precisely at the
weak-strong pairing transition point, and has gapless excitations in the bulk}.

Further evidence for the criticality of the HR state comes {}from the ground
states on the torus. For the quasiparticle effective Hamiltonian $K_{\rm eff}$,
the presence of two spin states means that the $\bk=0$ states, which occur
only for $++$ boundary conditions (see Section \ref{spinless}), can be
unoccupied in the strong-pairing phase, and doubly-occupied in the
weak-pairing phase. Thus there is a total of four ground states, all with $N$
even, and none with $N$ odd, in both phases. However, at the critical point,
$E_{\bk={\bf 0}}=0$, and the $\bk={\bf 0}$ state can be occupied zero, one, or
two times, with no energy penalty. Hence for $++$ boundary conditions,
there are two ground states for $N$ even, and two for $N$ odd. The latter pair,
in which $\bk={\bf 0}$ is singly occupied, form a spin-$1/2$ doublet. Thus
(on including the three ground states in the other sectors) there is a
total of five ground states for $N$ even, two for $N$ odd. This is exactly what
was found for the ground states of the hollow-core model of HR \cite{hr},
(for which the HR state is the exact ground state on the sphere), both
numerically \cite{hrunpub,rr} and analytically \cite{rr}. The long
distance behavior of the wavefunctions implied by the present approach agrees
with that in Ref.\ \cite{rr}, as in earlier cases. Also, we cite the energy
spectrum of the hollow-core model, which was obtained numerically for $N=8$
particles on the sphere in Ref.\ \cite{rr}. No clear gap can be seen in the
spectrum. In view of these results, other numerical work on this model should
also be reconsidered. Analytical results on zero-energy ``edge'' and
``quasihole'' states of the hollow core model \cite{milr,rr}, remain valid, but
the earlier interpretation assumed a fully-gapped bulk ground state,
and so if the bulk is gapless, the questions about the conformal field theory
pictures of the bulk wavefunctions and the fermionic edge excitations
\cite{mr,wwh,milr,gfn} are presumably moot. It is actually quite interesting
that the hollow-core model (like the Ho plus three-body
model in Section \ref{331}) is critical. In this it resembles also certain
other special Hamiltonians for which the exact ground states are known
\cite{rrapp,grr}; the latter cases involve pairing of composite bosons.

%%%%%%%%%%%%%%%%%%%%%%%%%%%%%%%%%%%%%%%%%%%%%
\subsection{Structure of the weak-pairing abelian phase}

It is now of interest to find the properties of the d-wave weak-pairing phase
(the strong pairing phase has the same rather trivial properties as the others
discussed previously). This phase has been discussed recently
\cite{senthil2}. Here we wish to consider its application as a paired state in
the FQHE. There are differences in the symmetry here compared with Ref.\
\cite{senthil2}, which necessitate a certain amount of discussion.
As we will see, the method of analysis of the edge and vortex states used
previously does not seem satisfactory in the present case. Therefore,
it seems necessary to use a more devious approach, which we now describe.

It is not difficult to see that, in the weak-pairing phase, the map {}from
$\bk$ to $u_\bk$, $v_\bk$ is topologically non-trivial and has $m=2$, that is,
it wraps around the sphere twice. Because of this and its d-wave ($l=-2$)
symmetry, it has double zeroes at $\bk={\bf 0}$ and $\bk=\infty$. While such
behavior at $\infty$ can be regarded as fixed by requirements of convergence
and finite particle number, as mentioned earlier, that at $\bk={\bf 0}$ is
non-generic {}from the topological point of view. More generically, the map
could pass over the south pole $\cal S$ in Anderson pseudospin space at two
different $\bk$ values, but this would require that the rotational symmetry be
broken.

It will be useful to analyze such generic behavior in order to find the
properties even of the d-wave case, as we will argue below.
A convenient way to break rotational symmetry is to introduce an s-wave
component $\Delta_{\bk s}$ of the gap function $\Delta_\bk$, in addition to the
d-wave part $\Delta_{\bk d}$ (s-wave is the simplest choice, and
the gap function must remain even in $\bk$, to retain the spin-singlet ground
state). Then at small $\bk$ the behavior is
\be
\Delta_\bk\simeq\Delta_s+\hat{\Delta}(k_x-ik_y)^2,
\ee
and $\Delta_s$ and $\hat{\Delta}$ are both complex. In this
case, when $\Delta_{\bk s}$ is a sufficiently small perturbation on the
weak-pairing d-wave phase, the effect is that $E_\bk$ has a minimum at just
two nonzero values of $\bk$, of the form $\pm\bk_0$ by symmetry. This does not
change the topology of the $u_\bk$, $v_\bk$, and since the system is fully
gapped as far as the quasiparticles are concerned, it should not change any
physical properties. However, if $\Delta_{\bk s}$ is large and dominates
$\Delta_{\bk d}$ for all $\bk$, then the system is essentially in an s-wave
state which must be topologically trivial. This is the same phase as the
strong-pairing d-wave phase, even though the rotational symmetry is different;
these limits can be connected as $\Delta_{\bk s}$, $\Delta_{\bk
d}$, and $\mu$ are varied, without crossing another transition. Therefore
there must be a phase transition as $|\Delta_{\bk s}/\Delta_{\bk d}|$ varies at
fixed $\mu>0$, similar to those discussed above. In
this case, the dispersion relation $E_\bk$ vanishes linearly at two points
$\pm\bk_0$ at the transition, and the map {}from $\bk$ to $u_\bk$, $v_\bk$ is
discontinuous at these two points. These will be points where
$\xi_\bk=k^2/2m^\ast-\mu=0$, as well as $\Delta_\bk=0$, and so occur at some
$\mu>0$. As $\mu\rightarrow 0$, $\bk_0\rightarrow0$, and these points
coalesce to give the previous discussion in which $\Delta_{\bk s}=0$ for all
$\bk$.

The two points $\pm\bk_0$ at which $E_\bk$ has a conical form give a spectrum
similar to a spin-$1/2$ doublet of Dirac fermions. This is similar to behavior
well-known in other condensed-matter models, including fermions on a lattice in
a magnetic field, and d-wave pairing with a ${\rm d}_{x^2-y^2}$ structure
(perhaps induced by a square lattice) \cite{senthil2}. By concentrating on the
degrees of freedom near these points, and shifting them in $\bk$-space to the
origin (which produces oscillating factors in real space correlation functions
of the fermions), the fermion excitations can be mapped onto a complex (i.e.
Dirac, not Majorana) spin-$1/2$ doublet of relativistic fermions for each pair
of such points in $\bk$ space. Near the transition point, or by varying
parameters in the other models mentioned (the ${\rm d}_{xy}$ part of the gap
function in the second model), the Dirac fermions gain a mass. If we now apply
to this a similar analysis for edges to that we used previously
when the minima were at $\bk=0$, then we find that the edge excitations form a
spin-$1/2$ doublet of chiral Dirac (or Weyl) fermion fields \cite{senthil2},
and we can give a similar analysis for the vortices.

For our purposes, we are interested in unbroken rotational symmetry in the
bulk, so the preceding remarks do not seem to apply directly. However, if we
analyze the edge excitations using the method of previous sections applied to
the d-wave model, by simply considering the effect of $\mu$ changing sign on a
domain wall, we do not find any edge modes. But we are suspicious of this
result because of the nongeneric form of $u_\bk$, $v_\bk$, and previously
we were relying on the assumption that the results are robust because of the
topological nature of the phenomena. In the present case, the results should be
the same as if we examine a domain wall caused by varying $\Delta_{\bk s}$ so
as to cross the transition, because the phases are the same, and therefore the
edge state and vortex properties should be those of the Dirac fermion doublet.
We expect that what happens, even if the bulk phases are rotationally
invariant, is that the edge or domain wall breaks rotational symmetry, and
induces a splitting of the zeroes of $\Delta_\bk$, in so far as this function
in $\bk$ space is meaningful. With the symmetry broken, the previous analysis
can be applied. Of course, what should be done in all cases is a full solution
of the BdG equations with the reduced symmetry, and of the gap equation
self-consistently. This is clearly difficult, though it has sometimes been
attempted, and a deeper analysis that explains why arguments of the type we
have given yield the correct results would be preferable. We will attempt to
give such an argument below. We note that, if we consider perturbations
analogous to the s-wave component in the cases in earlier sections, we do not
find any change in those results.

A feature of the Dirac-like nature of the fermion edge spectrum is that there
are two doublets of fermion modes, particles and antiparticles. These
correspond to the two points $\pm\bk_0$ in $\bk$ space {}from which they arose.
This can be described by saying that there is a U(1) quantum number or
``charge'', which we will call $M$, and that the particle and antiparticle
carry opposite values of $M$. It may be that the additional U(1) symmetry,
which must not be confused either with any part of the SU(2) of spin, or with
the U(1) of
underlying particle-number conservation, which is already broken spontaneously
by the pairing, is in fact broken by the dynamics, since there seems to be no
symmetry to protect it. However, as in the case of the 331 state plus
tunneling, even when the symmetry is broken, the counting of edge excitations,
and the statistics properties, etc, should be unchanged.

If we continue to assume the U(1) symmetry exists, for the sake of the latter
analysis, then the doublet of chiral Dirac fermions on an edge in fact has a
larger symmetry (note that there are no interactions to consider in the
theory). The fields are equivalent to four Majorana fermions, and there is
SO(4) symmetry. We note that, as Lie algebras, SO(4)$\cong$ SU(2)$\times$SU(2),
and here the first SU(2) can be identified with the spin-rotation symmetry
group, while the second contains the U(1) symmetry just discussed as a
subalgebra, generated by, say, rotation about the $z$ axis in the second space.
Thus the Dirac field and its conjugate can be viewed as carrying spin
$M=\pm 1/2$
under the second SU(2). Alternatively, viewing the theory just as two Dirac
fields, these can be bosonized, and we obtain two chiral bosons. The allowed
``charge'' states for the edge, which take values in the Cartan subalgebra of
SO(4), lie on a two-dimensional lattice. This lattice is a direct product of
two copies of the weight lattice of SU(2). Points in the lattice simply
describe the total $S_z$ of spin and the total $M$ on the edge. The same
desciption applies to a vortex, since we can view it as an edge rolled up into
a small circle. In the latter case, the half-flux quantum we assume in the
vortex corresponds to changing to periodic boundary conditions for the fermions
on the straight edge, for all components of the fermions. In addition to these
different boundary condition sectors, there are also selection rules {}from the
global quantum numbers. These are similar to the rules described in detail in
\cite{milr}. Specifically, for $N$ even, one can have the ground state with no
edge excitations, or one can create fermions on the edge, but only in even
numbers. In the present situation, each of these fermions can be in any of the
four states in the representation of spins $(1/2,1/2)$ under
SU(2)$\times$SU(2). For odd total fermion number, there must be one unpaired
fermion, which we can put on the edge to obtain a low-energy state. Then the
charge (or particle number) differs by one {}from the ground state, in addition
to the non-trivial SO(4) quantum numbers. Additional fermion pairs can be
excited in this case also. If we consider two parallel edges, as for a system
on a cylinder, then we build up the full spectrum by applying these rules to
the two edges together, and there will be different sectors corresponding to
the presence of either zero or one-half of a flux quantum threaded through the
cylinder. In the FQHE application, there are also chiral bosons for charge
excitations on the edge, and there are fractional charge sectors for each edge,
though the total charge must be integral, and the total charge (or particle
number) is correlated with the fermion excitations through its parity,
as already explained.

The quantum numbers of the vortices (or FQHE quasiparticles) correspond closely
to the possible quantum numbers for the edges, and are obtained in a similar
way. As a simple example, we again consider two vortices. The situation is
equivalent to the existence of a single fermion zero mode for the pair, which
can be occupied by any of four types of fermions. Thus there is a total of 16
states, of which half have $N$ even, half have $N$ odd. These should be
analyzed as a product of four states possible for each vortex. The states for
a vortex transform as $(1/2,0)\oplus(0,1/2)$ under SO(4), that is they either
carry spin 1/2 and no $M$, or vice versa. These correspond to the states for
two edges with a half flux quantum through the cylinder, in which the zero mode
shared by the two edges can be occupied by any of the four types of fermions.
For any even number of vortices, the results are similar, and the counting of
degenerate low-energy states is fully accounted for by the four states for each
vortex, and hence there is no nonabelian statistics. Thus a d-wave paired
state, or even a superconductor, in two dimensions, possesses vortices that
may carry spin 1/2, but not simply because a fermionic quasiparticle can
sit on the vortex. In the FQHE, these vortices also have well-defined
charge of $\pm 1/(2q)$ for filling factor $\nu=1/q$, where $q$ must be even
when the particles are fermions (such as electrons). This is also true in all
other cases discussed in this paper. It arises {}from the effective half
quantum of flux that the vortices carry, and they can exist only in even
numbers if the system has no edges.

We hope that the above discussion gives a sufficient impression of the quantum
numbers of the vortex states. For readers familiar with the general theory of
abelian FQHE states, we will now give a precise definition of the structure of
the state, using the Gram (or K) matrix language, which specifies the lattice
formed by the possible quantum numbers (including charge) of the FQHE
quasiparticles, as well as their statistics, and the order parameters and
chiral algebra of the edge theory \cite{read90,blokwen,wenrev,milr}.

The full lattice of possible quantum numbers of a vortex, or the total quantum
numbers of a set of multiple vortices, is denoted $\Lambda^\ast$
(as in Ref.\ \cite{read90}). We will describe it first as a set
of vectors in ${\bf R}^3$, using an orthonormal basis with the standard inner
product. Then $\Lambda^\ast$ consists of the set of vectors of the form
\be
{\bf v}=(r_1/\sqrt{2},r_2/\sqrt{2},r_3/[2\sqrt{q}])
\ee
where $r_1$, $r_2$, $r_3$ are integers obeying $r_1+r_2+r_3=0$ (mod 2).
Also the statistics of the excitation is $\theta/\pi={\bf v}^2$, where
$\theta$ is the phase picked up by exchanging two identical such
quasiparticles, and the conformal weight of the corresponding operator in the
edge theory is ${\bf v}^2/2$.
The basis has been chosen so that the three quantum numbers carried by the
excitations are proportional to the components in this basis. In fact, the spin
$S_z=r_1/2$, $M=r_2/2$, and the charge is $Q=r_3/(2q)$. {}From these rules
we see that the smallest possible vortex (the one with the smallest nonzero
${\bf v}^2$) carries either $S_z=\pm 1/2$ or $M=\pm 1/2$, and charge
$\pm1/(2q)$, as stated above, and has statistics $\theta/\pi=1/2 + 1/(4q)$ (mod
$2$). It is easy to verify that the full lattice $\Lambda^\ast$ is obtained
as the set of all linear combinations, with integer-valued coefficients, of
the vectors describing the quantum numbers of the basic (smallest) vortices,
and thus represents the possible quantum numbers of any collection of vortices.

The excitation lattice $\Lambda^\ast$ is the dual lattice to the condensate
lattice $\Lambda$, and $\Lambda$ is a sublattice of $\Lambda^\ast$. In the
present case, $\Lambda$ is the set of vectors $\bf w$ that are linear
combinations with integer coefficients of the vectors $(\pm 1/\sqrt{2},
\pm1/\sqrt{2},\pm\sqrt{q})$ (with all three signs independent) which represent
the underlying particles. These represent possible states for a particle
(electron) tunneling into an edge, and these electron operators, as we may term
them in spite of the emergent $M$ quantum number, are usually part of the
condensate lattice, as they are in the hierarchy theory. (An exception to this
is the strong-pairing phases, where only operators of charge a multiple of two
appear.) The fact that $\Lambda$ and $\Lambda^\ast$ are dual means that for any
${\bf v}\in\Lambda^\ast$, we have ${\bf v}\cdot{\bf w}\in{\bf Z}$ for all
${\bf w}\in\Lambda$, and vice versa. It suffices to check this for the
$\bf w$'s representing the electron operators.

Both lattices possess neutral sublattices, that is lattices of vectors such
that $Q=0$. The neutral sublattice of $\Lambda$, denoted $\Lambda^\perp$,
consists of vectors with $r_1$ and $r_2$ even. Thus these form a direct sum
${\bf Z}\oplus{\bf Z}$ of one-dimensional lattices. Each of the latter can be
viewed as the root lattice of SU(2) in Lie algebra theory, and
$\Lambda^\perp$ is the root lattice of SO(4). The neutral sublattice of
$\Lambda^\ast$, denoted $\Lambda^{\ast\perp}$, is the set of vectors with
$r_3=0$, and so $r_1+r_2=0$ (mod 2). This is a sublattice of the weight
lattice of SO(4), which would be the dual of $\Lambda^\perp$ as a
two-dimensional lattice. The simplest nontrivial neutral vector is of the form
$(\pm 1/\sqrt{2},\pm 1/\sqrt{2},0)$ (with independent plus and minus signs),
and these represent the neutral fermions, that is the BCS quasiparticles
considered in this paper. These cannot be created individually on a single
edge; only excitations lying in $\Lambda$, such as even numbers of such
fermions, can be \cite{milr}.

An integral basis for a lattice is a set of vectors in the lattice that are
linearly independent (over $\bf R$), such that all vectors in the lattice can
be written as linear combinations of those in the set, with integer
coeffecients. Such a basis cannot be an orthogonal set of vectors, unless
the lattice is a direct sum of one-dimensional lattices. In our case,
a convenient integral basis for $\Lambda$ (other than a suitable set of three
of the electron operators above), is ${\bf e_1} =(1/\sqrt{2},1/\sqrt{2},
\sqrt{q})$, ${\bf e_2}=(\sqrt{2},0,0)$, ${\bf e_3}=(0,\sqrt{2},0)$. The Gram
matrix of the lattice is the set of inner products of these vectors,
$G_{ij}={\bf e}_i\cdot{\bf e}_j$, and in this case is
\be
G=\left(\begin{array}{ccc}
         q+1&1&1\\
         1&2&0\\
         1&0&2    \end{array}\right).
\ee
The diagonal structure of the lower-right $2\times2$ block reflects the direct
product nature of the SO(4) root lattice. The determinant of $G$, $\det G=4q$,
determines the index of $\Lambda$ as a subgroup of $\Lambda^\ast$, that is the
number of equivalence classes of vectors in $\Lambda^\ast$ modulo $\Lambda$.
It gives the number of ground states of the system on a torus, or the number of
sectors of edge states. Factoring off the center of mass degeneracy $q$,
we find that there are four, in agreement with the analysis based on the
quasiparticle effective Hamiltonian. The Gram matrix, together with
distinguished vectors that specify the charge $Q$ and spin $S_z$ quantum
numbers, is sufficient information {}from which to reconstruct the
lattices $\Lambda$ and $\Lambda^\ast$, and hence the universal aspects of
the phase, such as ground state degeneracies, quasiparticle statistics,
and the theory of the edge states. In this context, the Gram matrix is
often called the K matrix.

This completes the analysis we will give of the phase. It is a generalized
hierarchy state in the sense of Ref.\ \cite{read90}, and resembles the 331
phase. The latter lacks the SU(2) of spin and hence has a two-dimensional
lattice; its structure was described in detail in Ref.\ \cite{milr,rr}.

We may now compare the universal properties of this state with others analyzed
previously. We find that several other constructions of this spin-singlet phase
have already been given. In Ref.\ \cite{read90}, this was mentioned briefly as
the structure of both a spin-singlet state for $\nu=1/2$ proposed by Jain
\cite{jain} (see also Ref.\ \cite{belkhir}), and one proposed by Lee and
Kane \cite{leekane}. A more detailed analysis was given in Ref.\ \cite{milr2},
where it was also identified with a hierarchical construction. In the latter,
one starts with the Halperin spin-singlet 2/5 state \cite{halp83}, which is a
$332$ state in the Coulomb plasma language, and then one makes a finite
density of spinless quasielectron excitations of that state, each carrying
charge 2/5. The quasielectrons are then put in a Laughlin 1/2 state, to obtain
a singlet state with $\nu=1/2$. The hierarchical step implies that the
resulting state has a three-dimensional lattice. In Ref.\ \cite{milr2}, this
and the Jain construction were shown to coincide. Unfortunately, the formulas
there contain a small mistake: the final basis vector in Eq.\ (4.8) in that
reference should be reduced by a factor of two, as should the entries in the
first row and column of the Gram matrix in Eq.\ (4.9) there. The resulting
matrix is then identical to $G$ above, with $q=2$, after permuting the basis
vectors. This basis is the natural one for the hierarchy approach. In our $G$
above, the top left $2\times2$ block (with $q=2$) describes the Halperin 2/5
state, and reflects its origin. (The $2/5$ state itself has the same $G$ as the
spin-polarized hierarchy $2/5$ state, which descends {}from $\nu=1/3$, as
reflected by the $q+1=3$ at the top left.)

The Jain $\nu=1/q$ spin-singlet state was proposed as a trial wavefunction,
namely
\be
\Psi=\chi_1^{q-1}\chi_2\chi_{1,1}.
\ee
Here $\chi_m$ stands for the wavefunction for $m$ filled Landau levels of
spinless particles, (so $\chi_1$ is the Vandermonde determinant or
Laughlin-Jastrow factor) and $\chi_{1,1}$ is the lowest Landau level filled
with particles of both spins. We ignore the Gaussian factors in these
wavefunctions, and have omitted the projection to the LLL. The filling factor
is again $1/q$. This wavefunction can be loosely viewed as a Coulomb plasma
of particles carrying charge, spin and another quantum number $M=\pm 1/2$ to
represent the two Landau levels in $\chi_2$. The exponents in the wavefunction
correspond to inner products of corresponding vectors, which are just those of
the four electron operators with $Q=1$. This establishes the equivalence,
as for the hierarchy states in Ref.\ \cite{read90}. In fact, the extra SU(2)
that appears here acts on the LL indices in $\chi_2$ in Jain's function,
just like the SU$(n)$ that appeared for $n$ Landau levels in the
spin-polarized Jain states \cite{jain} and the corresponding hierarchy states
with $n$ levels in the hierarchy. The state is a combination of the Halperin
spin-singlet structure with the spinless composite-fermion/hierarchy
structure. We should point out that the number of flux for Jain's state on the
sphere is $N_\phi=q(N-1)-2$, the same as for the HR state, which follows {}from
our analysis, and was noticed previously \cite{belkhir}. This is of course
essential in order for it to be possible to vary parameters smoothly to reach
the transition.

Thus the same $\nu=1/2$ (more generally, $\nu=1/q$) spin-singlet
phase has arisen in four different ways. We want to emphasize that
the equivalence of the universal long-distance properties does not
mean that the trial wavefunctions in different approaches are the
same. For example the paired wavefunction found here and Jain's
above do not look alike. It may be that one is a much better
description (has a much larger overlap with an exact ground state)
for medium size systems than the other, even though they describe
the same phase. The equivalence found here is analogous to that
between the 331 state and the A-phase p-wave state, however in
that case there was an exact equivalence of certain wavefunctions
through the Cauchy determinant identity. We may still expect some
equivalence in the long-distance form of the wavefunctions.

%%%%%%%%%%%%%%%%%%%%%%%%%%%%%%%%%%%%%%%%%%%%%
\subsection{Induced Chern-Simons actions and analogs of the Hall conductivity}
\label{spinhall}

The arguments given so far for the edge states and for zero modes on vortices,
on which the identifications of the weak-pairing phases have been based, may
appear not to be very well-founded, as they have been based on analyzing
the BdG equations for special forms of the gap function and variation of the
parameter $\mu$, though we did argue by continuity that the states found must
persist as the equations are varied while staying in the same phase. In this
Subsection, we argue that the results we have obtained are in fact very
robust, because the edge states, and the form of the bulk theory described by
the Gram matrix of the condensate lattice, are required as a consequence of
transport properties of the bulk weak-pairing phases. These transport properties
are the quantization of the spin and heat analogs of the Hall conductivity,
which we prove explicitly for the spin case. They imply the existence of
certain edge state structures, just as in the case of charge transport in the
QHE, and when the weak-pairing phases correspond to abelian FQHE states, the
Hall spin conductivity is actually a part of the Gram matrix description. The
remainder of this Subsection discusses these points, but the technical details
are relegated to the Appendix. We should point out that the form of the
argument has already appeared in Ref.\ \cite{senthil2} for the cases with SU(2)
or U(1) spin symmetries, though the explicit derivation of the Hall
conductivities was not given there.

For the cases of spin-singlet pairing, and of p-wave pairing with
an unbroken U(1) symmetry, we derive in the Appendix the Hall spin
conductivity $\sigma_{xy}^s$, and show that in any fully-gapped
translationally-invariant superconducting phase it is given by a
topological invariant, which within the BCS approximation is
proportional to the same winding number we discussed earlier in
Sec. \ref{spinless} and subsequently. Some similar statements
appeared earlier, but were for the charge Hall conductivity
\cite{volovik88,volovik92}, and also for the existence of a Hopf
term and of a nonabelian CS term, in the two-dimensional A-phase
with SU(2) symmetry broken spontaneously \cite{volyak}. In this
paper we restrict ourselves to the Hall conductivies for conserved
quantities for which the corresponding symmetry is not broken
spontaneously in the paired state, such as spin and energy. For
quantities for which this is not true, such as charge, we do not
find quantized Hall conductivities in superconductors (though of
course we do find quantized Hall conductivity in the QHE systems).
Since our point of view differs somewhat {}from that in, for
example, Ref.\ \cite{volyak}, we give a self-contained discussion.

To be precise, for the Hall spin conductivity in a spin-singlet paired state,
we find that if the particles are viewed as carrying spin $1/2$ (we usually set
$\hbar=1$), the Hall response to an applied spin analog of the electric field,
such as a gradient in the Zeeman splitting, is
\be
\sigma_{xy}^s = m \frac{(\hbar/2)^2}{2\pi\hbar},
\ee
where $m$ is the winding number, which is $\pm 2$ in a d-wave weak-pairing
phase. We have written the Hall spin conductivity in this form to emphasize the
similarity to the usual $e^2/h$, with $e$ replaced by $\hbar/2$ here. In
$\hbar=1$ units, we obtain in our d-wave weak-pairing phase $\sigma_{xy}^s
=1/4\pi$. We chose the $+$ sign, since in the FQHE applications of the $l=-2$
state, the edge modes propagate in the same direction as the charge modes.
This agrees with Refs.\ \cite{chalker,senthil2}, where different arguments
were used. For the p-wave case, we view the fermions as carrying isospin
$\pm 1$, and hence the Hall spin conductivity we obtain is
\be
\sigma_{xy}^s = m \frac{1}{2\pi\hbar}
\ee
where the winding number $m$ is $\pm 1$ in the weak-pairing p-wave phases,
again with $m=1$ for the $l=-1$ case.

These results agree with the descriptions we have already given of
the weak-pairing FQHE phases using the Gram matrix or lattice
theory. In fact we should point out that a quantized Hall spin
conductivity is not unusual in FQHE systems, though it is not
always emphasized. It occurs for example in any spin polarized
state, such as the integer and Laughlin states with $\nu=1/q$,
because the electrons carry spin 1/2 as well as charge, and so the
two Hall conductivities are proportional. It also occurs in some
spin-singlet Hall states (abelian examples were discussed in
\cite{milr2}), including the $\nu=2$ state with the LLL filled
with both spins, and the Halperin $mmn$ states with $m=n+1$, $m$
odd, which are a generalization of the $\nu=2$ state. In these
cases, we obtain the full SU(2) version of the Hall spin
conductivity, with $\sigma_{xy}^s$ taking the same value as in the
d-wave weak-pairing phase. The same quantized Hall spin
conductivity was also found in certain spin-liquid states for
lattice antiferromagnets \cite{haldgirv}. The wavefunctions of
these states are the same as that of the Halperin state, with the
charge degree of freedom removed (i.e.\ the wavefunction is
$1/2,1/2,-1/2$), and the spin-1/2 particles restricted to lattice
sites, explaining this result.

The result that the Hall conductivity, in units of (quantum
number)$^2/h$, is a topological invariant given by an integral
over $\bk$ space, is similar to one form of the ordinary charge
Hall conductivity, found originally for a noninteracting periodic
system with a rational number of flux quanta per unit cell, as an
integral over the Brillouin zone \cite{tkndn}. In systems where
$\bk$ is not a good quantum number, such as the same system with
irrational flux, or when disorder is present, or in other
geometries, including those with edges, that break translational
invariance, such a topological invariant is apparently not
available. Yet the idea of quantization as resulting {}from the
conductivity being a topological invariant that measures an
intrinsic local property of the ground state seems too good to
give up. For the noninteracting QHE with disorder, the topological
invariant has been extended using noncommutative geometry, so that
the Hall conductivity is an integral over a ``noncommutative
Brillouin zone'', and in this way quantization has been proved
even for the physically relevant case of a nonzero density of
localized states at the Fermi energy \cite{bel}. It would be
interesting to extend this to other cases, including the paired
states with disorder, which we discuss in the next Section. In the
Appendix, we give arguments that the form we obtain is exact to
all orders in the interactions, but only for a
translationally-invariant system.

Now we can use the results on the Hall spin conductivity to argue that the
edge state properties we have obtained are indeed correct. In the Appendix,
we derive the Hall spin conductivities by obtaining the
induced action for an external gauge field that couples to the spin or isospin.
The actions that result (cf.\ Ref.\ \cite{rahw}) are CS terms for an
SU(2) gauge field in the spin-singlet (d-wave) case, and for a U(1) gauge
field in the p-wave case. Now, using either the Hall conductivities and
arguing as in the QHE literature \cite{halp82} (see also \cite{hald95,xrs}),
or using the induced actions and arguing as in the field theory literature
\cite{calhar} (quite similar arguments appear in Ref.\ \cite{vol98} and
references therein), we can conclude that on a domain wall between phases
with different $\sigma_{xy}^s$'s (one of the phases might be the vacuum outside
an edge, with $\sigma_{xy}^s=0$) there must be chiral edge excitations. In the
presence of a uniform spin-electric field, a spin-current is induced in a
region with $\sigma_{xy}^s\neq0$, and the normal component of this at a domain
wall has a discontinuity, representing a net inflow of spin onto the wall.
To avoid violating the continuity equation for the spin density and spin
current density, there must be chiral modes on the wall, and a ``gauge
anomaly'' in the conservation of spin on the wall alone. The tangential field
induces a nonzero divergence (i.e., an anomaly) of spin current along the
wall, which cancels the net inflow {}from the bulk. Such an anomaly can
occur only if gapless chiral excitations exist on the domain wall. The minimal
chiral theory required to produce the anomaly is the usual chiral Luttinger
liquid (or chiral Gaussian model) in the U(1) case, with the value of the
coupling that corresponds to a free chiral Dirac (Weyl) fermion in $1+1$
dimensions, and the SU(2) chiral Wess-Zumino-Witten (WZW) theory \cite{wit84}
with $k=m/2=1$ in the SU(2) case. All of this applies even within the BCS mean
field framework we used before, and then the edge excitations must be free
fermions. In the p-wave case, we therefore expect simply a single chiral Dirac
fermion to propagate on the edge. In the d-wave case we must have an SU(2)
doublet of chiral Dirac fermions, which can be represented by the $k=1$ chiral
WZW model, together with an additional chiral U(1) degree of freedom,
which we argued earlier must exist, and called $M$. We have therefore
reproduced the claimed results about the edge states, neglecting the
charge degree of freedom.

We note that, when formulating such arguments for the nature of the chiral
edge theories, we can presumably assume that the theories are
unitary, conformal fields with local current operators for physical conserved
quantities (as usual, Lorentz invariance may be spoiled by the presence of
different velocities for different excitations, but this will not matter for
the statistics and other universal properties in which we are interested).
Previously, we might not have assumed this, because of the example
of the Haldane-Rezayi (HR) state in particular. But we have learned that the HR
state is at a critical point, and previous discussions of the edge and
quasiparticle properties of that state are irrelevant. Thus, with the demise
of the HR state as a bulk phase, it becomes attractive to believe that the edge
theories of incompressible FQHE phases are always unitary conformal field
theories. With this assumption, in theories with SU(2) Hall spin
conductivities, unitarity of the edge theory requires quite generally that
$k$ be an integer \cite{wit84,wit89}, and so $\sigma_{xy}^s=k/(4\pi)$ in the
above conventions.

The preceding arguments do not apply to the spinless p-wave case, in which
there is no continuous unbroken symmetry. This is unfortunate in view of the
great interest in the nonabelian properties of the weak-pairing phase.
But there is another Hall-type conductivity, which exists in all cases,
including those without a continuous symmetry. This is the Leduc-Righi (LR)
conductivity, which is the $xy$ component $\kappa_{xy}$ of the thermal
conductivity, and is of course related to the transport of energy, a conserved
quantity. Like the Hall conductivities for charge and spin, this is a
non-dissipative transport quantity that can only be nonzero when parity and
time-reversal symmetries are broken. In systems with a gap for all bulk
excitations and with chiral edge excitations, it can be argued that the LR
conductivity is nonzero \cite{kflr,senthil2}. Thus this applies to QHE systems,
and to superconductors (paired systems) if there is no gapless collective
charge mode. The value of $\kappa_{xy}$ can be obtained \cite{kflr,senthil2}
by considering a sample with two edges and a small temperature difference
between the edges (strictly speaking, the following argument yields the
LR conductance, not the conductivity). The chiral excitations on each edge
are excited by different temperatures, and this produces a larger heat current
on one edge in one direction than that on the other edge in the other, and
hence a net heat current. This shows that the current is related to the heat
capacity of each mode on an edge, times the velocity of the mode, summed over
modes. It is known that the heat capacity for each mode is related to the
Virasoro central charge $c$ in the conformal field theory of the edge
excitations \cite{ca}. In $\kappa_{xy}$, the velocities cancel, and the
remaining number is proportional to the total central charge of the edge
theory (including the charge modes in the FQHE); precisely,
\be
\kappa_{xy}=c\frac{\pi^2 k_B^2 T}{6\pi\hbar}.
\ee
{\em This is the fundamental relationship governing the LR conductivity in all
quantum Hall problems.} Here we assumed that all modes on an edge propagate in
the same direction; if not, then $c$ should be replaced by the difference of
the central charges for the right and left-moving theories. Related to this,
the two-probe thermal conductance of such a system is also equal to
$\kappa_{xy}$ (for a case with all edge modes propagating in the same
direction), just as the two-probe conductance is equal to the quantized
$\sigma_{xy}$ of the bulk. It has been found that the two-probe thermal
conductances of the Laughlin states at various $\nu=1/q$ are independent
of $q$, even though the theories of the chiral Luttinger liquids contain
$q$ as a parameter \cite{kirczenow}. This is because the central charge
is $c=1$, independent of $q$.

The formula for $\kappa_{xy}$ is similar in structure to that for the Hall spin
conductivity. However, the central charge $c$ does not have to be an integer,
and indeed for a single Majorana fermion, $c=1/2$ \cite{bpz}, so that for
free fermions in general, $c$ is a multiple of $1/2$. Nonetheless, we
do expect it to be quantized, in the sense of invariant under small
deformations of the theory (including adding weak disorder).

The argument above for the LR conductivity made use of the edge states.
However, we want to use it to back up our results on the fermionic edge modes
of the MR state, in analogy with the arguments for the other cases, which used
the Hall spin conductivities. Hence we need an independent argument for the
value of the LR conductivity. We believe that it should be possible to derive
such a result, analogously to the Hall spin conductivities, by considering the
system in external {\em gravitational} fields. Here the Christoffel symbols, or
the spin connection, play the role of the external gauge fields we used in the
spin case, but should be viewed as determined by the metric of spacetime,
which we treat as the independent variable and set to the usual Minkowski
metric after calculating responses. The role of the currents, to which the
vector potentials couple, is played by the energy-momentum tensor, which
includes the energy flux among its components. The significance of changing the
metric should be clear if we recall that equilibrium systems can be represented
in imaginary time, with the imaginary time direction periodic, the period being
$1/k_B T$. Thus a temperature gradient might be viewed as changing the geometry
of space and (imaginary) time. A more rigorous derivation would avoid imaginary
time, but should still involve the response to changes in the metric.
The leading term in the induced action for the external gravitational field in
$2+1$ dimensions will in general be the gravitational CS term, which can be
written in terms of the spin connection in close analogy to the usual CS terms:
\be
\frac{1}{4\pi}\frac{c}{24}\int d^3r \epsilon_{\mu\nu\lambda}{\rm
tr}
(\omega_\mu\partial_\nu\omega_\lambda+\frac{2}{3}\omega_\mu\omega_\nu\omega_\nu)
\ee where we view the spin connection $\omega$ as a $3\times3$
matrix, determined by the metric. (Such a term was also proposed
earlier for He$^3$-A films in Ref.\ \cite{vol90lett}.) The
coefficient contains $c/24$, which shows the relation to the
central charge $c$ in a corresponding chiral conformal field
theory on a boundary \cite{wit89}, is needed to cancel the anomaly
in energy-momentum conservation on the boundary by an inflow (LR
``Hall'' current) {}from the bulk \cite{calhar}, as for the charge
and spin Hall conductivities.

We have not completed a calculation (analogous to those in the
Appendix) of the induced action or LR conductivity for the bulk
{}from first principles. It would involve coupling the underlying
system to an arbitrary metric, then using Ward identities to
relate the response to a topological invariant. However, if we
assume a spectrum of relativistic fermions at low energies, with a
minimal coupling to the gravitational field, then the calculation
can be carried out, and is known in the literature \cite{alwit}.
The coefficient of the induced gravitational CS term again has the
form of the same topological invariant as in the Appendix, but
integrated over only half the sphere, as in the similar treatment
of the spin cases. For a single Majorana, this yields the above
form with $c$ replaced by $\pm1/4$ on the two sides of the
transition. Inclusion of the contribution of a Pauli-Villars
regulator then produces $c=0$ on one side, $c=1/2$ on the other.
We expect that for our nonrelativistic system, where the large
$\bk$ behavior is explicitly known, we would obtain such a result
also, with $c$ in a general system of paired fermions being
proportional to the same topological invariant (winding number) we
have seen already, and (allowing for factors of two associated
with spin degeneracy) we would find $c=1/2$ in the spinless p-wave
weak-pairing phase, and also $c=1$ ($2$) in the triplet p-wave
(singlet d-wave) cases. (These are the results for the paired
fermions, and in the FQHE would have to be supplemented by adding
$c=1$ for the charge degrees of freedom.) Therefore we believe
that the existence of the edge states and vortex degeneracies in
the MR phase can be placed on a firm footing.

A somewhat related issue is to obtain effective actions describing the
weak-pairing phases. We emphasize that the CS actions discussed in this
Subsection are {\em induced} actions for external fields which act as
sources, and should not be confused with {\em effective} actions, which contain
fields that should be functionally-integrated over, and represent the dynamics
of the system at low energies and long wavelengths. For the abelian FQHE
states, the effective actions fall into the framework of the known theory,
based on the Gram (K) matrix \cite{wenrev}. For the MR, and other similar,
phases, something different is required. For the abelian phases, there is an evident
similarity between the induced and effective actions. For example, in the
d-wave case, neglecting the charge degree of freedom, one would expect the
bulk effective action to be an SU(2)$\times$U(1) CS theory, with $k=1$ for
the SU(2) part (here, in the effective action, $k$ must be an integer to
maintain gauge invariance). This theory is determined by the requirement
that it produce the correct edge theory \cite{wit89}. (The U(1) part
could possibly be extended to a second SU(2) with $k=1$, to agree with
the SO(4) edge theory discussed above.) It is known that such an
effective theory also produces the desired induced gravitational CS term
in all cases \cite{wit89}. By analogy, we are led to expect that the MR
phase, where the edge theory involves only the $c=1/2$ representations of
the Virasoro algebra, can be described by an effective theory which is
some sort of gravitational CS theory (similarly, there were earlier
proposals for gravitational-type effective actions for three-dimensional
paired states \cite{volovik90}). Quantization of such a theory should
yield Virasoro conformal blocks in the same way that quantization
of CS theory yielded current algebra blocks \cite{wit89}, and thus be closely
related to the wavefunctions discussed in Sec.\ \ref{spinless} and in
\cite{mr,rr}. This hope is encouraged by the identification of the parameter
$\hat{\Delta}$ and the vector potential $A$, which should be functionally
integrated over in the full treatment, as the vielbein and spin connection
of $2+1$ gravity, or at least as the part relating to SO(2) rotations of space
only, as discussed in Sec.\ \ref{spinless}. The hope of producing conformal
blocks in this manner {}from a gravitational analog of CS theory has been
around for a long time \cite{wit89}, but does not seem to have reached
fruition, in spite of an interesting attempt by Verlinde \cite{herver}.
Such a theory would be an interesting, possibly more natural, alternative
to the ``conventional'' approach, along lines anticipated in Ref.\ \cite{mr},
of a CS analog of a coset construction \cite{frad}.

We also wish to comment on whether our results imply that fractional and
nonabelian statistics occur in paired superfluids, as opposed to FQHE systems.
For example, the weak-pairing d-wave and spin-triplet p-wave phases have
nontrivial Hall spin conductivity, and the smallest possible vortices carry
spin (or $M$) quantum numbers. Hence the spin degrees of freedom contribute
a fractional amount to the Berry phase on exchanging such vortices. Similarly,
exchanging vortices in the weak-pairing spinless p-wave phase should produce a
matrix action on the space of degenerate states we have identified, which we
may be tempted to term nonabelian statistics. However, although these
contributions {}from the spin (or fermion number) sector are well-defined,
in a neutral superfluid the charge degree of freedom is gapless, and the
vortices act on the charge (particle number) variables
also, so as with vortices in a simple neutral superfluid, which carry no
well-defined particle number, the contribution to the total Berry phase is
not well-defined, due to the charge sector \cite{halwu}. Nonabelian
statistics is still meaningful, modulo phase factors. In the
incompressible FQHE phases, this problem disappears, and the statistics
properties have been characterized above, in detail for the abelian cases.
Also, in a superfluid with a Coulomb interaction, there is again no
problem, even if the interaction is $\sim 1/r$, which does not produce a
plasmon gap in two dimensions. Vortices are neutral because of screening,
and so nonabelian statistics, or fractional statistics in the triplet
p-wave case, can occur, with no contribution {}from the charge sector.
There are also ``neutral'' vortex excitations with no net (spin-independent)
vorticity acting in the charge sector, which would not be subject to the
problem in compressible superfluids, but these are found not to have
fractional statistics. Note that a similar problem as for the charge
sector (in the compressible case) occurs in the spin sector in connection
with any subgroup of SU(2) that is spontaneously broken.

%%%%%%%%%%%%%%%%%%%%%%%%%%%%%%%%%%%%%%%%%%%%%%%%%%%%
\section{Effects of disorder on the transitions}
\label{disorder}

In this Section, we discuss the effects of disorder on the phases
above, and on the transitions between them. We consider the
phase-coherent, zero-temperature case, and neglect all
interactions between the quasiparticles (including the gauge field
fluctuations). The problem then reduces again to the quasiparticle
effective Hamiltonians $K_{\rm eff}$, this time with $\xi_\bk$ and
$\Delta_\bk$ replaced by operators that are local in position
space, with short-range correlations of the disorder. We consider
the problems above in reverse order, starting with the d-wave
case, which has the most symmetry (SU(2) of spin rotations), then
d- or p-wave with only U(1) symmetry, and finally the spinless
p-wave case, with no continuous symmetry. The first case has been
studied recently in the context of disordered superconductors
\cite{az,sfbn,bundschuh,senthil1,chalker,senthil2,glr}, while we
will argue that the second maps onto the usual noninteracting QHE
transition, and the last includes an unusual intermediate phase
where the heat transport is similar to that in a disordered {\em
metal}. In all cases, we expect the qualitative results to be
unaffected by interactions (or quantum fluctuations around the
mean field theories used), though the universality classes may be
changed.

The problem of the noninteracting BCS quasiparticles in disordered paired
fermion systems was discussed by Altland and Zirnbauer (AZ) \cite{az},
where a symmetry classification of random matrix ensembles was proposed,
that is analogous to the
familiar classification for ordinary one-particle Hamiltonians into
orthogonal, unitary, and symplectic ensembles (or symmetry classes). In these
ensembles, no particular value of the energy is singled out as special,
so that they apply to phenomena near generic energy (or Fermi energy) values,
where the average density of states is nonzero. The disordered paired systems
differ first in that the fermion number is not conserved, because of the
pairing terms. The full Hamiltonian of course conserves number, but this
involves the collective response of the condensate. In applying these models
in the FQHE, this is again true, but leads to the Hall response, not a
superfluid response. Therefore we disregard particle number transport,
and concentrate on conserved quantities carried by the quasiparticles only.
The latter quantities include spin, when this symmetry is not broken
spontaneously by the pairing. Then the classification of ensembles by AZ is
according to whether time-reversal symmetry (T) is broken or not, and whether
or not there is an unbroken SU(2) spin rotation symmetry; the latter is
unbroken in spin-singlet paired states. By making certain transformations, such
as a particle-hole transformation on the $\down$ spins in the cases where $S_z$
(at least) is conserved, the quasiparticle Hamiltonians can be related to
number-conserving Hamiltonians, and thus to single-particle or random matrix
problems. In this way, AZ identified four classes of random-matrix ensembles
for disordered paired systems, which they labelled C, CI, D, DIII.
For the quasiparticle Hamiltonians, the zero of quasiparticle energy $E$ is
a special point, where in most cases the average density of states of the
quasiparticles in the disordered system vanishes. Thus these four classes are
distinct {}from the usual three mentioned above.

The case of spin-singlet paired states with disorder was
considered in more detail \cite{oppermann,sfbn,bundschuh}. The
symmetry classes are CI (with time reversal symmetry), and C
(without time reversal). In the particular case of class C, it is
natural to consider in two dimensions the possibility of a Hall
spin conductivity $\sigma_{xy}^s$, as in the weak-pairing d-wave
phase \cite{chalker,senthil2}. There are nonlinear sigma model
formulations for class C, either using replicas, which lead to
target manifold Sp$(2n)$/U$(n)$ (with $n\rightarrow0$) in the
compact formulation \cite{sfbn}, or supersymmetry, which leads to
the target supermanifold Osp$(2n|2n)$/U$(n|n)$ (with $n>0$
arbitrary) \cite{az}. The Hall spin conductivity shows up in that
a topological term with coefficient $\theta$ proportional to
$\sigma_{xy}^s$ can be included in these models, when the
dimension of space is two \cite{sfbn}. Analysis of the model, and
numerical work on a network model with the symmetries appropriate
to class C \cite{chalker}, has shown that this system has phases
with quantized $\sigma_{xy}^s$ which are the same, or multiples
of, those we discussed in the pure system in Subsec.\
\ref{spinhall}. These phases can therefore be viewed as the
disordered analogs of the weak- and strong-pairing phases. In
these phases, the BCS quasiparticle (fermion) excitations at low
energies have a non-zero density of states which vanishes
quadratically as $E\rightarrow0$, and these states are localized
\cite{senthil1}. The situation is thus similar to the usual QHE,
in which the low-energy excitations are localized and have
nonvanishing density of states at the Fermi energy. Localization
is necessary to obtain the quantized Hall conductivity. In the
weak-pairing phase, we therefore expect that {\em the results for
the edge and statistics properties obtained in Sec.\ \ref{dwave}
are still valid when disorder is included}. We do not expect this
conclusion to be affected by the inclusion of interactions in the
analysis.

\begin{figure}
\epsfxsize=3.375in
\centerline{\epsffile{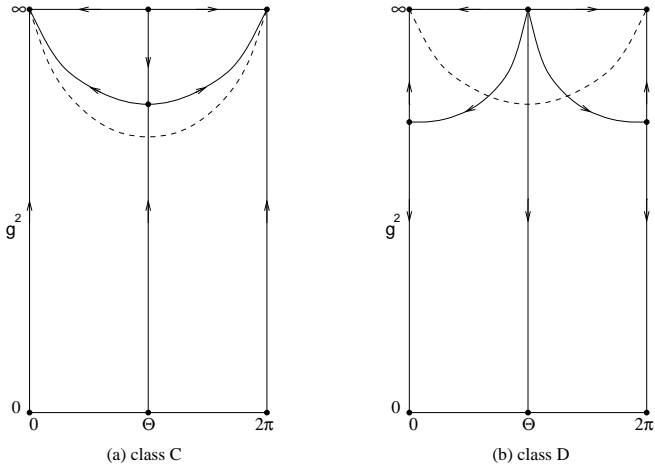}}
\vspace{0.1in}
\caption{Proposed renormalization group flow diagrams for (a)
the unitary ensemble (IQHE), as in \protect\cite{pruisken}, and class C,
and (b) class D. The dashed curves represent schematically the (nonuniversal)
bare values of the coupling parameters. Other features are universal when the
renormalized couplings are defined using the as-measured conductivity
parameters, as explained in the text, and repeat periodically in the
$\theta$ variable.}
\label{flow}
\end{figure}

The transition between these two phases, and the role of $\sigma_{xy}^s$, can
be understood via a renormalization group (RG) flow diagram, shown
schematically in Fig.\ \ref{flow} (a). The flows can be thought of as
representing the values of the local conductivity parameters $\sigma_{xx}^s$,
$\sigma_{xy}^s$ that would be measured at a given length scale, and how they
change with this scale. In the nonlinear sigma model, similar parameters
appear in the action of the quantum field theory. It is natural to define the
renormalized values of these couplings at any scale to be the conductivities
that would be measured at that scale, in which case the flows are the same as
the RG flows within the field theory model. The conductivities
should be understood in this way {}from here on, instead of as the bare values,
with which they coincide only when the scale is of order the mean free path.
Quantized values refer to the renormalized values at very large scales.
In the case of class C, the form of the flows is
identical to that in the IQHE for noninteracting electrons \cite{pruisken}.
Similarly to the usual IQHE, the transition occurs because of a nontrivial
fixed point, at which $\sigma_{xy}^s$ is midway between adjacent quantized
values, and $\sigma_{xx}^s$ will take some nontrivial universal value.
However, the spin quantum Hall transition in class C is in a different
universality class {}from the IQHE. Recent numerical work has obtained some
of the critical exponents for this transition, both {}from the network model
\cite{chalker}, and {}from a mapping of the network model to a supersymmetric
vertex model and a superspin chain \cite{senthil1,glr,senthil2}, which was
then analyzed numerically \cite{senthil2}. The results are in excellent
agreement with exact values of some exponents, which were proposed using
the relation of the supersymmetric vertex model to classical percolation
\cite{glr}. For example, the localization length exponent is $\nu=4/3$,
and at the critical point the density of states vanishes as $E^{1/7}$.
The effect of interactions on these results is presently unknown.

Inclusion of a Zeeman splitting $\propto h$, which was neglected so far,
will split the transition, and the phase diagram as a function of $\mu$
and $h$ will be similar to Fig. 4 in Ref.\ \cite{senthil2}. There will be an
intermediate phase in which $\sigma_{xy}^s$ is quantized and halfway between
the quantized values in the phases on either side. (This is somewhat like the
intermediate MR phase produced by tunneling $t$ in the p-wave case.) In the
present case, the Zeeman term leaves unbroken a U(1) subgroup of the SU(2)
symmetry present in the spin-singlet paired state; for ${\bf h}$ parallel to
the $z$ axis in spin space, this U(1) is generated by $S_z$. The pairing
is still between spin $\up$ and spin $\down$, and so the symmetry is the same
as in the p-wave case: with disorder, the distinction between p-wave and
d-wave, and between spin-singlet and spin-triplet, is lost. Therefore
we expect that in this intermediate phase, there is a single chiral Dirac
fermion mode on the edge, and that the statistics properties are the same
as those of the 331 state. Also, the transitions {}from the weak-pairing abelian
phase, in which the statistics properties will be unchanged even though the
SU(2) symmetry is broken (similar to the effect of $t$ on the 331 state),
to this phase, and {}from this phase to the strong-pairing phase with
$\sigma_{xy}^s=0$, are expected to be in the universality class of the usual
IQHE \cite{chalker,senthil2,glr}. Here the unbroken U(1) of spin is playing
the role of the particle number in the usual IQHE case; the real particle
number is of course still not conserved in the paired state. The appearance of
this symmetry class in a disordered superconductor was seemingly overlooked by
AZ; in fact, there are two such possible classes, in which there is an
unbroken U(1) [not SU(2)], and T may be either unbroken or broken. By applying
the methods used by AZ in the case of broken T and unbroken U(1) symmetry, one
is led back to the usual unitary ensemble, with the spin $\up$ and spin
$\down$ quasiparticles playing the role of particles and holes, respectively.
It follows that these phases at non-zero $h$ possess a
nonvanishing density of (localized) states for the quasiparticles at
$E\rightarrow0$. In the two-dimensional case, the unitary class admits a
topological term with coefficient $\propto\sigma_{xy}^s$ in our case, and
thus an IQHE transition. This is consistent with the results in Ref.\
\cite{chalker}. Note that the relevant interaction effects in the case of
paired states and the IQHE may turn out to be different, however, so the
equivalence might not hold when interactions are taken into account.

Turning to the p-wave states relevant to the double-layer system in the FQHE,
for $t=0$ and with non-zero disorder we are once again in the situation just
discussed of broken T and unbroken U(1). Hence for this case, we again expect,
within our model of noninteracting quasiparticles, a transition between
quantized $\sigma_{xy}^s$ phases (analogs of the weak and strong-pairing
phases) that is in the universality class of the noninteracting IQHE.
With $t\neq0$, we saw in the pure case that the transition splits into two, each
in the class of the spinless p-wave case, and the intermediate phase had the
properties of the MR state. The non-zero tunneling breaks the
U(1) symmetry. We consider this case next, and then return to its application
to the double-layer system.

The symmetry classes of pairing of spinless fermions, or with
SU(2) symmetry fully broken by the Hamiltonian, are denoted DIII
(with unbroken T) and D (with broken T) by AZ. These cases were
not analyzed in two dimensions previously, but some of our results
have been found independently in Ref.\ \cite{sfnew} (see also a
remark in Ref.\ \cite{bundschuh}). Our interest is in class D,
with broken T. In this case the nonlinear sigma model target
(super-) manifold is SO$(2n)$/U$(n)$ (using replicas in the
compact formulation, with $n\rightarrow0$), or
Osp$(2n|2n)$/U$(n|n)$ using supersymmetry (with $n>0$ arbitrary)
\cite{az}. In the supersymmetric formulations, classes C and D
differ in that, while the bosonic submanifolds are real forms of
SO$(2n)$/U$(n)\times$Sp$(2n)$/U$(n)$ in both cases, in class C the
first factor is noncompact and the second compact, and in class D
it is the other way round (corresponding to the compact replica
formulation) \cite{az}. (These statements are for $n>1$. For
$n=1$, the first factor is a single point in both cases.) In the
two-dimensional case, class D admits a topological term in the
nonlinear sigma model, like class C. In the case of class D, there
is no continuous symmetry in the underlying fermion problem. The
only candidates for the physical meaning of the couplings in the
nonlinear sigma model are in terms of thermal conductivities,
since energy is still a conserved quantity \cite{sfbn}. The
diagonal (dissipative) thermal conductivity
$\kappa_{xx}=\kappa_{yy}$, and the off-diagonal LR conductivity
$\kappa_{xy}$ have the dimensions of $k_B^2 T/h$ ($h$ is Planck's
constant, $=2\pi$ in our units) at low temperatures. We define
\bea \kappa_{xx}&=&\frac{\pi^2 k_B^2 T}{3h}\ktxx,\non\\
\kappa_{xy}&=&\frac{\pi^2 k_B^2 T}{3h}\ktxy, \eea where the
numerical factor of $\pi^2/3$ is that which arises in the
quantized values of $\kappa_{xy}$ (see Sec.\ \ref{spinhall}), and
so may be conveniently included here. Then we expect that the
sigma model couplings are (similarly to the charge and spin
transport cases, where no factors of $k_B^2 T$ are involved)
$1/g^2\sim \ktxx$, and $\theta=4\pi\ktxy$. Since the quantized
values of $\ktxy$ are $\ktxy=c$, which is a multiple of $1/2$ for
Majorana fermions, we have arranged that the quantized values of
$\theta$ would be multiples of $2\pi$. We expect that if the
nonlinear sigma model for class D is derived microscopically at
weak coupling, which we will not attempt here, then the above
relation for $\theta$ will hold. We note here that in class D, the
density of states at $E\rightarrow0$ in the localized (quantized
$\ktxy$) phases is expected to approach a nonzero nonuniversal
constant, as one can see {}from the random matrix expressions in
AZ \cite{az}, using an argument in Ref.\ \cite{senthil1}. We want
to emphasise that the nonlinear sigma model in class D describes
only the case of very generic disorder, and not necessarily more
restricted forms of disorder. We will return to this point below.

We may now consider the form of the RG flows for class D in two
dimensions. We begin with perturbation theory at weak coupling,
that is large $\ktxx$. This was considered previously for classes
C and CI \cite{oppermann,sfbn}, and for class D is mentioned in
Ref.\ \cite{bundschuh}. We define the RG beta function as
\be
\beta_{xx}(g^2)\equiv\frac{d(g^2)}{d\ln L},
\ee
where $g^2$ is the nonlinear sigma model coupling squared, and $g^2\sim
1/\ktxx$, and $L$ is the length scale on which the renormalized coupling is
defined. In two dimensions this has the form
\be
\beta_{xx}(g^2)=a g^4 +O(g^6) \ee at weak coupling (small $g^2$);
effects of the topological term involving $\theta$, if the model
admits one, are nonperturbative in $g^2$, of order $e^{-b/g^2}$,
and contain the $\theta$ dependence. (Also, there will be another
beta function $\beta_{xy}\equiv d\theta/d\ln L$ for $\theta$,
which will be entirely nonperturbative.) Here $a$ and $b$ are
constants. In classes C and CI, $a$ is positive
\cite{oppermann,sfbn}, and the flow is towards strong coupling,
that is localization, as shown in Fig.\ \ref{flow} (a) for class
C. However, in classes D and DIII, $a$ is negative, and in fact
equal to minus its values in C and CI respectively
\cite{bundschuh,glrunpub}. The reason lies in the relation of the
target (super-)manifolds in the nonlinear sigma models, described
above. This relation is similar to that between the manifolds in
the symplectic (spin-orbit scattering) and orthogonal (potential
scattering) ensembles of the usual random matrix or localization
problems, where the $a$'s also have opposite signs. The origin of
this is that the $a$ term in the beta function of any nonlinear
sigma model is related to the Ricci curvature tensor of the target
manifold. When we compare these for the compact and noncompact
versions of the ``same symmetry'', such as Sp$(2n)$/U$(n)$ (at $n$
a positive integer), we find that they are of opposite sign: the
noncompact case has negative curvature, the compact positive.
These geometric phenomena for symmetric Riemannian (non-super-)
manifolds are discussed by Helgason \cite{helgason}. The
noncompact space Sp$(2n,{\bf R})$/U$(n)$ at $n\rightarrow0$
represents class D in perturbation theory, as does the compact
space SO$(2n)$/U$(n)$ in the same limit, and so their perturbative
$\beta$ functions are equal. This establishes the result using
replicas. Likewise for the target supermanifolds, the factors in
the bosonic parts each have opposite curvature in the two cases,
and this presumably extends to the full supermanifold.
Consequently, $a$ has the opposite sign in the two cases. A
similar result also holds in the principal chiral models with
target spaces Sp($2n$) and SO($2n$) in the compact replica
approach, with $n\rightarrow0$, which describe classes CI and DIII
respectively. Note that in the unitary case the target
supermanifold is U$(n,n|2n)$/[U$(n|n)\times$U$(n|n)$], and the
bosonic part contains just the compact and noncompact forms of
U$(2n)$/[U$(n)\times$U$(n)$], so the model maps to itself under
interchange of compact and noncompact, and hence the net
coefficient $a$ in the unitary case vanishes, as is well-known.

We now try to find the simplest possible flow diagram compatible with the
weak-coupling behavior and some other simple requirements. In the unitary case,
the flows in Fig.\ \ref{flow} (a) can be considered to be the simplest
possible, on including nonperturbative $\theta$-dependent effects that cause
the attractive fixed points to be at $\theta\equiv0$ (mod $2\pi$). These flows
are in fact obtained if one takes the dilute instanton gas calculations of
Pruisken, which are the leading nonperturbative effects at weak coupling,
together with the perturbation theory result just discussed, and uses these
forms for all coupling values \cite{pruisken}. The nontrivial fixed point at
$\theta=\pi$ and at some universal $\sigma_{xx}$ controls the transition
between the quantized fixed points in this case. The picture obtained {}from
these flows seems to be in qualitative agreement with what is known {}from
numerical work for this transition in the unitary ensemble, and for class C
\cite{chalker}.

For class D we can try to guess the nonperturbative form of the flows without
calculation. In view of the weak-coupling result, we could try reversing the
arrows on the flows for unitary and class C. However, we also expect that the
stable, attractive fixed points, which will represent quantized values of
$\ktxy$ at $\ktxx=0$, will be at $\theta\equiv0$ (mod $2\pi$) again, not at
$\pi$ (mod $2\pi$). In particular, this means that an insulating phase with
quantized $\ktxy=0$ is possible. It seems reasonable that sufficiently small
$\ktxx$ can produce localization when $\ktxy=0$, in spite of the flow to
$\ktxx=\infty$ in the weak-coupling region, just as in other localization
problems, including the case of spin-orbit scattering, and this should be
stable against small changes in $\ktxy$. In order to achieve this,
we also shift the flows by $\pi$ along the $\theta$ direction. The result
is the flow diagram shown in Fig.\ \ref{flow} (b). The interesting non-trivial
fixed points now occur on the lines $\theta\equiv0$ (mod $2\pi$). These flows
could be checked in the weak-coupling region by comparing them with a dilute
instanton gas calculation as in the unitary case. Indeed, if the
latter calculation is assumed to give the same form as in the unitary case
\cite{pruisken}, as is plausible, then the competition with the perturbative
terms will give the flows as shown.

In order to use the RG flow diagram to make predictions about the effects of
disorder on the pure transition {}from weak to strong pairing with T broken and
no spin-rotation symmetry, we need to know where the bare values of
$\ktxx$ and $\ktxy$ lie on the diagram. In the usual IQHE unitary case, and
also for class C, the values are shown as the dashed curve in Fig.\ \ref{flow}
(a). If one uses the self-consistent Born approximation to obtain the values in
the IQHE case, for disorder weak compared with the cyclotron energy, then
one obtains a semicircle in the $\sigma_{xx}$-$\sigma_{xy}$ plane
\cite{pruisken}. The precise position of the curve is unimportant, but it
associates the transition, at which $\sigma_{xy}$ is half an odd integer, with
the middle of the disorder-broadened Landau bands. Similar behavior occurs for
class C.

In the present case of class D, we again expect the bare values of $\ktxx$ and
$\ktxy$ to lie on an arc, as shown in Fig.\ \ref{flow} (b). These values pass
through the quantized points at $\ktxx=0$ ($g^2=\infty$) and $\theta\equiv0$
(mod $2\pi$) (quantized $\ktxy$). This is reasonable, as these are the values
in the two phases in the pure case, and disorder that is weak compared with
the gap $|\mu|$ in the spectrum should produce only small corrections to these
values. Connecting these regions with the dashed arc, we always produce a
curve of the form shown for topological reasons. This curve intersects the
separatrices shown, which flow into the non-trivial fixed points. We see that
the regions near the quantized fixed points flow into those fixed points, so
that {\em the quantized phases, one of which corresponds to the nonabelian
statistics phase in the pure case, still exist in the presence of
disorder} according to our proposed flow
diagram. But there is an intermediate set of bare values near $\theta\equiv\pi$
(mod $2\pi$), which flow to weak coupling, and at large scales they map onto
the entire interval of $\ktxy$ values between the two quantized values in
question, with a $\ktxx$ that increases logarithmically with $L$, according to
the weak coupling beta function $\beta_{xx}$ above. This is therefore an {\em
intermediate phase with metallic behavior of the thermal conductivities,
between the two quantized phases}. The intermediate phase is separated by phase
boundaries {}from the quantized phases, and the critical behavior at these
transitions is governed by the nontrivial fixed points. At these fixed points,
$\ktxy$ is equal to the quantized value in the neighboring quantized phase.
The critical exponents for these transitions are unknown at present.
Experimentally, one would see plateaus in $\ktxy$, separated by intermediate
regions, and the width of the latter will stay nonzero as the system size goes
to infinity, and as the temperature goes to zero. In the intermediate regions,
$\ktxy$ will vary continuously to interpolate its neighboring quantized values,
and $\ktxx$ will have a peak, the height of which will grow logarithmically
with increasing system size or inverse temperature.
We emphasize that the charge transport properties are still either
superconducting or quantized Hall, depending on the system considered, and
unaffected by the transition in the quasiparticles (there would be a collective
mode (phonon-like) contribution to thermal transport in a neutral superfluid
case, such as a He$^3$ film).

We should respond to one possible objection to our claim that there will
generically be an intermediate metallic phase in the class D problem
(this point is raised in Ref.\ \cite{sfnew}). This
objection begins with the pure problem, in which (since we assume
noninteracting quasiparticles) the critical theory is a Majorana fermi field
with a mass term that changes sign, and then considers weak disorder as a
perturbation of this continuum theory. The similar problem of a Dirac field has
been analyzed in recent years \cite{ludwig}, and a central argument is that for
$E=0$, one can consider the problem as a Dirac field in two (Euclidean)
dimensions. It is then argued that there are only a few possible random terms
that can be included in the 2D Dirac action that are marginal or relevant by
power counting at the pure fixed point. These terms, which are bilinears in the
Dirac field since the problem is noninteracting, are a random U(1) vector
potential and two types of random mass term. For the random Majorana fermion,
no U(1) vector potential is possible, and there is a unique mass term. It is
further known that the mass term is marginally irrelevant for weak disorder;
this type of randomness arises when one considers the 2D Ising model with
disorder in the intersite Ising couplings \cite{rbim}. If one assumes
that the disordered paired system we consider here must fall into this
scheme, then the only possible randomness is irrelevant, and there will be
a transition directly between the disordered versions of the weak- and
strong-pairing phases, with the critical properties of the pure system, up
to logarithmic corrections, in disagreement with our prediction of an
intermediate phase and a different universality class.

There is, however, a form of disorder not considered in this
argument. In Sec. \ref{spinless}, we discussed vortices in the gap
function, and in Subsec.\ \ref{geom} related their description in
terms of the gap function and the vector potential of the
underlying problem to the vielbein and spin connection that appear
in the general Majorana action. We should consider the possibility
that these are random; indeed, the general analysis of AZ requires
generic randomness, even though the Dirac or Majorana actions do
not explicitly appear there. Not all of the random fluctuations of
these quantities are necessarily relevant. For example, small
random fluctuations of the magnitude of the vielbein (i.e., of
$|\hat{\Delta}|$) around its non-zero mean are irrelevant by power
counting. Also, if the gap function is nonzero the vector
potential can mostly be gauged away because the superconductor is
a Meissner phase. However, this is clearly not true at a vortex,
and for general probability distributions of the disorder,
vortices will be present. In the FQHE application, there is
underlying potential disorder which couples to the particle
density, and for unbounded distributions of disorder, or for
sufficiently strong disorder with a bounded distribution, it will
nucleate vortices (FQHE quasiparticles) in the ground state, which
can occur in isolation {}from other vortices since they have
finite energy. It is not difficult to see, either intuitively or
formally, that a small density of randomly placed vortices will be
highly relevant at the massless Majorana critical point.
Intuitively, they introduce destructive phase interference.
Formally, in the 2D Majorana action, the effect is to insert the
spin fields of the Majorana theory (so-called because they
represent the Ising spins in the related Ising model), which (in
the same gauge choice as earlier) cause square-root branch points
in the Majorana field \cite{bpz}. Such random vortices do not seem
to have been considered in previous work on random Dirac fields.
They are relevant because, while the spin field in the critical
Majorana theory has scaling dimension 1/8 (which corresponds to a
relevant perturbation even in the non-random case), on replicating
or supersymmetrizing the system, the spin fields act on all the
components simultaneously, and hence their dimension is then
$0\times1/8=0$. Thus the coefficient with which these fields
appear in the action after averaging has scaling dimension 2,
showing they are strongly relevant \cite{glrunpub}. This will
cause an RG flow away {}from the pure Majorana fixed point, and we
expect the generic behavior of class D, with the intermediate
metallic phase, to result. Another possible form of disorder is
that which violates the $l=-1$ symmetry, that is general
p$_x+i$p$_y$ pairing with random coefficients. If both were
completely random, the average would restore parity, that is the
symmetry of reflection in any line, and prevent the existence of
nonzero $\ktxy$, so this form of disorder is not completely
acceptable in our physical systems; we must allow for a net
violation of parity. Thus, the generic disorder that defines class
D should include all of these relevant effects, and randomly
placed, isolated vortices alone are a relevant perturbation that
leads to a flow away {}from the pure Majorana critical theory.
Notice that, with random vortices included, the ordered phase we
are describing is no longer a superconductor in the strict sense,
due to the random phases frozen into the gap function. Instead it
is what has been called a gauge glass, since the order in the gap
function is similar to that in a spin glass.

In contrast, in the random bond Ising model (a subject also raised in
Ref.\ \cite{sfnew}), negative bonds produce frustrated
plaquettes, and a string of negative bonds produces two frustrated plaquettes
at the ends of the string. The insertion of a semi-infinite string of
negative bonds is the definition of the Ising disorder operator of Kadanoff
and Ceva \cite{kadanoff}, which is dual to, and at the critical point has the
same scaling dimension as, the Ising spin field. However, in this case the
analysis of weak randomness in terms of a random mass is justified, because
for a low probability of negative bonds the disorder variables (vortices)
appear only in closely-spaced pairs, not in isolation. In the continuum
critical field theory, the operator product of two disorder operators, which
represents the close pair, produces the Majorana mass operator, and thus the
randomness generates the random mass term. We see that the distribution of the
randomness in the Ising problem with weak bond disorder differs substantially
{}from the problem we wish to consider. While the random bond Ising model
in 2D does have a direct transition between the two quantized $\ktxy$
phases, and a multicritical point, there is no reason to suppose that
these occur in class D.

The properties of the disordered version of the MR phase, which
has a nonzero quantized $\ktxy$, are subtle. Since we have assumed
that isolated vortices are possible, which are localized FQHE
quasiparticles analogous to those in the usual states on the
plateaus in the FQHE, these will carry zero modes, and there will
be $2^{n-1}$ many-particle states when the system contains $2n$
vortices. These are very nearly degenerate when the vortices are
well-separated, since the energy splittings are expected to
decrease exponentially in the separation of the vortices.
Nonabelian statistics of additional vortex (FQHE quasiparticle)
excitations should be considered in terms of exchanges of such
quasiparticles separated by many times the typical separation of
the vortices in the ground state. Then the fermion zero modes of
the ground state can interfere with those on the added
quasiparticles, complicating the nonabelian statistics properties.
Further study of these effects is beyond the scope of this paper.

Finally, we return as promised to the case of the double-layer system with
tunneling $t$. Then the phase boundaries at non-zero $t$ between the
weak-pairing, abelian phase, the MR phase, and the strong-pairing phase
will be broadened and replaced by an intermediate region in which
metallic thermal properties will occur, again with sharp phase boundaries
between this and the other phases. As $t\rightarrow0$, this intermediate phase
will shrink in width to become a single point at $t=0$, where we have already
explained that a direct transition in the IQHE universality class occurs. Thus
this transition broadens to become the intermediate metallic phase at finite
$t$. We expect that at sufficiently small $t$, there is a single region of
the metallic phase, which interpolates between $\ktxy$ values differing by two
steps, which are the values in the $t=0$ weak and strong-pairing phases. As $t$
increases, a point is reached at which another plateau in $\ktxy$ appears,
which is the MR-type phase. Such behavior is allowed by our flow diagram, if we
plot it {}from $\theta=0$ to $\theta=4\pi$, and the initial values lie on an arc
between the quantized fixed points at those $\theta$ values that avoids the
basin of attraction of the quantized fixed point at $\theta=2\pi$ completely
for small $t$, but not for larger $t$.

%%%%%%%%%%%%%%%%%%%%%%%%%%%%%%%%%%%%%%%%%%%%%%%%%%
\section{Conclusion}
\label{conclusion}

In this paper, we have considered exotic properties of P and T-violating
paired states of fermions in two dimensions, and the relation to the FQHE
using pairing of composite fermions. The results have been summarized in
the Introduction.

To conclude, our main results are: (i) p-wave pairing in spinless or
spin-polarized fermions in the weak-pairing phase leads to the properties
also found in the FQHE in the MR states, and supports the ideas of
nonabelian statistics as a robust property, at least in the case of a pure
system. Such statistics will also occur for the vortices in such a p-wave
state in general in charged superfluids, and in neutral superfluids
modulo U(1) phase factors that arise {}from the compressible charge sector;
(ii) in an A-phase type p-wave phase, statistics may be abelian, though
tunneling or a Zeeman term may lead to a transition to a MR phase; (iii)
in the d-wave spin-singlet case, the HR state corresponds to the
transition point, and, {}from now on, may be disregarded in considering
generic spin-singlet FQHE systems, which will most likely be in the weak-pairing
phase. The latter is abelian, but has a nonzero Hall spin conductivity, and
spin-1/2 chiral Dirac fermions on the edge \cite{senthil2}; (iv) disorder
does not destroy the phases in question, but may modify the MR phase in an
essential way. In the spinless p-wave case, randomly-placed vortices are a
relevant perturbation of the pure transition, and there is an intermediate
phase with metallic thermal conductivity properties due to delocalized BCS
quasiparticles.

Issues which remain to be addressed include the full effective theories
for the states, and for the transitions between them, and the effect of
interactions on the random problems. Also, a direct derivation of
nonabelian statistics in terms of pairing of fermions in the MR case is
desirable.

One further comment on tunneling into the edge is in order. Such tunneling
could provide a useful diagnostic for the paired states in the FQHE.
Since the fermion excitations on the edge in the weak-pairing phases are
now always Dirac (or Majorana), their contribution to the exponent is
always the same. Thus at filling factor $\nu=1/q$ (where $q$ will be 2,
4, \ldots, for fermions such as electrons), the current will scale
as $I\sim V^\alpha$ with $\alpha=q+1$ in all the weak-pairing or MR
phases. In contrast, in the Halperin-type paired states, we
have $\alpha=4q$ \cite{rr}. The former result is the same as in the
compressible Fermi-liquid-like states \cite{slh}. Assuming that edge
theories of fully-gapped bulk states must always be unitary conformal
field theories, as argued in Sec.\ \ref{dwave}, the exponent for tunneling into
an edge on which all modes propagate in the same direction must in fact always
be an odd integer, as a consequence of the Fermi statistics of the electrons.
For the 5/2 state, the $I\sim V$ contribution of the LLL will presumably
dominate.

{\it Note Added:} In view of a suggestion which has circulated,
that the Fermi-liquid-like phase of Ref.\ \cite{hlr} may
generically have an instability to pairing in some channel, albeit
at extremely low energy or temperature scales, we will consider
here the case of a weak-pairing phase of spin-polarized fermions
in an arbitrary angular momentum $l$ eigenstate ($l$ must be odd).
Similar arguments as before show that there will be $|l|$ chiral
Majorana fermion modes on an edge, and correspondingly
$2^{|l|n-1}$ degenerate states for $2n$ vortices. Since $|l|$ must
be odd, this always leads to nonabelian statistics, with the same
monodromy properties as for $l=-1$ up to Berry phase factors
(because each added pair of Majoranas makes a Dirac fermion which
contributes only abelian effects).

%%%%%%%%%%%%%%%%%%%%%%%%%%%%%%%%%%%%%%%%%%%%%%%%%
\acknowledgements

We thank M.P.A. Fisher, I.A. Gruzberg, F.D.M. Haldane, B.I. Halperin,
T.-L. Ho, A.W.W. Ludwig, M.V. Milovanovi\'{c}, G. Moore, E.H. Rezayi, and
T. Senthil for informative comments and helpful discussions. N.R. is
grateful for the hospitality of the Institut Henri Poincar\'{e}, Paris,
where this paper was completed. This work was supported by NSF grants,
numbers DMR-9157484 and DMR-9818259.

%%%%%%%%%%%%%%%%%%%%%%%%%%%%%%%%%%%%%%%%%%%%%%%%%
\appendix

\section{Quantized Hall conductivity for spin}
\label{spinqhe}

In this Appendix, we provide a detailed derivation of the Hall conductivity in
(iso-) spin transport in the d- and p-wave (A-phase) cases. This is equivalent
to showing that the induced action for the system in an external gauge field
that couples to the (iso-) spin is a CS term. In the d-wave case, the system is
spin-rotation invariant, so we obtain an SU(2) CS term, while in the p-wave
case, there is only a U(1) symmetry, so we find a U(1) CS term. In both cases,
the Hall spin conductivity is given in suitable units by a topological
invariant. Within the BCS mean field approach, using a suitable conserving
approximation, this topological invariant is the winding number of
$(u_\bk,v_\bk)$ discussed in Sec.\ \ref{spinless}, and is therefore an
integer, which is the statement of quantization. We argue that the
quantization in terms of a topological invariant is more general than the
approximation used.

Considering first the spin-singlet paired states, we use the Nambu basis where
the symmetries are transparent. Define
\be
\Psi=\frac{1}{\sqrt{2}} \left(\begin{array}{c}
       c  \\
       i\sigma_y c^\dagger  \end{array}\right),
\ee
with
\be
c=\left(\begin{array}{c}
           c_\up\\
           c_\down \end{array}\right),
\ee
so that $\Psi$ transforms as a tensor product of particle-hole and
spin-space
spinors. We consider an interacting system and approximate it as in BCS
theory, then with a minimal coupling to the gauge field, we use a conserving
approximation to obtain the spin response. In Fourier space, we should note
that
\be
\Psi_\bk=\frac{1}{\sqrt{2}} \left(\begin{array}{c}
       c_\bk  \\
       i\sigma_y c_{-\bk}^\dagger  \end{array}\right).
\ee
In the Nambu basis, the kinetic energy becomes (again, $K=H-\mu N$)
\bea
K_0&=&\sum_\bk\xi_\bk^0(c_{\bk\up}^\dagger c_{\bk\up}+
c_{\bk\down}^\dagger c_{\bk\down})\non\\
&=&\sum_\bk\xi_\bk^0\Psi_\bk^\dagger(\sigma_z\otimes I)\Psi_\bk,
\eea
where $\xi_\bk^0=|\bk|^2/(2m)-\mu$ is the kinetic energy, containing the bare
mass $m$, and the products in the spinor space are understood. Products like
$\sigma_z\otimes I$ act on the Nambu spinors, with the first factor acting in
the particle-hole factor, the second in the spin-space factor.
The interaction term, for a spin-independent interaction $V$, is
\be
K_{\rm int}=\frac{1}{2}\sum_{\bk\bk'{\bf q}}V_{\bf q}:\Psi_{\bk+{\bf
q}}^\dagger(\sigma_z\otimes I)\Psi_\bk\Psi_{\bk'-{\bf q}}^\dagger
(\sigma_z\otimes I)\Psi_{\bk'}:.
\ee
Here the colons $:\ldots:$ denote normal ordering, that is all the $c^\dagger$s
are brought to the left. In the BCS-extended Hartree-Fock approximation, the
effective quasiparticle Hamiltonian (for later reference) is
\bea
K_{\rm eff}&=&\sum_\bk \Psi_\bk^\dagger\left[\xi_\bk(\sigma_z\otimes I)
   +{\rm Re}\,\Delta_\bk(\sigma_x\otimes I)\right.\non\\
&&\left.\mbox{}-{\rm Im}\,\Delta_\bk(\sigma_y\otimes I)\right]\Psi_\bk.
\eea
This is for singlet pairing, where $\Delta_{-\bk}=\Delta_\bk$, and not
just for d-wave. Here $\xi_\bk$
is $\xi_\bk^0$ plus the Hartree-Fock corrections. If we define a vector
\be
\bE_\bk=({\rm Re}\,\Delta_\bk,-{\rm Im}\,\Delta_\bk,\xi_\bk)
\ee
then the quasiparticle energy $E_\bk=|\bE_\bk|$, and
\be
K_{\rm eff}=\sum_\bk\Psi_\bk^\dagger(\bE_\bk\cdot\hbox{\boldmath$\sigma$}
\otimes I)\Psi_\bk.
\ee

In the Nambu notation, it is clear that $K=K_0+K_{\rm int}$, and $K_{\rm eff}$,
are invariant under global SU(2) rotations that act on the spin-space, that is
the second factor in the tensor products. The spin density, the integral of
which over all space is the total spin and generates such global
transformations, and the spin current densities are given by
\bea
J_0^a({\bf q})&=&\frac{1}{2}\sum_\bk\Psi_{\bk-{\bf q}/2}^\dagger(I\otimes
\sigma_a)\Psi_{\bk+{\bf q}/2}\\
J_i^a({\bf q})&=&\frac{1}{2}\sum_\bk\frac{k_i}{m}\Psi_{\bk-{\bf q}/2}^\dagger
(\sigma_z\otimes \sigma_a)\Psi_{\bk+{\bf q}/2},
\eea
where $i=x$, $y$ is the spatial index, and $a=x$, $y$, $z$ is the spin-space
index. Spin conservation implies the continuity equation, as an
operator equation,
\be
\partial J_\mu^a/\partial x_\mu =0,
\ee
where $\mu=0$, $x$, $y$, and the summation convention is in force.

So far we have not introduced a gauge field for spin. Since the spin is
conserved locally, we can turn the symmetry into a local gauge symmetry by
introducing an SU(2) vector potential, and making all derivatives covariant.
The effect on $K$ is to add the integral of
\be
A_\mu^a J_\mu^a +\frac{1}{8m}A_i^a A_i^a \Psi^\dagger(\sigma_z\otimes I)\Psi.
\ee
The gauge field is to be used solely as an external source, with which to probe
the spin response of the system, and then set to zero.

If we now consider integrating out the fermions, then we can obtain an action
in the external gauge fields, which can be expanded in powers of $A_\mu^a$. The
zeroth-order term is the free energy density, times the volume of spacetime,
and the first-order term vanishes by spin-rotation invariance. The second-order
term corresponds to linear response: the second functional derivative with
respect to $A_\mu^a$, at $A_\mu^a=0$, yields the (matrix of) linear response
functions. In particular, the spatial components yield the conductivity tensor
in the usual way. Therefore we consider the imaginary-time time-ordered
function,
\be
\Pi_{\mu\nu}^{ab}=-i\langle J_\mu^a(q)J_\nu^b(-q)
\rangle,
\ee
where time-ordering is understood, and {}from here on we use a convention that
$p$, $q$, etc stand for three-vectors $p=(p_0,{\bf p})$, and further
$p_0=i\omega$ is imaginary for imaginary time.
For $\mu=\nu=i=x$ or $y$, an additional ``diamagnetic'' term
$\bar{n}\delta^{ab}/4m$ is present in $\Pi_{\mu\nu}^{ab}$, which we do not show
explicitly.
As consequences of the continuity equation and the related gauge invariance,
$\Pi_{\mu\nu}$ must be divergenceless on both variables, $q_\mu\Pi_{\mu\nu}
=q_\nu\Pi_{\mu\nu}=0$. To maintain these when using the BCS-Hartree-Fock
approximation for the equilibrium properties, one must use a conserving
approximation for the response function, which in this case means summing
ladder diagrams (compare the charge case in Ref.\ \cite{schrieffer},
pp.\ 224--237).

One begins with the BCS-Hartree-Fock approximation, which can be written in
terms of Green's functions as (we consider only zero temperature, and $\int
dp_0$ is along the imaginary $p_0$ axis throughout)
\bea
G^{-1}(p)&=&p_0-\xi_{\bf p}^0\sigma_z\otimes I -\Sigma(p),\\
\Sigma(p)&=&i\int\frac{d^3k}{(2\pi)^3}(\sigma_z\otimes I)G(k)(\sigma_z\otimes I)
V(k-q).
\eea
Note that $G(p)$ and $\Sigma(p)$ are matrices acting on the tensor product
space. The equations are solved by
\be
G^{-1}(p)=p_0-\bE_{\bf p}\cdot\hbox{\boldmath$\sigma$}\otimes I,
\ee
(we write $1$ for $I\otimes I$) as one can also see {}from the effective
quasiparticle Hamiltonian $K_{\rm eff}$, and $\Delta_{\bf p}$ obeys the
standard gap equation.

In the response function, the ladder series can be summed and included by
dressing {\em one} vertex, to obtain (again not showing the diamagnetic term)
\bea
\Pi_{\mu\nu}^{ab}(q)&=&-i\int\frac{d^3p}{(2\pi)^3}
{\rm tr}\left[\gamma_\mu^a(p,p+q)G(p+q)\right.\non\\
&&\left.\times\Gamma_\nu^b(p+q,p)G(p)\right],
\eea
where $\gamma_\mu^a$ is the bare vertex,
\bea
\gamma_0^a(p,p+q)&=& \frac{1}{2}I\otimes\sigma_a,\\
\gamma_i^a(p,p+q)&=& -\frac{(p+\frac{1}{2}q)_i}{2m}\sigma_z\otimes\sigma_a,
\eea
and $\Gamma_\mu^a$ is the dressed vertex satisfying
\bea
\Gamma_\nu^b(p+q,p)&=&\gamma_\nu^b(p+q,p)+i\int\frac{d^3k}{(2\pi)^3}
\sigma_z\otimes I G(k+q)\non\\
&&\mbox{}\times\Gamma_\nu^b(k+q,k)G(k)\sigma_z\otimes I V(p-k).
\eea
At small $q$, we can obtain useful information about this function {}from the
Ward identity that results {}from the continuity equation. The particular
Ward identity we use here is an exact relation of the vertex function to the
self-energy, and the conserving approximation (the ladder series) was
constructed to ensure that it holds also for the approximated vertex and self
energy functions.

Following Schrieffer's treatment \cite{schrieffer}, we consider the vertex
function with external legs included:
\be
\Lambda_\mu^a(r_1,r_2,r_3)=\langle J_\mu^a(r_3)\Psi(r_1)\Psi^\dagger(r_2)
\rangle,
\ee
for spacetime coordinates $r_1$, $r_2$, $r_3$. Applying $\partial/\partial
r_{3\mu}$ to both sides and using the operator continuity equation, we obtain
the exact identity in Fourier space
\be
q_\mu\Gamma_\mu^a(p+q,p)=\frac{1}{2}I\otimes\sigma_a G^{-1}(p)
    -\frac{1}{2}G^{-1}(p+q)I\otimes\sigma_a.
\ee
Since $G^{-1}$ is trivial in the spin-space indices, it commutes with
$I\otimes\sigma_a$. Hence at $q\to0$, the right-hand side vanishes, so
$\Gamma(p+q,p)$ has no singularities as $q\to0$. This differs {}from the charge
case, for example, where this calculation (using the ladder series
approximation) leads to the discovery of the collective mode \cite{anderson58}.
Since the spin symmetry is unbroken, no collective mode is necessary to restore
this conservation law, and so there is no singularity in the vertex function
for spin.

One can verify that the Ward identity is satisfied using the BCS-Hartree-Fock
$G^{-1}$ and the ladder series for $\Gamma$. At $q=0$, this yields the
important results
\be
\Gamma_\mu^a(p,p)=-\frac{1}{2}\partial_\mu G^{-1}(p)I\otimes\sigma_a,
\label{qzerovert}
\ee
or explicitly,
\bea
\Gamma_0^a(p,p)&=&\frac{1}{2}I\otimes\sigma_a,\non\\
\Gamma_i^a&=&-\frac{1}{2}\partial_iG^{-1}(p)I\otimes\sigma_a,
\eea
where $\partial_i$ and $\partial_\mu$ stand for $\partial/\partial p_i$,
$\partial/\partial p_\mu$ {}from here on, and the extra minus in the first
relation is consistent because implicitly $q_\mu\Gamma_\mu=q_0\Gamma_0
-q_i\Gamma_i$.

We now calculate $\Pi$ at small $q$. To zeroth order, use of the Ward identity
shows that the $J$-$J$ function gives zero, except when $\mu=\nu=i$.
In that case, it reduces to a constant that cancels the diamagnetic term
also present in just that case. Hence we require only the part first-order in
$q$. In the expression for $\Pi$ above, we first shift $p\to p-\frac{1}{2}q$,
so that $q$ no longer appears in any bare vertices, but does appear in the
Green's functions on both sides of the ladder, between the rungs which are the
interaction lines. Hence to first order, we obtain a factor
$\pm\frac{1}{2}\partial_\mu G=\mp\frac{1}{2} G\partial G^{-1} G$ in place of
$G$ in one position in the ladder. Since there may be any number of rungs
(including zero) between this and either of the vertices at the ends, the
terms can be summed up into a ladder dressing each vertex, evaluated at $q=0$.
Hence we obtain to first order
\bea
\Pi_{\mu\nu}^{ab}(q)&=&-\frac{i}{2}\int\frac{d^3p}{(2\pi)^3}{\rm
tr}\left[\Gamma_\mu^a(p,p) q_\lambda\partial_\lambda G
\Gamma_\nu^b(p,p)G(p)\right.\non\\
&&\left.\mbox{}-\Gamma_\mu^a(p,p)G(p) \Gamma_\nu^b(p,p) q_\lambda
\partial_\lambda G \right].
\eea
Using the Ward identity, this becomes
\bea
\Pi_{\mu\nu}^{ab}(q)&=&\frac{i}{8}q_\lambda\int\frac{d^3p}{(2\pi)^3}{\rm tr}
\left\{(I\otimes\sigma_a)(I\otimes\sigma_b)G\partial_\mu G^{-1}\right.\non\\
&&\left.\times[G\partial_\lambda G^{-1},G\partial_\nu G^{-1}]\right\}
\label{3form}
\eea
Since the $G$'s are independent of the spin-space indices, the explicit
$\sigma$'s factor off, and the result is $\delta^{ab}$ times a
spin-independent part. The latter can be simplified using the BCS-Hartree-Fock
form of $G$, by writing the latter as
\be
G(p)=\frac{p_0+\bE_{\bf p}\cdot\hbox{\boldmath$\sigma$}\otimes I}
{p_0^2-E_{\bf p}^2}.
\ee
The spin-independent factor contains $\epsilon_{\mu\nu\lambda}$ since it is
antisymmetric in these labels. Keeping track of the signs, we find for the
quadratic term in the induced action
\be
\frac{1}{4\pi}\frac{\cal M}{4}\int d^3r A_\mu^a\frac{\partial A_\nu^a}{\partial
r_\lambda}\epsilon_{\mu\nu\lambda},
\ee
with $\cal M$ given by the topological invariant
\be
{\cal M}=\int \frac{d^2p}{8\pi}\epsilon_{ij}\bE_{\bf p}\cdot(\partial_i
\bE_{\bf p}\times\partial_j\bE_{\bf p})/E_{\bf p}^3.
\ee
The right hand side is exactly the winding number $m$ discussed in Sec.\
\ref{spinless}, and is an integer as long as $\bE$ is a continuous,
differentiable function of $\bf p$; it is $2$ for the d-wave case.

To ensure SU(2) gauge invariance, the CS term should include also a term cubic
in $A$, with no derivatives. For this term we evaluate the triangle one-loop
diagrams with three insertions of $J$, with each vertex dressed by the ladder
series. Setting the external momenta to zero, the Ward identity can be used
for all three vertices, and the result can be seen to be
\bea
\Pi_{\mu\nu\lambda}^{abc}(0,0)&=&\mbox{}-\frac{1}{24}\int\frac{d^3p}{(2\pi)^3}
{\rm tr}\left[(I\otimes\sigma_a) G\partial_\mu G^{-1}\right.\non\\
&&\times\left.\{(I\otimes\sigma_b) G\partial_\nu G^{-1},(I\otimes\sigma_c)
G\partial_\lambda G^{-1}\}\right].\non\\
&&
\label{triangle}
\eea
The anticommutator $\{\,,\,\}$ arises since the result must be symmetric under
permutations of the index pairs $\mu$, $a$, etc.
The product $\sigma_a\sigma_b\sigma_c$, when traced over the spin-space
indices, yields a factor $2i\epsilon_{abc}$, which is antisymmetric, and so the
remainder must contain $\epsilon_{\mu\nu\lambda}$ to maintain symmetry; the
rest of the structure is the same as before.
Hence the full result is the SU(2) CS term, which we write in terms of the
$2\times2$ matrix vector potentials $A_\mu=\frac{1}{2}\sigma_aA_\mu^a$,
\be
\frac{k}{4\pi}\int d^3x\epsilon_{\mu\nu\lambda}{\rm tr}(A_\mu\partial_\nu
A_\lambda+\frac{2}{3}A_\mu A_\nu A_\lambda).
\ee
Here $k$ is the conventional notation for the coefficient of such a term, in
this same normalization; if we wished to quantize the theory by functionally
integrating over $A$, we would need $k=$ an integer. In our case
$k={\cal M}/2=1$ for d-wave.

For the spin-triplet case with an unbroken U(1) symmetry, we must use
the fact that $\Delta_{-\bk}=-\Delta_\bk$. For example, in
the two-dimensional A-phase, as occurs in the 331 state in the double-layer
FQHE system with zero tunneling, discussed in Sec.\ \ref{331}, the pairs are
in the isospin $S_z=0$ triplet state $\up_i\down_j+\down_i\up_j$, and the U(1)
symmetry generated by $S_z$ is unbroken; we recall that the underlying
Hamiltonian is not assumed to have a full SU(2) symmetry. The effective
quasiparticle Hamiltonian (\ref{331ham}) becomes, in the Nambu-style notation,
\bea
K_{\rm eff}&=&\sum_\bk \Psi_\bk^\dagger\left[\xi_\bk(\sigma_z\otimes I)
   +{\rm Re}\,\Delta_\bk(\sigma_x\otimes \sigma_z)\right.\non\\
&&\left.\mbox{}-{\rm
Im}\,\Delta_\bk(\sigma_y\otimes \sigma_z)\right]\Psi_\bk.
\eea
The U(1) vector potential $A_\mu$ couples to $S_z$, and the vertex functions
contain $I\otimes\sigma_z$, which commutes with the BCS-Hartree-Fock Green's
function $G$. The tensors appearing in the three terms in $K_{\rm eff}$
obey the same algebra as the three in that for the spin-singlet case
(where they were
trivial in the second factor), and as in that case commute with
$I\otimes\sigma_z$. Consequently, the derivation for the induced action to
quadratic order in $A_\mu$ is similar to that for the SU(2) singlet case
above, and the traces in the Nambu indices can be carried out with the same
result as before, to obtain the abelian CS term
\be
\frac{1}{4\pi}{\cal M}\int d^3r A_\mu\frac{\partial A_\nu}{\partial
r_\lambda}\epsilon_{\mu\nu\lambda},
\ee
and no cubic term. In this case, $\cal M$ is again given by the winding
number $m$ which is $0$ or $\pm 1$ in the p-wave strong and weak-pairing phases
(respectively) discussed in this paper.

We note that the effect of the vertex corrections we included as
ladder series is to renormalize the $q=0$ vertices as shown in eq.\
(\ref{qzerovert}) for the spin-singlet case, and use these in one-loop
diagrams with no further corrections. This corresponds to the minimal
coupling $p \to p-A$ in the action, as one would expect by gauge
invariance. If we assume such a coupling, and treat the low-energy,
long-wavelength theory near the weak-strong transition as Dirac fermions
with relativistic dispersion and minimal coupling to the external gauge
field, then the expression for $\cal M$ as an integral over $\bf p$
covers only half the sphere in $\bf n$ space, and we would get $\pm 1$
(d-wave), $\pm 1/2$ (p-wave). The missing part results {}from the
ultraviolet regulator in the field theory version of the calculation
\cite{rahw}, or {}from a second fermion with a fixed mass in some lattice
models \cite{ludwig}. In our calculation, the remainder is provided by
the ultraviolet region, where $\Delta_\bk\to0$ as $\bk\to\infty$. At the
transition, $\mu=0$, the map is discontinuous and covers exactly half the
sphere in the p-wave case, so ${\cal M}=1/2$, as in other problems. In the
d-wave case with rotational symmetry, the value of $|v_\bk/u_\bk|$ as
$\bk\to0$ is nonuniversal, as noted in Sec.\ \ref{hr}, and hence so is
the value of $\sigma_{xy}^s$ at the transition. This is a consequence of
the non-relativistic form of the dispersion relation of the
low-energy fermions in this case. We may also note that for a paired system
on a lattice, as in models of high $T_c$ superconductors, a similar
calculation will give an integral over the Brillouin zone, which is a torus
$T^2$, instead of the $\bk$ plane which can be compactified to $S^2$. But maps
{}from $T^2$ to $S^2$ are again classified by the integers, and the
integer winding number is given by the same expression, so quantization
is unaffected.

We can also argue that the quantization result away {}from a transition is
exact in a translationally-invariant system, at least in all orders in
perturbation theory. For this we use the form in eq.\ (\ref{3form}) or
(\ref{triangle}), where the Ward identity for the vertex has been used.
Diagrammatically, it is clear that the exact expression can be similarly
written, using the exact (i.e., all orders in perturbation) Green's
function and vertex function.
(This is also true when the CS gauge field interaction is included.) The
Ward identity that relates them is exact, and the result for $\sigma_{xy}^s$
is of the same form as shown. The next step, the frequency integrals,
cannot be done explicitly in this case, because the precise form of the
Green's function is unknown, and the analogs of $\xi_\bk$, $\Delta_\bk$ (or of
$u_\bk$, $v_\bk$) do not exist. The latter do not exist because in general the
poles in the Green's function, which would represent the quasiparticles, are
broadened by scattering processes, except for the lowest energies for
kinematical reasons. However, the form in eq.\ (\ref{3form}) is itself a
topological invariant, as we will now argue. As long as there is a gap in the
support of the spectral function of $G$, $G(p)$ is continuous and
differentiable on the {\em imaginary} frequency axis, and tends to
$I\otimes I/p_0$ as $p_0\to\pm i\infty$. Thus $G^{-1}$ exists and never
vanishes. Considering the spin-singlet case for convenience, the spin-space
structure is trivial, so we may perform the corresponding traces, and then $G$
or $G^{-1}$ is a $2\times 2$ matrix, with the same reality properties on the
imaginary $p_0$ axis as in the BCS-Hartree-Fock approximation. (The
spin-triplet case should work out similarly, because of the
algebraic structure already mentioned.) It thus represents a real non-zero
4-component vector, in ${\bf R}^4-{0}$, which topologically is the same as
$S^3$. $S^3$ is obtained by dividing $G$ by its norm, $({\rm tr}G^\dagger G)
^{1/2}$, and the normalized $G$ is a $2\times 2$ unitary matrix with
determinant $-1$, so it lies in $S^3$. The $\bk$ space can be compactified
to $S^2$ as before, and the frequency variable can be viewed as an element of
the interval ${\cal I}=(-1,1)$, so the integral is over $S^2\times{\cal I}$.
However, since the limit of the Green's function as $p_0\to\pm i\infty$ for
fixed $\bk$ is independent of $\bk$, we can view this as simply $S^3$. Thus
we are dealing with maps {}from $S^3$ to $S^3$, the equivalence classes of
which are classified by the homotopy group $\pi_3(S^3)={\bf Z}$. The
integral we have obtained simply calculates the integer winding number or
Pontryagin index of the map, when properly normalized ($G$ can be
normalized to lie in SU(2) without affecting the integral). This
establishes the quantization of $\sigma_{xy}^s$ in a translationally-invariant
system with a gap, at least to all orders in perturbation theory, and
probably can be made fully non-perturbative (as the Ward identity is already).

%%%%%%%%%%%%%%%%%%%%%%%%%%%%%%%%%%%%%%%%%%%%%%%%%%

%%%%%%%%%%%%%%%%%%%%%%%%%%%%%%%%%%%%%%%%%%%%%%%%%%%


\begin{references}

\bibitem{bcs}
J. Bardeen, L.N. Cooper, and J.R. Schrieffer, Phys. Rev. {\bf 106}, 162
(1957); {\bf 108}, 1175 (1957).
\bibitem{schrieffer}
J.R. Schrieffer, {\it Theory of Superconductivity}, (Addison-Wesley, Reading
MA, 1964).
\bibitem{vollwolf}
For a review, see {\it e.g.}, D. Vollhardt and P. W\"{o}lfle, {\it The
Superfluid Phases of Helium 3}, (Taylor and Francis, London, 1990).
\bibitem{volovik88}
G.E. Volovik, Zh. Eksp. Teor. Fiz. {\bf 94}, 123 (1988) [Sov. Phys. JETP {\bf
67}, 1804 (1988)].
\bibitem{volyak}
G.E. Volovik and V.M. Yakovenko, J. Phys. Cond. Matter {\bf 1}, 5263 (1989).
\bibitem{volovik90}
G.E. Volovik, Physica B {\bf 162}, 222 (1990).
\bibitem{vol90lett}
G.E. Volovik, Sov. Phys. JETP Lett. {\bf 51}, 125 (1990).
\bibitem{volovik92}
G.E. Volovik, Sov. Phys. JETP Lett. {\bf 55}, 368 (1992).
\bibitem{book}
For a review, see, {\it e.g.}, {\it The Quantum Hall Effect},
edited by R.E.~Prange and S.M.~Girvin (Second Edition,
Springer-Verlag, New York, 1990).
\bibitem{laugh}
R.B. Laughlin, \prl {\bf 50}, 1395 (1983).
\bibitem{halp83}
B.I. Halperin, Helv. Phys. Acta, {\bf 56}, 75 (1983).
\bibitem{dmh}
R. Morf, N. d'Ambrunil, and B.I. Halperin, \prb {\bf 34}, 3037 (1986).
\bibitem{hr}
F.D.M.~Haldane and E.H.~Rezayi, \prl {\bf 60}, 956, 1886 (E) (1988).
\bibitem{mr}
G. Moore and N. Read, Nucl.\ Phys.\ B{\bf 360}, 362 (1991); N.~Read and
G.~Moore, Prog.\ Theor.\ Phys.\ (Kyoto) Supp.\ {\bf 107}, 157 (1992).
\bibitem{hald83}
F.D.M. Haldane, \prl {\bf 51}, 605 (1983), and in Ref. \cite{book}.
\bibitem{g}S.M.~Girvin in Ref.\ \cite{book}.
\bibitem{gm}S.M.~Girvin and A.H.~MacDonald, \prl {\bf 58}, 1252 (1987).
\bibitem{read87}N.~Read, Bull. Am. Phys. Soc, {\bf 32}, 923 (1987).
\bibitem{laughan}R.B.~Laughlin, \prl {\bf 60}, 2677 (1988).
\bibitem{zhk}S.C.~Zhang, T.H.~Hansson, and S.~Kivelson, \prl {\bf 62}, 82
    (1989).
\bibitem{read89}N.~Read, \prl {\bf 62}, 86 (1989).
\bibitem{jain}J.K.~Jain, \prl {\bf 63}, 199 (1989); \prb {\bf 40}, 8079
    (1989); {\it ibid.} {\bf 41}, 7653 (1990).
\bibitem{fishlee}D.-H.~Lee and M.P.A.~Fisher, \prl {\bf 63}, 903 (1989).
\bibitem{lopfrad}A.~Lopez and E.~Fradkin, \prb {\bf 44}, 5246 (1991).
\bibitem{hlr}B.I.~Halperin, P.A.~Lee, and N.~Read, \prb {\bf 47}, 7312
   (1993).
\bibitem{read94}N.~Read, Semicond. Sci. Technol. {\bf 9}, 1859 (1994)
   [cond-mat/9501090].
\bibitem{sm}R.~Shankar and G.~Murthy, \prl {\bf 79}, 4437 (1997);
    cond-mat/9802244.  \bibitem{dhlee}D.-H.~Lee, \prl {\bf 80}, 4745 (1998).
\bibitem{ph}V.~Pasquier and F.D.M.~Haldane, Nucl. Phys. B {\bf 516}, 719 (1998).
\bibitem{read98}N. Read, \prb {\bf 59}, 8084 (1998).
\bibitem{willett}R.L. Willett, J.P. Eisenstein, H.L. Stormer, D.C. Tsui, A.C.
Gossard, and J.H. English, \prl {\bf 59}, 1776 (1987).
\bibitem{gww}
M.~Greiter, X.-G.~Wen and F.~Wilczek, \prl {\bf 66}, 3205 (1991); Nucl. Phys.
B{\bf 374}, 567 (1992).
\bibitem{morf}
R. Morf, \prl {\bf 80}, 1505 (1998).
\bibitem{rhnew}
E.H. Rezayi and F.D.M. Haldane, cond-mat/9906137.
\bibitem{gww2}M. Greiter, X.-G. Wen, and F. Wilczek, \prb {\bf 46},
9586 (1992).
\bibitem{halpnewport} B.I. Halperin, Surface Sci. {\bf 305}, 1
(1994).
\bibitem{expt}
Y.W. Suen, L.W. Engel, M.B. Santos, M. Shayegan, and D.C. Tsui,
\prl {\bf 68}, 1379 (1992); J.P. Eisenstein, G.S. Boebinger, L.N. Pfeiffer, K.W.
West, and Song He, \prl {\bf 68}, 1383 (1992).
\bibitem{he}
S. He, X.-C.~Xie,
S.~Das~Sarma, and F.-C.~Zhang, \prb {\bf 43}, 9339 (1991); S.~He, S.~Das~Sarma,
and X.-C.~Xie, {\it ibid.}\ {\bf 47}, 4394 (1993).
\bibitem{wit89}
E. Witten, Commun. Math. Phys. {\bf 121}, 351 (1989).
\bibitem{wen3}
X.-G.~Wen, \prl {\bf 70}, 355 (1993).
\bibitem{wwh}
X.-G. Wen and Y.-S. Wu, Nucl. Phys. B{\bf 419}
[FS], 455 (1994);  X.-G.~Wen, Y.-S.~Wu and Y.~Hatsugai, Nucl. Phys. B{\bf 422}
[FS], 476 (1994).
\bibitem{milr}
M. Milovanovi\'{c} and N.~Read, Phys. Rev. B {\bf 53}, 13559 (1996).
\bibitem{nayak}
C.~Nayak and F.~Wilczek, Nucl. Phys. B{\bf 479}, 529 (1996).
\bibitem{rr}
N. Read and E. Rezayi, Phys. Rev. B {\bf 54}, 16864 (1996).
\bibitem{gfn}
V. Gurarie, M. Flohr and C. Nayak, Nucl. Phys. B {\bf 498}, 513 (1998).
\bibitem{gn}
V. Gurarie and C. Nayak, Nucl. Phys. B (1998); [cond-mat/9706227].
\bibitem{frad}
E. Fradkin, C. Nayak, A. Tsvelik, and F. Wilczek, Nucl. Phys. B
{\bf 516}, 704 (1998); E. Fradkin, C. Nayak, and K. Schoutens, Nucl. Phys. B
{\bf 546}, 711 (1999); D.C. Cabra, E. Fradkin, G.L. Rossini, and F.A.
Schaposnik, cond-mat/9905192; X.-G. Wen, cond-mat/9811111.
\bibitem{read90}
N. Read, Phys. Rev. Lett. {\bf 65}, 1502 (1990).
\bibitem{blokwen}
B. Blok and X.-G. Wen, \prb {\bf 42}, 8133, 8145 (1990); {\it ibid.}
{\bf 43}, 8337 (1991).
\bibitem{degennes}
P.G. de Gennes, {\it Superconductivity of Metals and Alloys}, (Addison-Wesley,
Reading, MA, 1989).
\bibitem{oppermann}
R. Oppermann, Physica A {\bf 167}, 301 (1990).
\bibitem{az}
A. Altland and M. R. Zirnbauer, Phys. Rev. B {\bf 55}, 1142 (1997);
M. R. Zirnbauer, J. Math. Phys. {\bf 37}, 4986 (1996).
\bibitem{sfbn}
T. Senthil, M.P.A. Fisher, L. Balents, and C. Nayak, Phys. Rev. Lett. {\bf 81},
4704 (1998).
\bibitem{bundschuh}
R. Bundschuh, C. Cassanello, D. Serban, and M.R. Zirnbauer, Phys.
Rev. B {\bf 59}, 4382 (1999).
\bibitem{senthil1}
T. Senthil and M.P.A. Fisher, cond-mat/9810238.
\bibitem{chalker}
V. Kagalovsky, B. Horovitz, Y. Avishai, and J.T. Chalker, \prl {\bf 82}, 3516
(1999).
\bibitem{senthil2}
T. Senthil, J.B. Marston, and M.P.A. Fisher, cond-mat/9902062.
\bibitem{glr}
I.A.~Gruzberg, A.W.W.~Ludwig, and N.~Read, \prl {\bf 82}, 4524 (1999).
\bibitem{laughlin98}
R.B. Laughlin, \prl {\bf 80}, 5188 (1998).
\bibitem{ish}
J. Goryo and K. Ishikawa, cond-mat/9812412.
\bibitem{readgreen}
N. Read and D. Green, Bull. Am. Phys. Soc. {\bf 44}, 304 (1999).
\bibitem{sfw}
Some of our results have also been found independently by T. Senthil and M.P.A.
Fisher (unpublished), and by X.-G. Wen (unpublished).
\bibitem{anderson58}
P.W. Anderson, Phys. Rev. {\bf 110}, 827 (1958); {\it ibid.}, {\bf 112}, 1900
(1958).
\bibitem{hald85}
F.D.M. Haldane, Phys. Rev. Lett. {\bf 55}, 2095 (1985).
\bibitem{hopriv}
We are grateful to T.-L. Ho for suggesting that we consider a
domain wall as a model for the edge.
\bibitem{jackreb}
R. Jackiw and C. Rebbi, Phys. Rev. D {\bf 13}, 3398 (1976).
\bibitem{boundstates}
C.~Caroli, P.G.~de~Gennes, and J. Matricon, Phys. Lett. {\bf 9}, 307 (1964);
J.~Bardeen, R.~K\"{u}mmel, A.E.~Jacobs, and L.~Tewordt, Phys. Rev. {\bf 187},
556 (1969); P.G.~de~Gennes, Sec. 5.2 in Ref.\ \cite{degennes}.
\bibitem{kopnin}
N.B. Kopnin and M.M. Salomaa, \prb {\bf 44}, 9667 (1991).
\bibitem{gsw}
See a general relativity text, or, {\it e.g.}, M.B. Green, J.H. Schwartz, and E.
Witten, {\it Superstring Theory, Vol. 2: Loop Amplitudes, Anomalies, and
Phenomenology}, (Cambridge University Press, Cambridge, England, 1987), Sec.\
12.1.
\bibitem{wn}
X.-G. Wen and Q. Niu, \prb {\bf 41}, 9377 (1990).
\bibitem{ho}
T.-L. Ho, \prl {\bf 75}, 1186 (1995).
\bibitem{belkhir}
L. Belkhir, X.-G. Wu, and J.K. Jain, \prb {\bf 48}, 15245 (1993).
\bibitem{milr2}
M. Milovanovi\'{c} and N. Read, \prb {\bf 56}, 1461 (1997).
\bibitem{leekane}
D.-H. Lee and C.L. Kane, \prl {\bf 64}, 1313 (1990).
\bibitem{hrunpub}
F.D.M. Haldane and E.H. Rezayi, unpublished.
\bibitem{rrapp}
See Appendix B of Ref.\ \cite{rr}.
\bibitem{grr}
D. Green, N. Read and E. Rezayi, (unpublished).
\bibitem{wenrev}
X.-G.~Wen, Int. J. Mod. Phys. B{\bf 6}, 1711 (1992).
\bibitem{haldgirv} F.D.M. Haldane and D.P. Arovas, \prb
{\bf 52}, 4223 (1995); K. Yang, L.K. Warman, and S.M. Girvin, \prl
{\bf 70}, 2641 (1993).
\bibitem{tkndn}
D.J. Thouless, M. Kohmoto, M.P. Nightingale, and M. Den Nijs, \prl {\bf 49}, 405
(1982).
\bibitem{bel}
J. Bellissard, A. van Elst, and H. Schulz-Baldes, J. Math. Phys. {\bf 35}, 5373
(1994).
\bibitem{rahw}
A.N. Redlich, \prl {\bf 52}, 18 (1984); I. Affleck, J. Harvey, and E. Witten,
Nucl. Phys. B {\bf 206}, 413 (1982).
\bibitem{halp82}
B.I. Halperin, \prb {\bf 25}, 2185 (1982).
\bibitem{hald95}
F.D.M. Haldane, \prl {\bf 74}, 2090 (1995).
\bibitem{xrs}
S. Xiong, N. Read, and A.D. Stone, \prb {\bf 56}, 3982 (1997).
\bibitem{calhar}
C. Callan and J. Harvey, Nucl. Phys. B {\bf 250}, 427 (1985).
\bibitem{vol98}
G.E. Volovik, Pis'ma ZETF {\bf 66}, 492 (1997); cond-mat/9709084.
\bibitem{wit84}
E. Witten, Commun. Math. Phys. {\bf 92}, 455 (1984).
\bibitem{kflr}
C.L. Kane and M.P.A. Fisher, \prb {\bf 55}, 15832 (1997).
\bibitem{ca}
H.W.J. Bl\"{o}te, J.L. Cardy, and M.P. Nightingale, \prl {\bf 56}, 742 (1986);
I. Affleck, {\it ibid.}, {\bf 56}, 746 (1986).
\bibitem{kirczenow}
L.G.C. Rego and G. Kirczenow, Bull. Am. Phys. Soc. {\bf 44}, 212 (1999).
\bibitem{bpz}
A. Belavin, A. Polyakov, and A. Zamolodchikov, Nucl.
Phys. B{\bf 241}, 33 (1984).
\bibitem{alwit}
L. Alvarez-Gaum\'{e} and E. Witten, Nucl. Phys. B {\bf 234}, 269 (1983); L.
Alvarez-Gaum\'{e}, S. Della Pietra, and G. Moore, Ann. Phys. (NY) {\bf 163}, 288
(1985).
\bibitem{herver}
H. Verlinde, Nucl. Phys. B {\bf 337}, 652 (1990).
\bibitem{halwu}
F.D.M. Haldane and Y.-S. Wu, \prl {\bf 55}, 2887 (1985).
\bibitem{pruisken}
See A.M.M. Pruisken, in \cite{book}, and references therein.
\bibitem{sfnew}
T. Senthil and M.P.A. Fisher, cond-mat/9906290.
\bibitem{glrunpub}
I.A.~Gruzberg, A.W.W.~Ludwig, and N.~Read, unpublished.
\bibitem{helgason}
S. Helgason, {\it Differential Geometry, Lie Groups, and Symmetric Spaces},
(Academic Press, New York, NY, 1978), Sections IV.4, V.2.
\bibitem{ludwig}
A.W.W.~Ludwig, M.P.A.~Fisher, R.~Shankar, and G.~Grinstein, \prb {\bf 50}, 7526
(1994).
\bibitem{rbim}
Vl.S Dotsenko and V.S. Dotsenko, Adv. Phys. {\bf 32}, 129 (1983);
R. Shankar, \prl {\bf 58}, 2466 (1987); A.W.W. Ludwig, \prl {\bf 61}, 2388
(1988).
\bibitem{kadanoff}
L. Kadanoff and H. Ceva, \prb {\bf 3}, 3918 (1971).
\bibitem{slh}
A.V. Shytov, L.S. Levitov, and B.I. Halperin, \prl {\bf 80}, 141 (1998).

\end{references}
\end{document}